\documentclass[a4paper,11pt]{article}
\pdfoutput=1 

\usepackage{jheppub} 
\usepackage{amsmath,amssymb}
\usepackage{graphicx}
\usepackage{subcaption}
\usepackage[percent]{overpic}
\usepackage{color}
\usepackage{bm}

\definecolor{purple}{rgb}{0.7,0,0.7}

\def\be{\begin{equation}}
\def\ee{\end{equation}}
\def\bea{\begin{eqnarray}}
\def\eea{\end{eqnarray}}
\def\bal{\begin{align}}
\def\eal{\end{align}}

\newcommand\bR{\mathbb{R}}

\newcommand\bS{\mathbb{S}}
\newcommand\bZ{\mathbb{Z}}
\newcommand\bC{\mathbb{C}}
\newcommand\bI{\mathbb{I}}

\newcommand\cN{\mathcal{N}}

\newcommand\cB{\mathcal{B}}

\newcommand\cR{\mathcal{R}}

\newcommand\cG{\mathcal{G}}

\newcommand\cS{\mathcal{S}}

\newcommand\dd{\mathrm{d}}
\newcommand\ex{\mathrm{e}}
\newcommand\ii{\mathrm{i}}

\newcommand\qqq{\qquad\qquad}
\newcommand{\nn}{\nonumber \\ {} }

\newcommand{\CS}{\mathcal{S}}
\newcommand{\CC}{C}

\newcommand{\IC}{\mathbb{C}}

\title{BPS Graphs: From Spectral Networks to BPS Quivers}

\author[a]{Maxime Gabella,}
\author[b]{Pietro Longhi,}
\author[c]{Chan Y.~Park,}
\author[d]{and Masahito Yamazaki}

\affiliation[a]{Institute for Advanced Study, Einstein Drive, Princeton, NJ, USA}
\affiliation[b]{Department of Physics and Astronomy, Uppsala University, Uppsala, Sweden}
\affiliation[c]{NHETC and Department of Physics and Astronomy, Rutgers University, NJ, USA}
\affiliation[d]{Kavli IPMU (WPI), University of Tokyo, Kashiwa, Chiba, Japan}

\emailAdd{gabella@ias.edu}
\emailAdd{pietro.longhi@physics.uu.se}
\emailAdd{chan@physics.rutgers.edu}
\emailAdd{masahito.yamazaki@ipmu.jp}

\preprint{UUITP-11/17, IPMU17-0055}

\abstract{
We define ``BPS graphs'' on punctured Riemann surfaces associated with $A_{N-1}$ theories of class $\mathcal{S}$. BPS graphs provide a bridge between two powerful frameworks for studying the spectrum of BPS states: spectral networks and BPS quivers. They arise from degenerate spectral networks at maximal intersections of walls of marginal stability on the Coulomb branch. While the BPS spectrum is ill-defined at such intersections, a BPS graph captures a useful basis of elementary BPS states.  The topology of a BPS graph encodes a BPS quiver, even for higher-rank theories and for theories with certain partial punctures. BPS graphs lead to a geometric realization of the combinatorics of Fock-Goncharov $N$-triangulations and generalize them in several ways.
}

\begin{document} 

\maketitle
\flushbottom

\begin{quote}
\textit{``Mille viae ducunt homines per saecula Romam \\
Qui Dominum toto quaerere corde volunt.''}%
\footnote{``A thousand roads lead forever to Rome the men who seek the Lord with all their heart.''}
\begin{flushright}
\sc{Alain de Lille}, \emph{Liber Parabolarum} (1175)
\end{flushright}
\end{quote}

\section{Introduction}

Studying the spectrum of BPS states has led to many exact non-perturbative results about four-dimensional gauge theories with $\cN=2$ supersymmetry.
BPS states are defined for a theory on the Coulomb branch $\cB$ of the moduli space of vacua, where the gauge symmetry is spontaneously broken to a Cartan torus $U(1)^r$.
A BPS state of charge $\gamma$ saturates the bound $M \ge |Z_\gamma|$, with $M$ its mass and $Z_\gamma$ its $\cN=2$ central charge.
The spectrum of BPS states is piecewise constant on $\cB$ but can jump across \emph{walls of marginal stability}, when BPS bound states appear from the interaction of two BPS states whose central charges have the same phase, $\arg Z_{\gamma_1} =\arg Z_{\gamma_2}$.
This wall-crossing phenomenon played an important role in the development of Seiberg-Witten theory~\cite{Seiberg:1994rs,Seiberg:1994aj}.

Powerful frameworks for the study of BPS spectra in 4d $\cN=2$ theories have been developed in recent years.
They split mainly into two categories: a geometric method based on \emph{spectral networks}~\cite{Gaiotto:2009hg,Gaiotto:2012rg,Gaiotto:2012db}, and an algebraic method based on \emph{BPS quivers}~\cite{Cecotti:2010fi,Cecotti:2011rv,Alim:2011ae,Alim:2011kw}.
In this paper, we introduce the concept of a \emph{BPS graph}, which bridges the gap between spectral networks and BPS quivers.

More specifically, we consider 4d $\cN=2$ theories of class~$\CS$~\cite{Gaiotto:2009we, Gaiotto:2009hg} that are constructed by compactifying the 6d $(2,0)$ superconformal field theory of type $A_{N-1}$ on a Riemann surface~$\CC$ with punctures. 
A spectral network on~$\CC$ consists of a collection of oriented paths, called $\cS$-\emph{walls} (or simply \emph{walls}),
associated with an $N$-fold branched covering~$\Sigma \to \CC$. 
The cover~$\Sigma$ is the spectral curve of the Hitchin system related to the theory and is identified with the Seiberg-Witten curve, whose geometry varies over the Coulomb branch~$\cB$.
The BPS spectrum in a generic vacuum $u\in \cB$ can in principle be worked out by looking at a family of spectral networks labeled by a phase $\vartheta$. 
At critical values $\vartheta_\gamma$, the topology of the spectral network degenerates as some of its walls overlap entirely, forming \emph{double walls}.
Such a double wall, which lifts to a closed one-cycle $\gamma$ on $\Sigma$, signals the presence of a stable BPS state with charge $\gamma$ and central charge of phase $\arg Z_\gamma= \vartheta_\gamma$. 
BPS states indeed arise from calibrated M2-branes ending along one-cycles on an M5-brane that wraps $\mathbb{R}^{1,3}\times \Sigma$.
However, for a generic point $u\in \cB$ of the Coulomb branch, spectral networks are too complicated to implement this procedure efficiently, especially for higher-rank theories with $N>2$.

While the conventional way to deal with wall-crossing phenomena is to avoid walls of marginal stability in the Coulomb branch,  we do the exact opposite and focus on the \emph{maximal intersection}~$\cR$ of walls of marginal stability, where the central charges of \emph{all} BPS states have the same phase $\vartheta_c$. It is then very easy to keep track of all double walls in the corresponding spectral network, since they all appear simultaneously. 
Of course, by moving to~$\cR$, we lose some information about the stability of BPS bound states, and the BPS spectrum becomes somewhat ill-defined. As we will argue, this nevertheless gives us a direct access to some essential wall-crossing invariant information about the associated theory of class~$\CS$. 

\begin{figure}[h!]
\centering
\begin{minipage}{.28\textwidth}
\centering
\includegraphics[width=\linewidth]{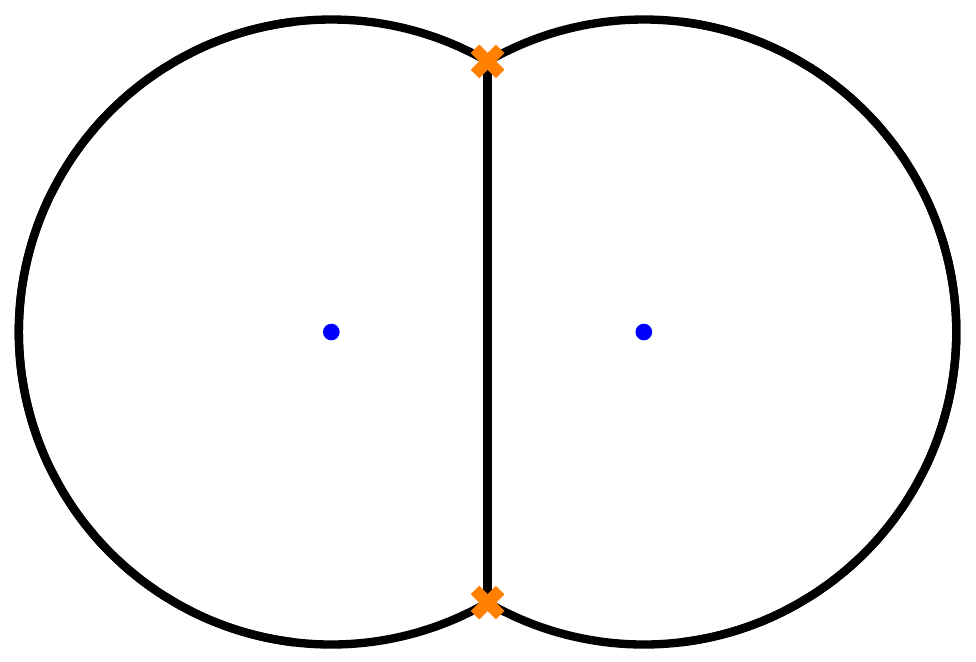}
\end{minipage}\hfill
\begin{minipage}{.32\textwidth}
\centering
\includegraphics[width=\linewidth]{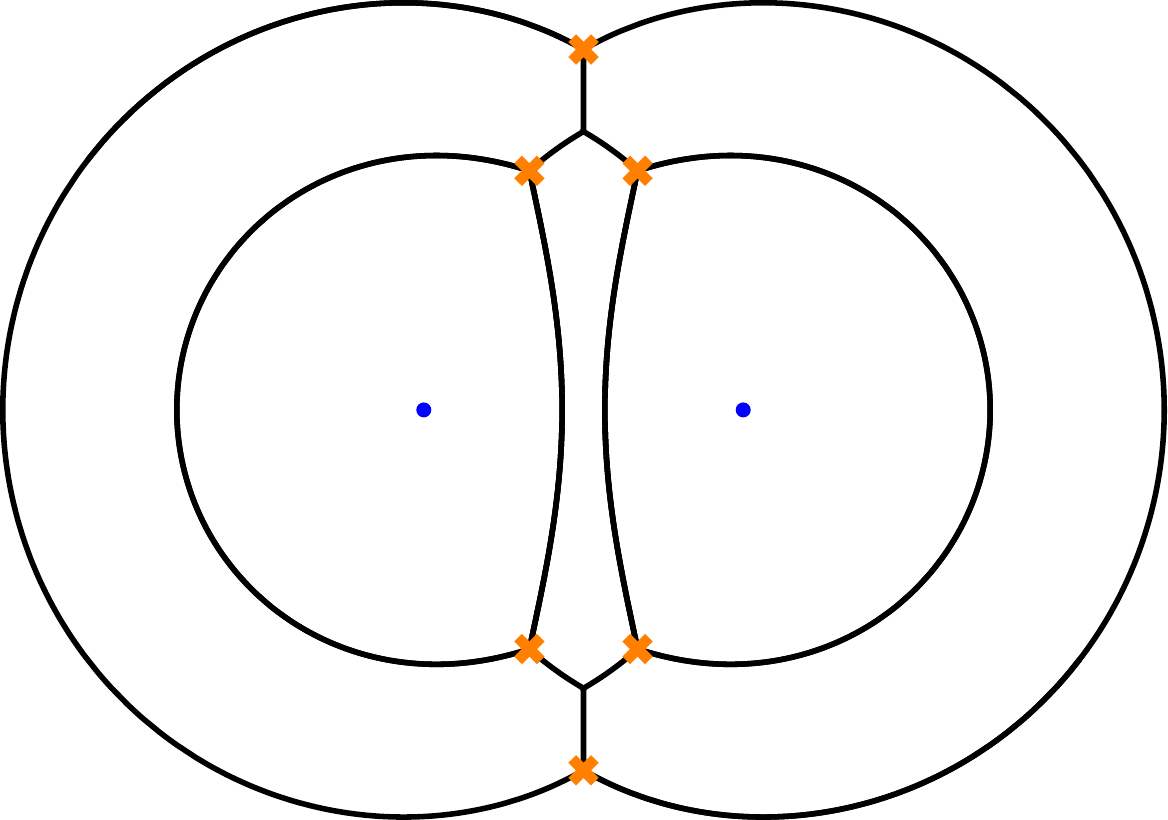}
\end{minipage}\hfill
\begin{minipage}{.36\textwidth}
\centering
\includegraphics[width=\linewidth]{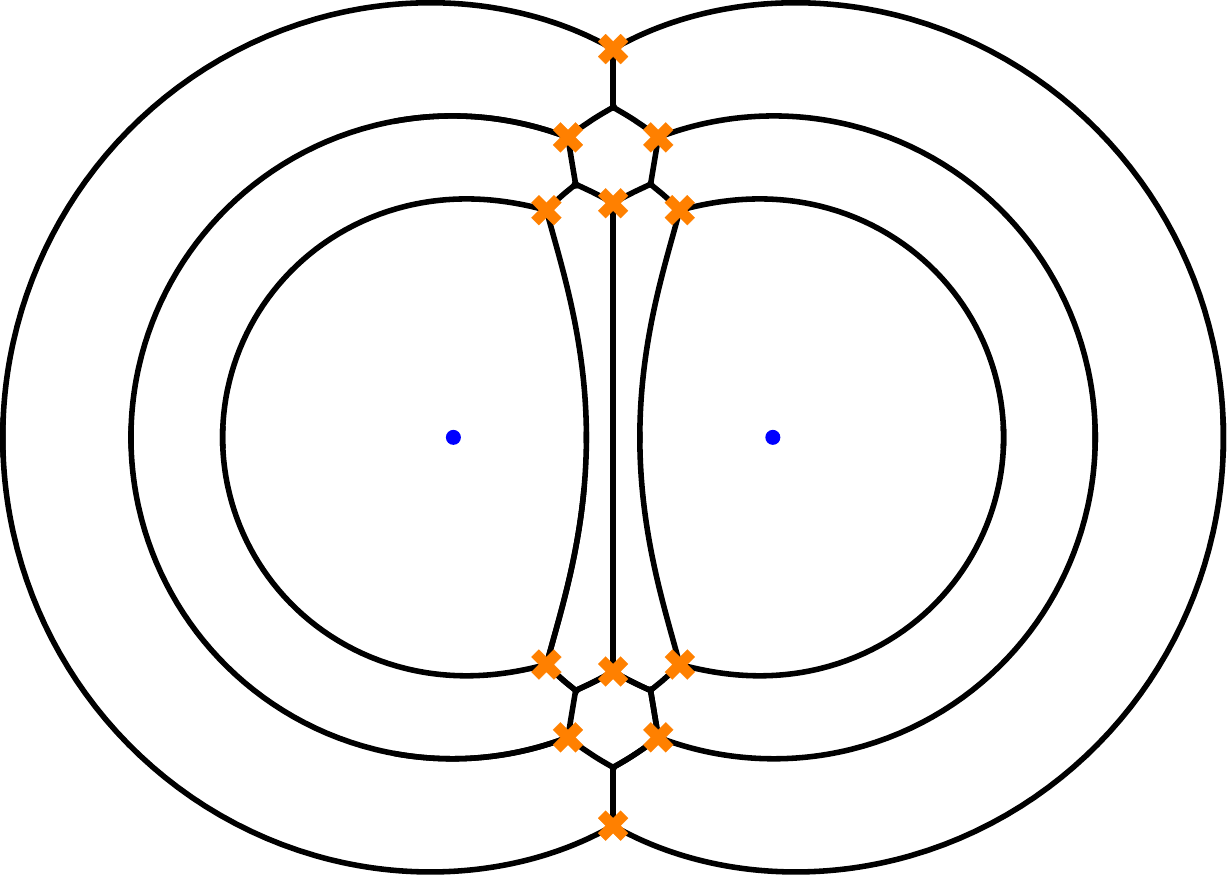}
\end{minipage}
    \caption{BPS graphs on the 3-punctured sphere for the $T_N$ theories with $N=2,3,4$.}
    \label{graphsT234}
\end{figure}

More specifically, among all the double walls that appear in the maximally degenerate spectral network at~$\vartheta_c$, there is a distinguished set that corresponds to \emph{elementary} BPS states, in the sense that they provide a \emph{positive basis} for the charge lattice~$\Gamma = H_1(\Sigma,\bZ)$.%
\footnote{The notion of ``positivity'' is related to the half-plane centered on~$\ex^{\ii\vartheta_c}$ in the complex plane of central charges~$Z_\gamma$.}
We define the \emph{BPS graph} as this distinguished set of double walls.
Some examples of BPS graphs on the 3-punctured sphere associated with the $T_N$ theories~\cite{Gaiotto:2009we} for $N=2,3,4$ are plotted in Figure~\ref{graphsT234}. 

An important wall-crossing invariant of an $\cN=2$ theory is the (mutation class of the) BPS quiver, from which the BPS spectrum in any chamber of the Coulomb branch~$\cB$ can in principle be computed by solving a quantum mechanics problem.
The \emph{topology} of the BPS graph naturally encodes the BPS quiver: edges corresponds to nodes, and intersections on the cover~$\Sigma$ to arrows. 
For $A_1$ theories of class~$\CS$, this relation between BPS graphs and BPS quivers is clear from their common relation to ideal triangulations of the Riemann surface~$\CC$.
For higher-rank $A_{N-1}$ theories, we show that it passes non-trivial checks and confirms the validity of conjectural BPS quivers~\cite{Alim:2011kw, Xie:2012dw}.

The existence of the maximal intersection~$\cR$ where BPS graphs are defined is guaranteed for ``complete'' $A_1$ theories~\cite{Cecotti:2011rv}, but we do not have a general proof for all $A_{N-1}$ theories of class~$\CS$. 
We will however present many explicit examples of $A_{N-1}$ BPS graphs on spheres with three and four punctures.
We are not limited to theories with \emph{full} punctures, but can also consider certain types of \emph{partial} punctures. 

Motivated by the pattern emerging from these examples, we make a conjecture about the topological structure of BPS graphs for all $A_{N-1}$ theories of class~$\CS$ with punctures of type $[k,1,\ldots,1]$ for any integer $k$. For theories with full punctures ($k=1$), certain candidate BPS graphs are dual to the $N$-triangulations of~$\CC$ used by Fock and Goncharov to study higher Teichm\"uller theory~\cite{FockGoncharovHigher}, which were shown to be related to spectral networks in~\cite{Gaiotto:2012db}. However, they are part of a much more general class of BPS graphs, related to each other by sequences of elementary local transformations (\emph{flip} and \emph{cootie} moves). We show that these BPS graphs are dual to the ``ideal bipartite graphs'' recently defined by Goncharov~\cite{2016arXiv160705228G}.
In the presence of some partial punctures, BPS graphs provide a geometric justification for the combinatorial patterns conjectured in~\cite{Xie:2012dw}.

\bigskip

The paper is organized as follows.
After a short review of spectral networks in Section~\ref{Spectral}, we define BPS graphs
in Section~\ref{BPSgraphs}, and discuss their existence and general properties.
We then present examples of BPS graphs in $A_1$ theories (Section~\ref{sec:A1-graphs}), 
$A_{N-1}$ theories with full punctures (Section~\ref{sec:N-lift-graphs}), and $A_{N-1}$ theories with certain partial punctures (Section~\ref{PartialP}). In all cases, we make conjectures about the topology of general BPS graphs, building on a relation to $N$-triangulations.
In Section~\ref{BPSquivers}, we explain how to obtain BPS quivers with superpotentials from BPS graphs.
Finally, we mention some further stimulating applications of BPS graphs in Section~\ref{secApplications}.

\section{Review of spectral networks} \label{Spectral}

We first review some basic aspects of spectral networks and BPS states in 4d $\cN=2$ theories of class~$\CS$, in preparation for our definition of BPS graphs in the next section.

\subsection{\texorpdfstring{Spectral curves for theories of class~$\CS$}{Spectral curves for class S}} \label{SECspectralCurve}

Let $C$ be an oriented real surface with genus $g$ and $n$ punctures (internal marked points), potentially with a boundary, each connected component of which is a copy of $S^1$ with at least one marked point.
The Seiberg-Witten curve $\Sigma$ of an $A_{N-1}$ theory of class~$\CS$ associated with~$\CC$ is the spectral curve of an $A_{N-1}$ Hitchin system~\cite{Hitchin}:
\bea\label{eq:spectral-curve}
\Sigma = \{\det \left( \lambda - \varphi \right) = 0 \}  \subset T^*\CC \; .
\eea
Here $\lambda$ is the tautological one-form on $T^*\CC$ whose pull-back to $\Sigma$ gives the Seiberg-Witten differential, and the Higgs field
$\varphi$ is a one-form on $C$ with values in the Lie algebra $\mathfrak{sl}_{N}$. 
The determinant in~\eqref{eq:spectral-curve} is taken in the fundamental ($N$-dimensional) representation of $A_{N-1}$, so that $\lambda$ is multiplied by the unit matrix and $\varphi$ is matrix-valued.
Expanding~\eqref{eq:spectral-curve} in $\lambda$ gives a presentation of $\Sigma$ as an $N$-fold branched cover of $C$:
\bea \label{SigmaNfold}
\Sigma = \Big\{ \lambda : \lambda^N + \sum_{k=2}^N \phi_k \lambda^{N-k} = 0 \Big\} \subset T^*\CC \;,
\eea 
where the $\phi_k$ are $k$-differentials corresponding to Casimir invariants of $\varphi$. The Coulomb branch $\cB$ of an $A_{N-1}$ theory of class~$\CS$ is parameterized by tuples $u = \left( \phi_2, \ldots, \phi_N\right)$ of $k$-differentials with prescribed singularities at the punctures.

The Seiberg-Witten curve $\Sigma$ has the property that its homology one-cycles $\gamma \in H_1(\Sigma,\bZ)$ are naturally identified with electromagnetic and flavor charges $\gamma\in\Gamma$ in the IR effective abelian gauge theory defined at $u\in \cB$.
Strictly speaking, the charge lattice $\Gamma$ is not quite $H_1(\Sigma,\bZ)$, but a certain sub-quotient of it~\cite{Gaiotto:2009hg}. 
The charge lattice is equipped with an antisymmetric electromagnetic pairing $\langle\cdot,\cdot \rangle: \Gamma \times \Gamma  \to \bZ$ given by the intersection form on $H_1(\Sigma,\bZ)$.
A 4d state with charge $\gamma$ arises from a string on the corresponding one-cycle, and its central charge is obtained by integrating the Seiberg-Witten one-form~$\lambda$ along~$\gamma$:  
\bea \label{ZMgamma}
Z_\gamma = \frac1\pi \oint_\gamma \lambda  .
\eea

The definition of a theory of class~$\CS$ requires a choice of boundary conditions for Hitchin's equations at the punctures of $C$~\cite{Gaiotto:2009we, Nanopoulos:2009uw}.
At a \emph{regular} puncture, the Higgs field has a simple pole
\be \label{eq:regular-puncture}
\varphi  \sim \frac 1 z \begin{pmatrix} m_1 & &  &\\ & m_2 & & \\ & & \ddots & \\ & & & m_N \end{pmatrix} \dd z + \cdots \;, 
\ee
where $\sim$ denotes equivalence up to conjugation by an $SL(N)$ matrix, and the omitted terms are non-singular. The parameters $m_i$ are identified as the complex UV mass parameters 
of the 4d $\mathcal{N}=2$ theory, and satisfy $\sum_{i=1}^N m_i = 0$, in order for $\varphi$ to be an element of $\mathfrak{sl}_N$.%
\footnote{The behavior of $\varphi$ at a massless regular puncture is slightly different.}
The condition \eqref{eq:regular-puncture} on the Higgs field translates into conditions for the $k$-differentials:
\bea
\phi_k \sim \frac{M_k}{z^k} (\dd z)^k+ \cdots \;, 
\eea
where $M_k$ are symmetric polynomials of order $k$ in the mass parameters $m_i$, given by 
\bea
M_k  = (-1)^k \sum_{i_1<\cdots <i_k}^N m_{i_1}   \cdots m_{i_k}   \;.
\eea
We can also consider \emph{irregular} punctures, where the $k$-differentials have singularities of order larger than $k$~\cite{Gaiotto:2009hg, Xie:2012hs, Gaiotto:2012db}.

The singularity in \eqref{eq:regular-puncture} is known as a full singularity when $m_i\neq m_j$,
and the associated puncture as a \emph{full} puncture.
More generally, when some of the eigenvalues $m_i$ coincide, the puncture is known as a \emph{partial} puncture. In $A_{N-1}$ theories, punctures are classified by Young diagrams with $N$ boxes: the height of each column of the diagram encodes the number of identical eigenvalues (in particular, a full puncture corresponds to a Young diagram with a single row of $N$ boxes). 
The flavor symmetry associated with a generic puncture is 
\be
	S\left[ U\left(n_1\right)\times \cdots\times U\left(n_k\right)\right] \! \;,
\ee
where $n_1,\dots, n_k$ count columns of the Young diagram with the same height.
We can therefore read off the rank of the flavor charge lattice to be
\be\label{eq:flavor-rank}
	f = \sum_{a\in \text{punctures}} (n^{(a)}_1 + \cdots + n^{(a)}_k - 1)\;.
\ee
Similarly, the rank of the gauge charge lattice is $2r$, where $r$ is the complex dimension of the Coulomb branch $\cB$. It is given by
\be\label{eq:gauge-rank}
	r = {\rm dim}_\IC\, \cB = \sum_{k=2}^{N} d_k \;,
\ee
where $d_k$ is the dimension of the moduli space of $k$-differentials with prescribed singularities at the punctures of~$\CC$, 
\be
	d_k = (2k-1)(g-1) + \sum_{a\in\text{punctures}} (k-h^{(a)}_k)\;,
\ee
with $h^{(a)}_k$ the height of the $k$-th box in the Young diagram associated with puncture $a$ 
(boxes are counted row-wise, beginning from the longest row, as in~\cite{Tachikawa:2013kta}).
The rank of the lattice $\Gamma$ of gauge and flavor charges is then
\be
\text{rank}\, \Gamma = 2r + f .
\ee

As a simple example, let us consider the Riemann sphere with three full punctures at $z_a,z_b,z_c \neq \infty$. 
The $k$-differentials then take the form
\bea\label{phik}
\phi_k = \frac{M_k^az_{ab}^{k-1}z_{ac}^{k-1}(z\! -\!z_b)(z\!-\!z_c)  +\text{cyclic} +P_{k-3}(z)\, (z\! -\! z_a) (z \!- \!z_b) (z\! -\! z_c)} {(z-z_a)^k(z-z_b)^k(z-z_c)^k  } (\dd z)^k \;, 
\eea
where $z_{ab} \equiv z_a-z_b$, and ``cyclic'' stands for the two additional terms obtained by cyclic permutations of $a,b,c$. 
The polynomials $P_{k-3}(z)$ are of order $k-3$ and contain $k-2$ Coulomb branch parameters (we have $P_{-1}\equiv0$).
This is determined by the requirement that there be no singularity at $z=\infty$, which imposes that $\phi_k$ decays as $z^{-2k}$ or faster at infinity. 
There is a total of $f=3(N-1)$ mass parameters $m_i^{a,b,c}$, and $r=\sum_{k=2}^N (k-2)=\frac12 (N-1)(N-2)$ Coulomb branch parameters, in agreement with~\eqref{eq:flavor-rank} and~\eqref{eq:gauge-rank}, which gives  $\text{rank}\, \Gamma= N^2-1$.

\subsection{Spectral networks}

\begin{figure}[tb]
\begin{center}
\begin{overpic}[width=0.3\textwidth]{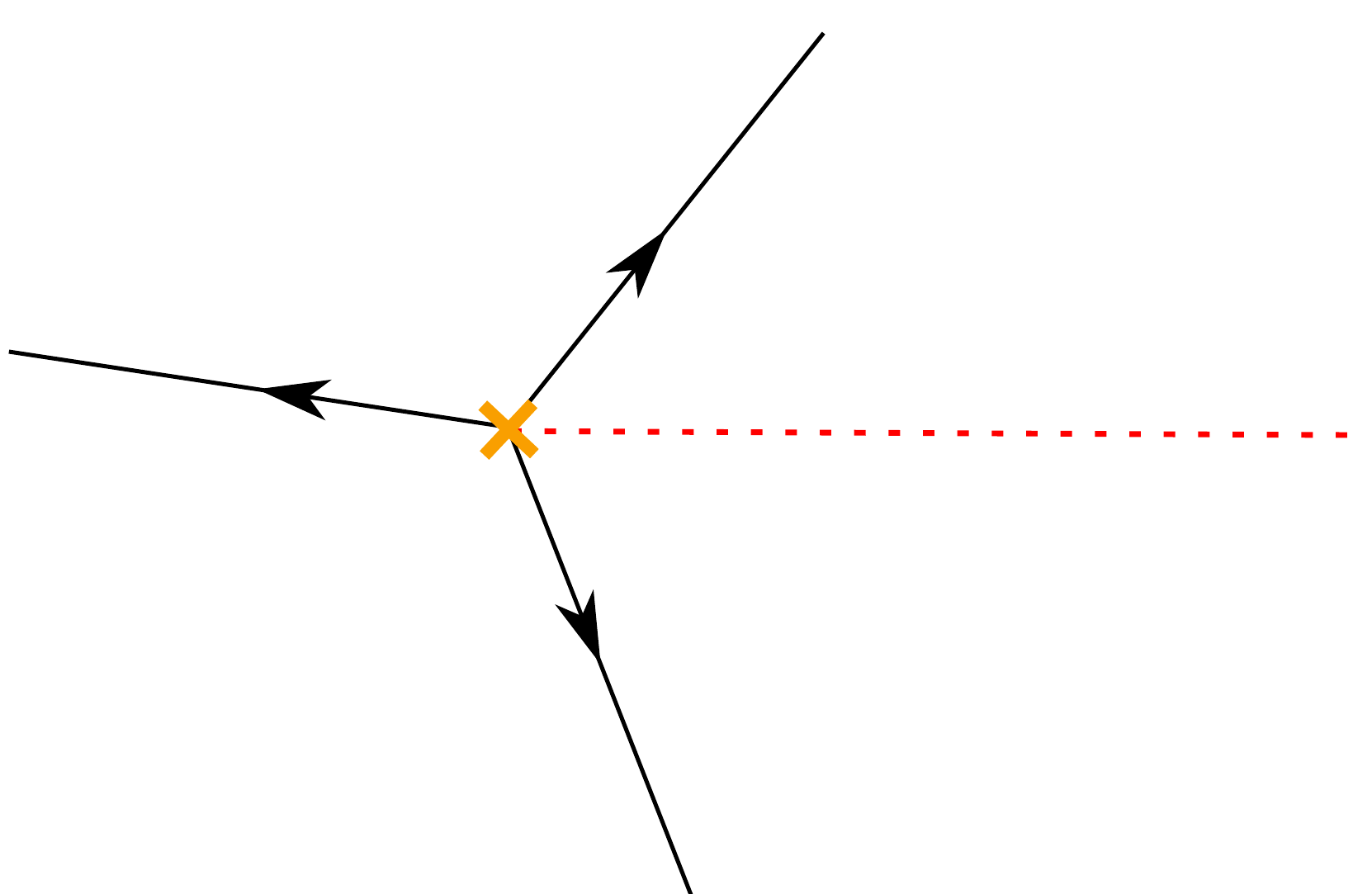}
 	\put (10,42) {$ij$}
	\put (50,10) {$ji$}
	\put (45,55) {$ji$}
\end{overpic}
\caption{Generic behavior of a spectral network near a branch point (orange cross): three walls emerge, with labels $ij$ or $ji$ corresponding to sheets $i$ and $j$ in a trivialization of the branched covering $\Sigma\to \CC$ (the orange dashed line indicates a branch cut).}
\label{fig:network-branchpoint}
\end{center}
\end{figure}

\begin{figure}[htb]
\begin{center}
\begin{overpic}[width=0.35\textwidth]{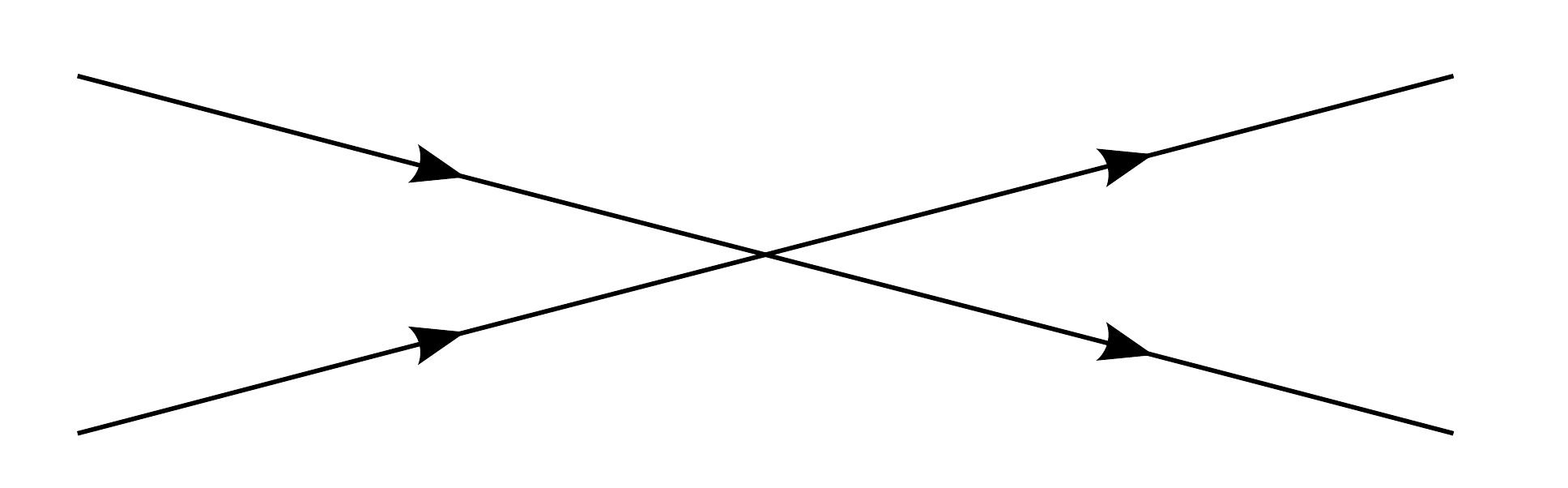}
 	\put (8, 8) {$ij$}
	\put (8, 27) {$kl$}
	\put (85, 8) {$kl$}
	\put (85, 27) {$ij$}
\end{overpic}
\qqq\qqq
\begin{overpic}[width=0.35\textwidth]{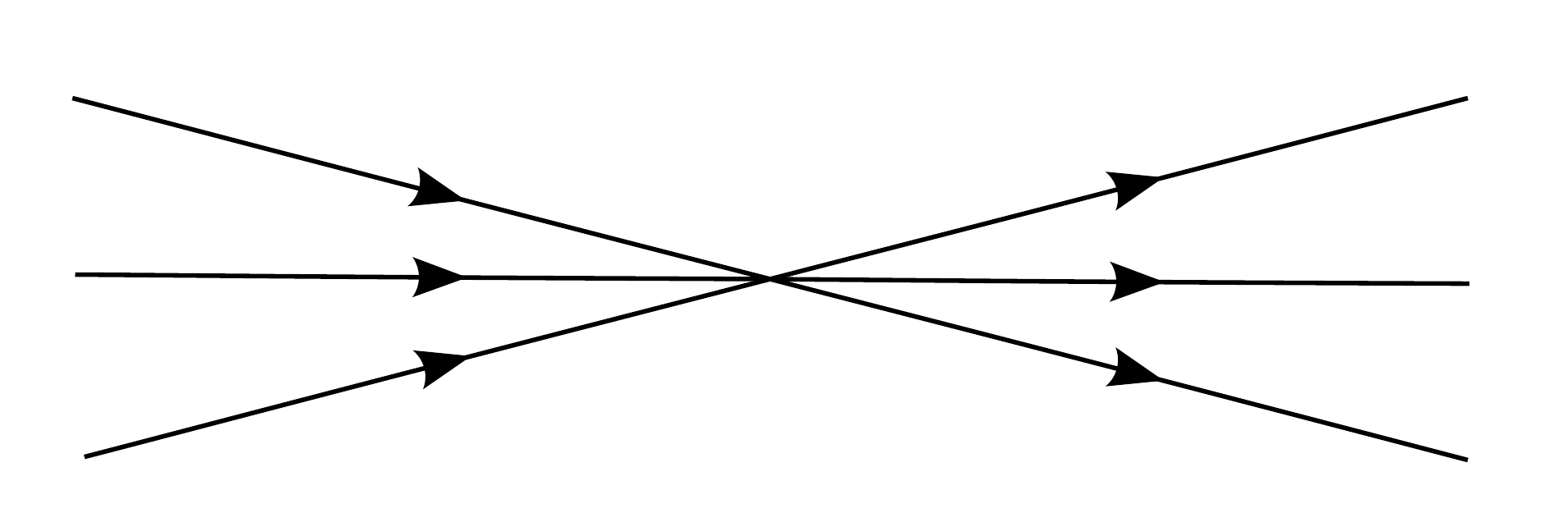}
 	\put (8, 8) {$ij$}
	\put (8, 27) {$jk$}
	\put (85, 8) {$jk$}
	\put (85, 27) {$ij$}
	\put (8, 17) {$ik$}
	\put (85, 17) {$ik$}
\end{overpic}
\caption{\emph{Left}: No new wall is created at the intersection of an $ij$-wall and a $kl$-wall with $i,j,k, l$ all different. \emph{Right}: An $ik$-wall can start or end
at the intersection of an $ij$-wall and a $jk$-wall.}
\label{fig:network-intersections}
\end{center}
\end{figure}
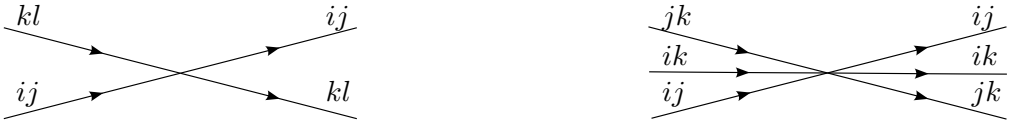

A \emph{spectral network} on a punctured Riemann surface~$\CC$ is a geometric object associated with an $N$-fold branched covering $\pi:\Sigma\to \CC$~\cite{Gaiotto:2012rg}. 
It consists of oriented paths, called \emph{$\cS$-walls} (or simply \emph{walls} hereafter), that generically begin at branch points and end at punctures or at marked points on the boundary of~$\CC$. 
Although not strictly necessary, it is often useful to choose a trivialization of the covering, that is a choice of branch cuts on~$\CC$ and a choice of labels for the $N$ sheets of $\Sigma$.
Each wall of the network is then labeled by an ordered pair of integers $ij$, corresponding to sheets $i$ and $j$. 
A branch point where sheets $i$ and $j$ coincide%
\footnote{Branch points are assumed to be of square-root type, so that the monodromy around a branch point exchanges two sheets of $\Sigma$.}
sources three walls of types $ij$ or $ji$, as shown in Figure~\ref{fig:network-branchpoint}. 
In addition, for $N>2$, a wall of type $ik$ can also start or end at the intersection of an $ij$-wall and a $jk$-wall, as shown in Figure~\ref{fig:network-intersections}. 

The shape of an $ij$-wall is determined by the following differential equation: 
\be\label{eq:S-wall}
	(\partial_t , \lambda_i-\lambda_j) \ \in \ \ex^{\ii\vartheta}\mathbb{R}^+\,,
\ee
where $t$ is a parameter along the wall, and $\lambda_i$, $\lambda_j$ denote two solutions of the degree-$N$ polynomial equation~\eqref{SigmaNfold} that defines $\Sigma$.
A spectral network thus depends both on a point $u= (\phi_2, \ldots, \phi_N)$ on the Coulomb branch $\cB$, and on a phase $\vartheta\in \bR/2 \pi \bZ$.

Each wall carries some combinatorial \emph{soliton data}, which captures the degeneracies of \emph{2d-4d BPS states} associated with a \emph{canonical surface defect} $\bS_z$, labeled by a point $z\in \CC$~\cite{Gaiotto:2011tf}.
The 2d-4d coupled system has $N$ massive vacua, in one-to-one correspondence with the solutions $\lambda_1,\dots ,\lambda_N$ of \eqref{SigmaNfold}.
The soliton data of an $ij$-wall passing through~$z$ counts 2d-4d BPS solitons interpolating between vacua $\lambda_{i}(z)$ and $\lambda_{j}(z)$ with central charge of phase~$ \vartheta$.

\subsection{BPS states}\label{sec:bps-states}

As the phase $\vartheta$ varies, the shape of a spectral network transforms according to \eqref{eq:S-wall}. 
At certain critical phases $\vartheta_c$ the topology of the network can be degenerate. This often signals the existence of a stable 4d BPS state in the Coulomb vacuum $u\in \cB$. 
 
\begin{figure}
    \centering
    \begin{subfigure}[b]{0.3\textwidth}
        \includegraphics[width=\textwidth]{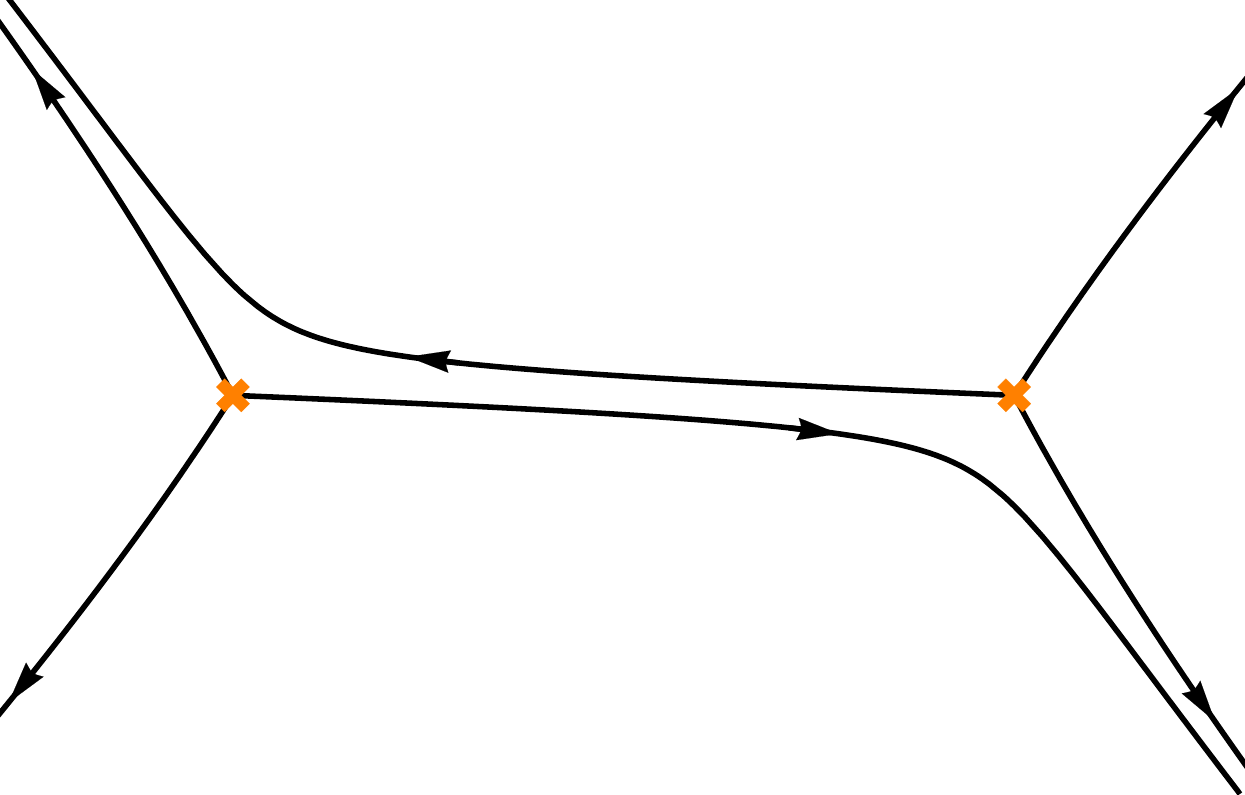}
        \caption{$\vartheta\lesssim\vartheta_c$}
        \label{DoubleWall0}
    \end{subfigure}
    \quad
    \begin{subfigure}[b]{0.3\textwidth}
        \includegraphics[width=\textwidth]{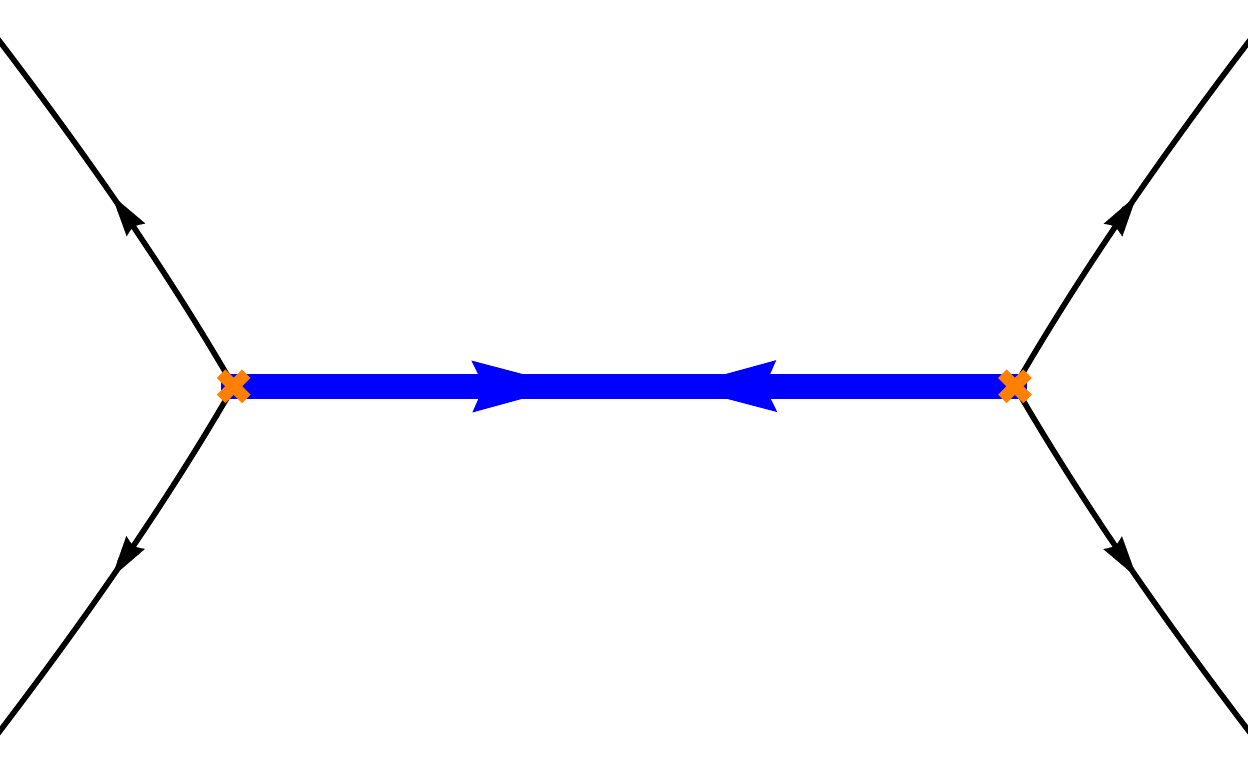}
        \caption{$\vartheta=\vartheta_c$}
        \label{DoubleWall1}
    \end{subfigure}
    \quad
    \begin{subfigure}[b]{0.3\textwidth}
        \includegraphics[width=\textwidth]{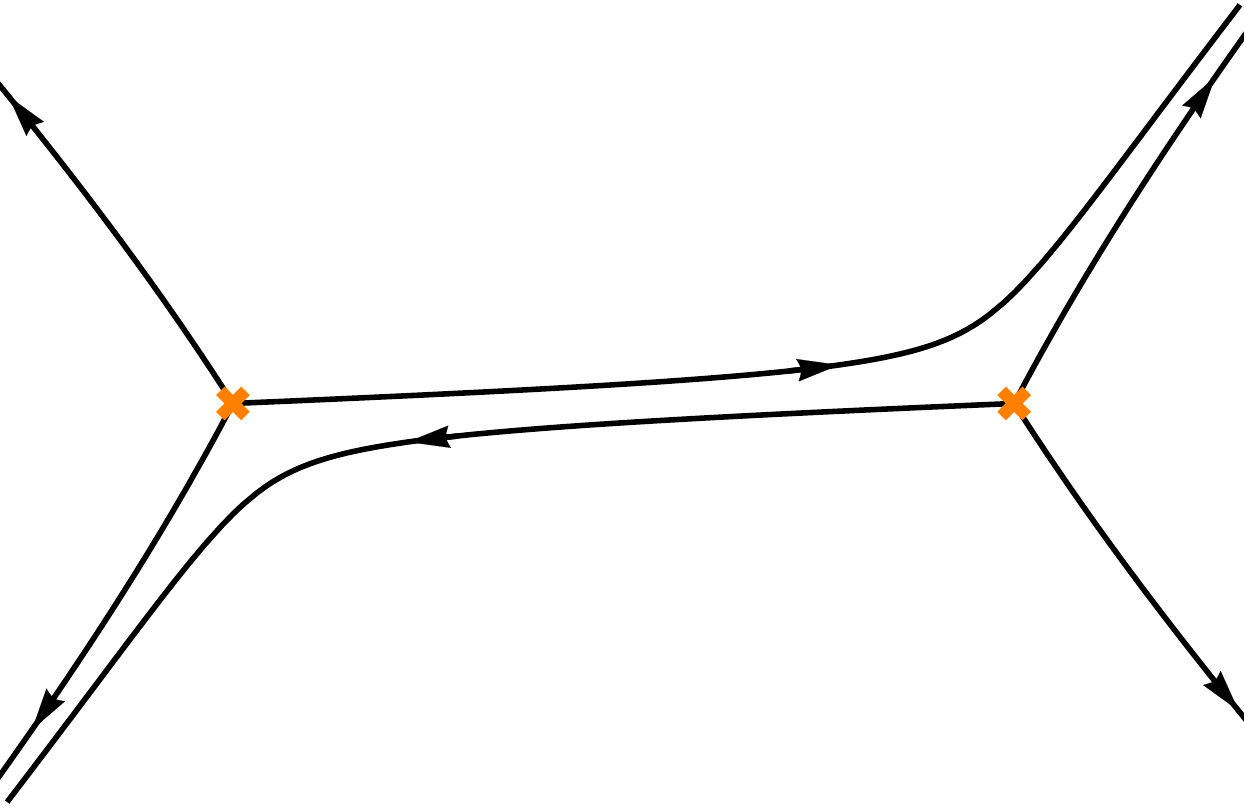}
        \caption{$\vartheta\gtrsim\vartheta_c$}
        \label{DoubleWall2}
    \end{subfigure}
    \caption{Double wall (thick blue) appearing at the critical phase $\vartheta=\vartheta_c$.}
    \label{DoubleWall}
\end{figure}

\begin{figure}
    \centering
    \begin{subfigure}[b]{0.3\textwidth}
        \includegraphics[width=\textwidth]{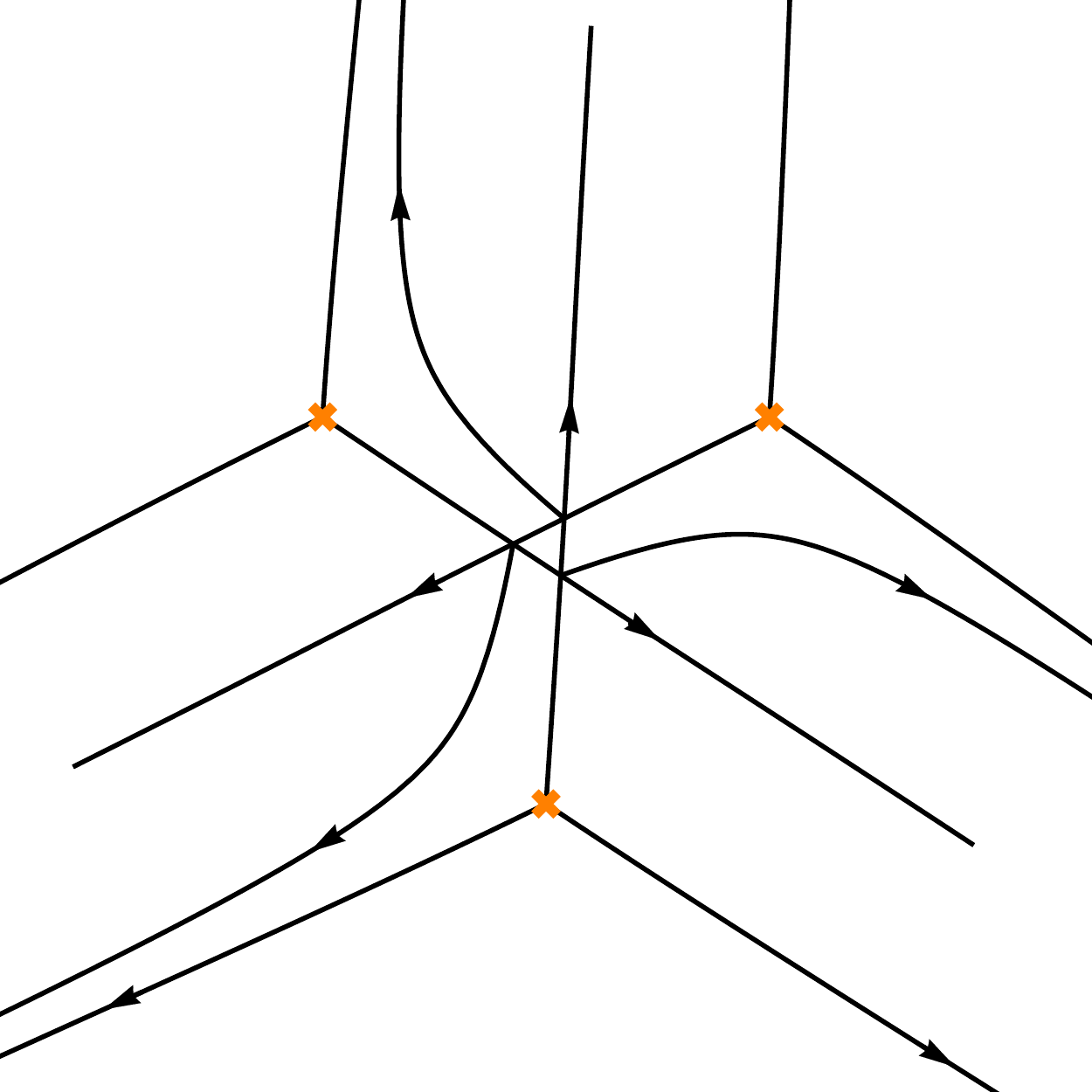}
        \caption{$\vartheta\lesssim\vartheta_c$}
        \label{Yweb1}
    \end{subfigure}
    \quad
    \begin{subfigure}[b]{0.3\textwidth}
        \includegraphics[width=\textwidth]{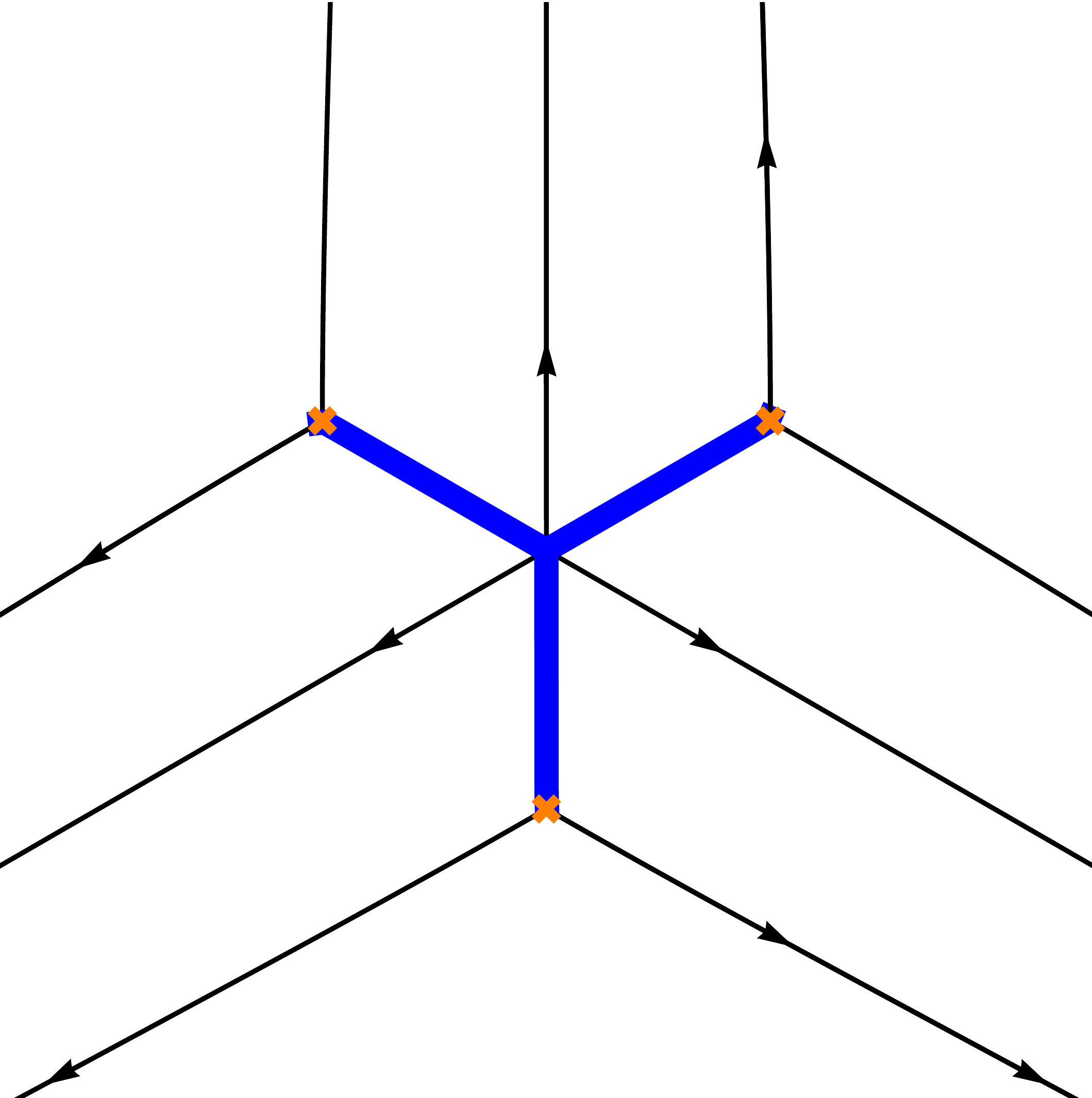}
        \caption{$\vartheta=\vartheta_c$}
        \label{Yweb2}
    \end{subfigure}
    \quad
    \begin{subfigure}[b]{0.3\textwidth}
        \includegraphics[width=\textwidth]{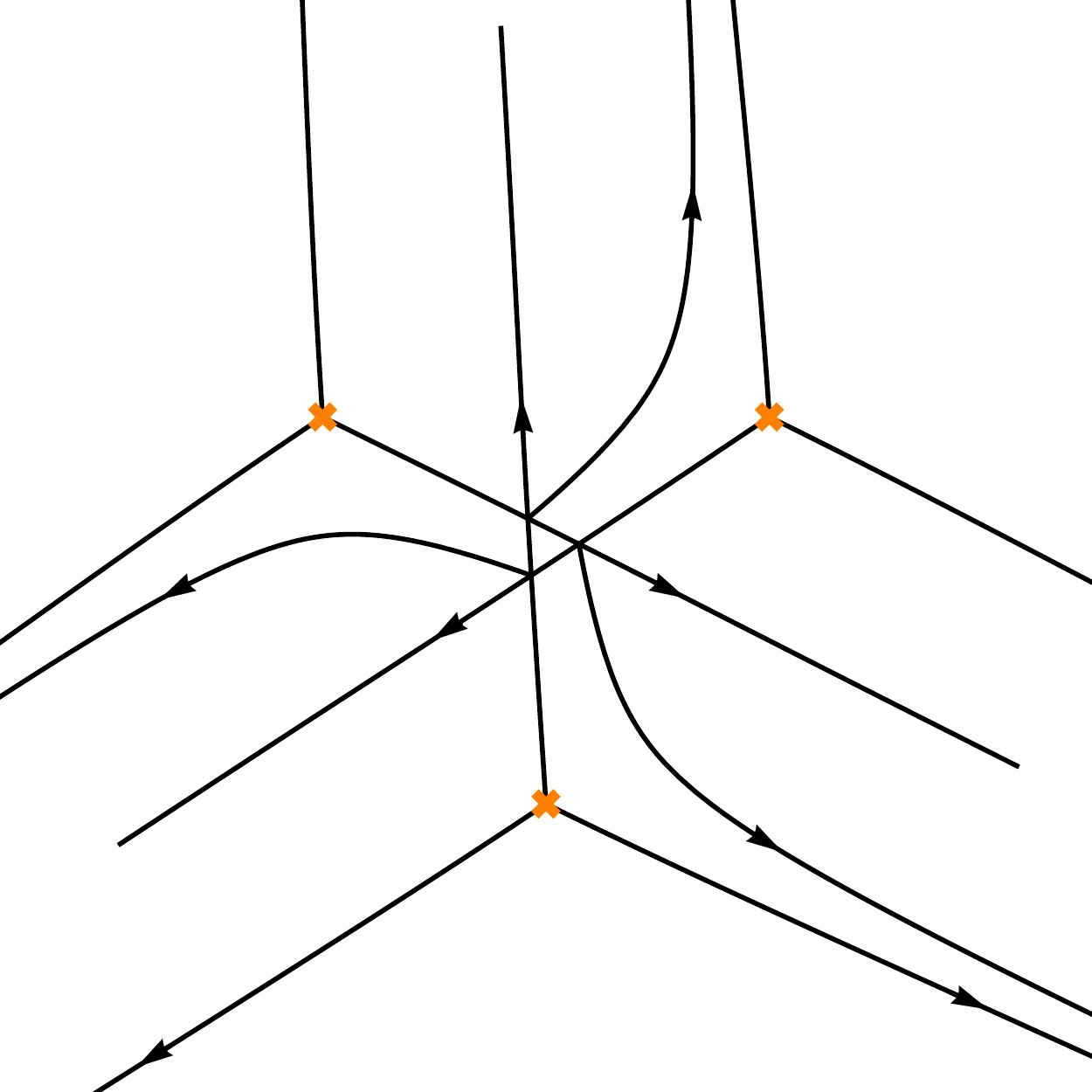}
        \caption{$\vartheta\gtrsim\vartheta_c$}
        \label{Yweb3}
    \end{subfigure}
    \caption{Y-web (thick blue) made out of three double walls meeting at a joint, appearing at the critical phase $\vartheta=\vartheta_c$.}
    \label{Yweb}
\end{figure}

The simplest instance of such a topological jump is when two walls running anti-parallel to each other merge into a \emph{double wall} connecting two branch points, as illustrated in Figure~\ref{DoubleWall}.%
\footnote{This figure and many others in this paper were plotted using the Mathematica package swn-plotter~\cite{swn-plotter}.}
More generally, double walls can have \emph{joints} like those shown in Figure~\ref{fig:network-intersections}, resulting in a web with endpoints on branch points. We refer to all such double walls and webs collectively as \emph{finite webs}.
Figure~\ref{Yweb} shows an example of a finite web running between three branch points and shaped like the letter Y, which we will call a \emph{Y-web}.
Each finite web lifts to a closed one-cycle $\gamma$ on $\Sigma$,%
\footnote{A finite web may admit several lifts to closed cycles, see e.g.~\cite{Galakhov:2013oja}. However, for a generic $u \in \cB$, all these lifts are multiples of a common primitive cycle.}
on which there can be a string. 
The differential condition~\eqref{eq:S-wall} implies that the central charge given in~\eqref{ZMgamma} and the mass are related via
\bea
Z_\gamma  = \ex^{\ii\vartheta_c} M_\gamma \;,
\eea
which saturates the BPS bound $|Z_\gamma|\leq M_\gamma$.
This is the geometric description of 4d BPS states of charge $\gamma$ in theories of class~$\CS$.
The degeneracy $\Omega(\gamma,u)$ of this BPS state can be computed from the combinatorial soliton data carried by the walls of the finite web, which is entirely determined by its topology. 

Because of CPT symmetry, a given BPS state at the phase $\vartheta_c$ is always accompanied by its anti-particle, another BPS state at $\vartheta_c+\pi$. So it is sufficient to vary $\vartheta$ in the region $[0, \pi)$ to gather information about the whole BPS spectrum, by studying all the finite webs that appear at the critical phases $\vartheta_{c,i}$ where the topology of the network jumps.
For a sampling of finite webs of varying degrees of complexity see Section~9 of~\cite{Gaiotto:2009hg}, Section~8 of~\cite{Gaiotto:2012rg}, and~\cite{Galakhov:2013oja}.

\begin{figure}
    \centering
    \begin{subfigure}[b]{0.3\textwidth}
        \includegraphics[width=\textwidth]{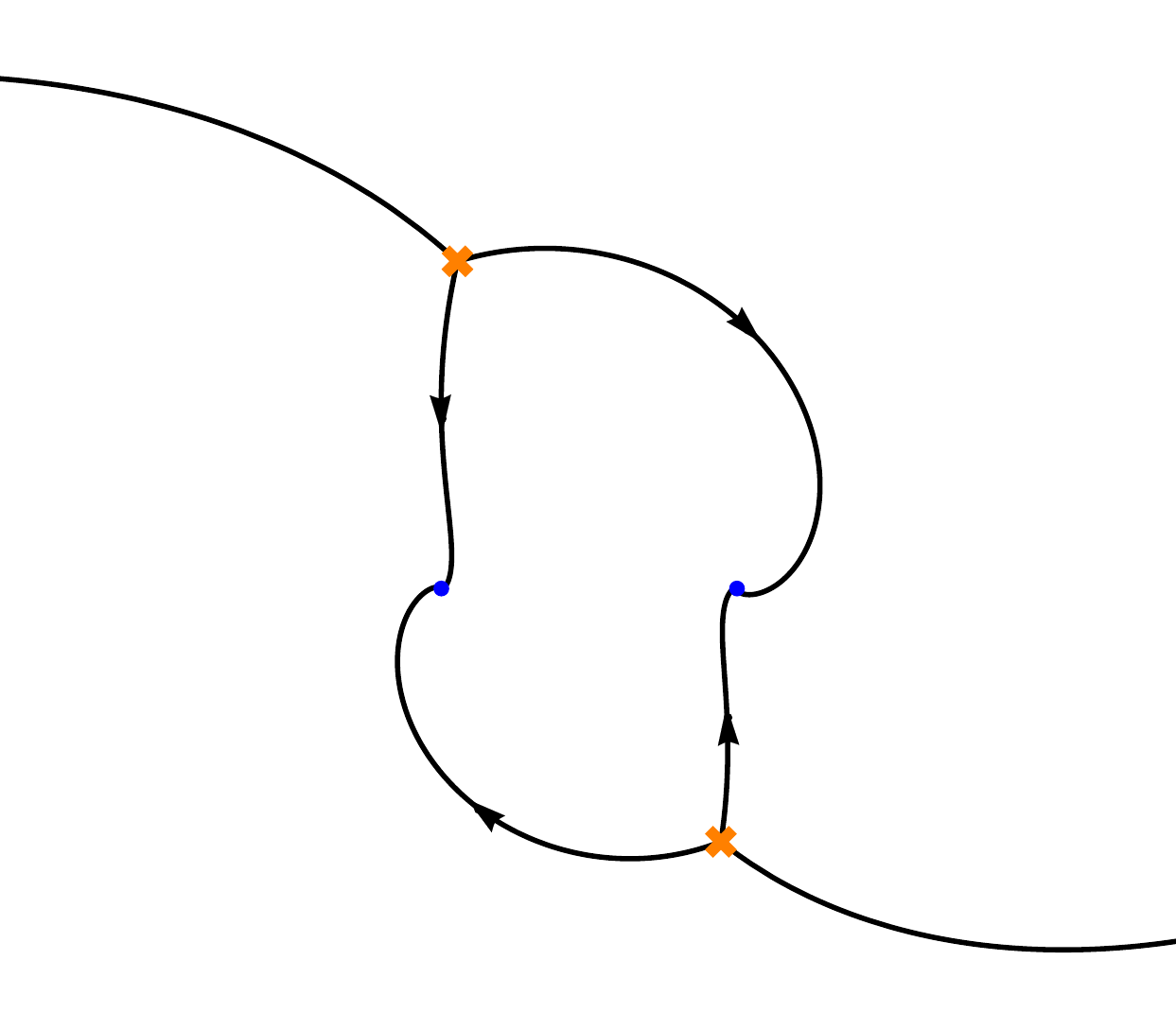}
        \caption{$\vartheta=0$}
        \label{T2BPS0}
    \end{subfigure}
    \quad
    \begin{subfigure}[b]{0.3\textwidth}
       \begin{overpic}[width=\textwidth]{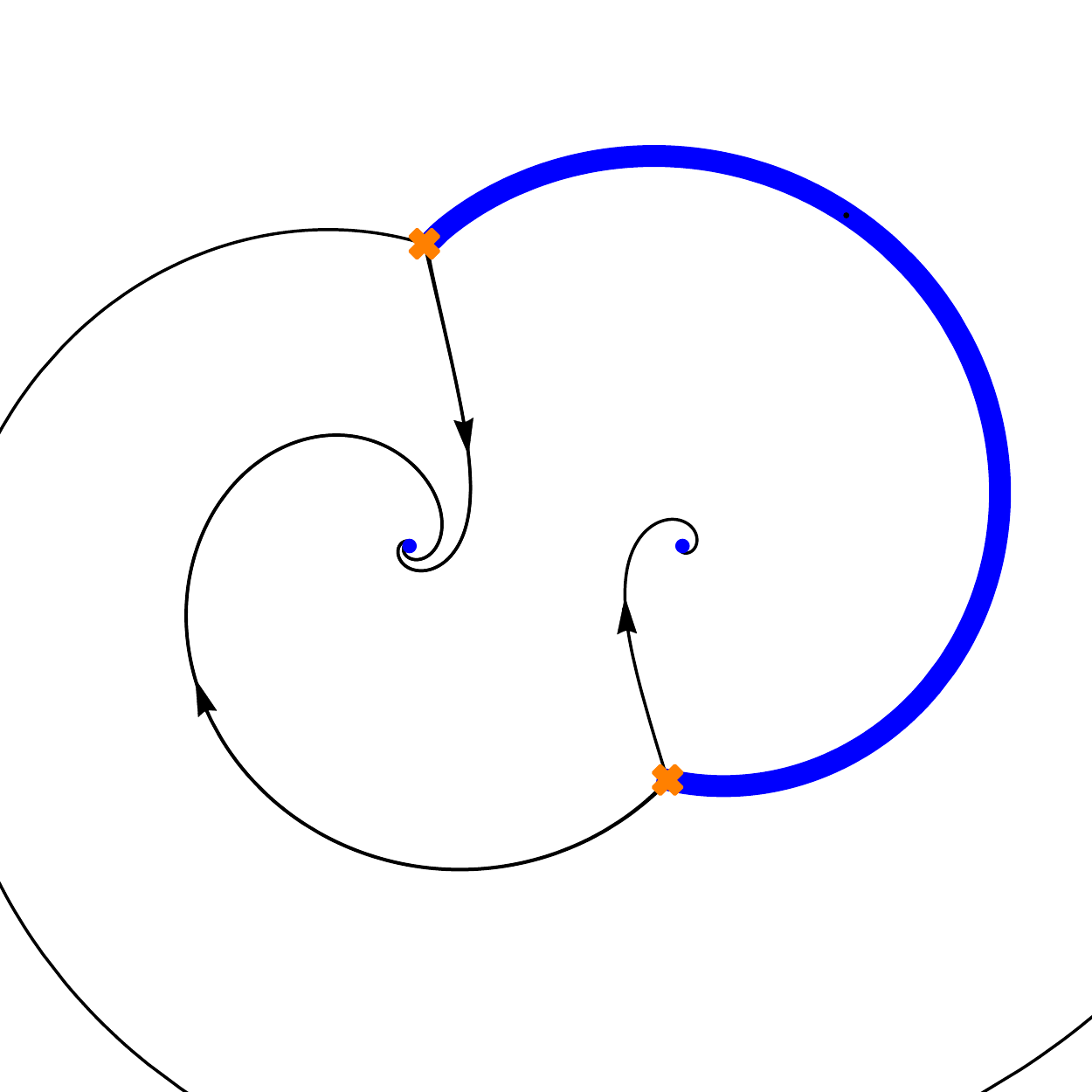}
 	\put (85, 80) {$\gamma_1$}
\end{overpic}
        \caption{$\vartheta_{c,1}=0.69$}
        \label{T2BPS1}
    \end{subfigure}
    \quad
    \begin{subfigure}[b]{0.3\textwidth}
       \begin{overpic}[width=\textwidth]{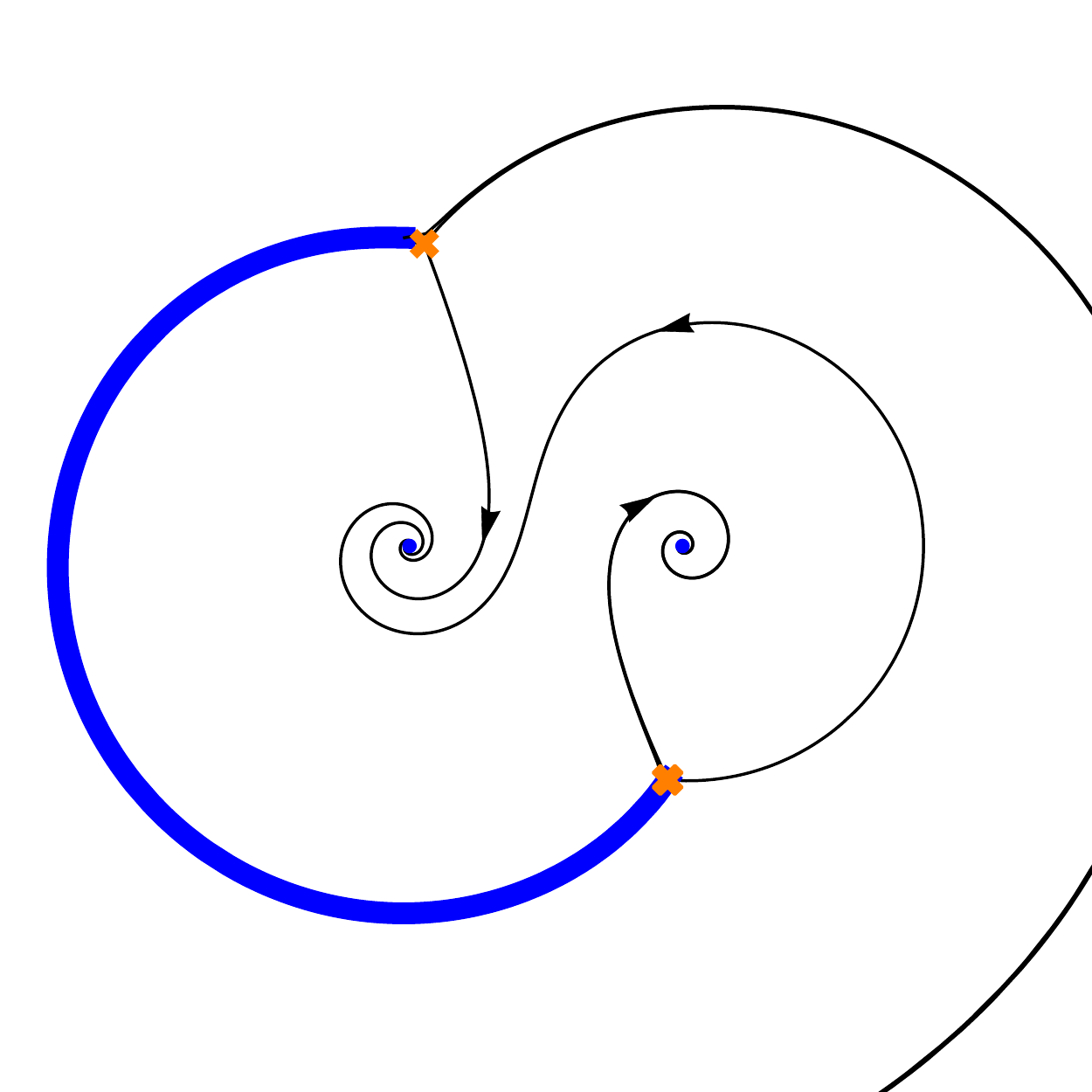}
 	\put (5, 75) {$\gamma_2$}
\end{overpic}
        \caption{$\vartheta_{c,2}=0.89$}
        \label{T2BPS2}
    \end{subfigure}
    
        \begin{subfigure}[b]{0.3\textwidth}
       \begin{overpic}[width=\textwidth]{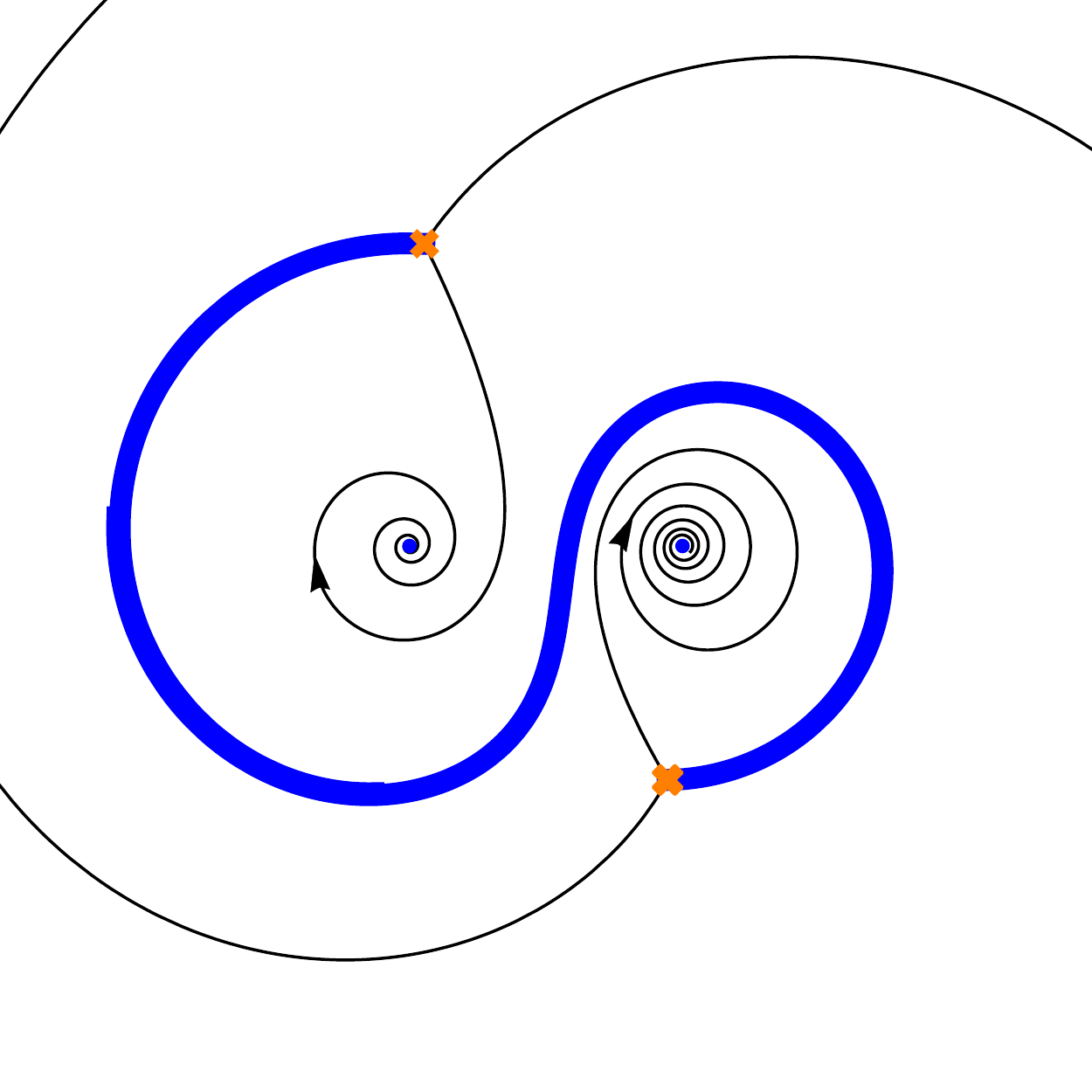}
 	\put (70, 70) {$\gamma_3$}
\end{overpic}
        \caption{$\vartheta_{c,3}=1.05$}
        \label{T2BPS3}
    \end{subfigure}
    \quad
    \begin{subfigure}[b]{0.3\textwidth}
       \begin{overpic}[width=\textwidth]{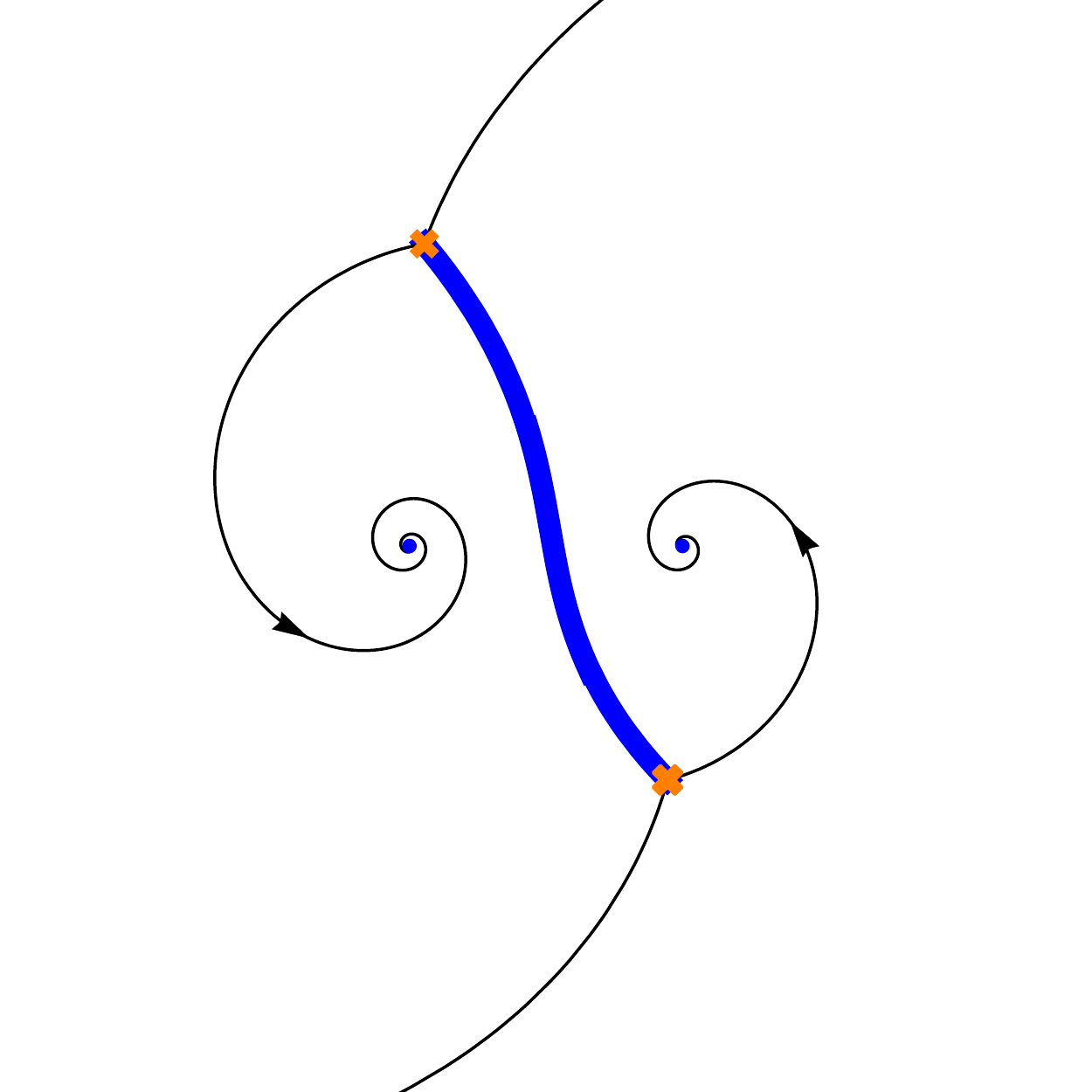}
 	\put (50, 70) {$\gamma_4$}
\end{overpic}
        \caption{$\vartheta_{c,4}=1.40$}
        \label{T2BPS4}
    \end{subfigure}
    \quad
    \begin{subfigure}[b]{0.3\textwidth}
        \includegraphics[width=\textwidth]{images/T2-BPS0}
        \caption{$\vartheta=\pi$}
        \label{T2BPS5}
    \end{subfigure}
    \caption{Spectral networks on a sphere with three punctures at $z=\pm1,\infty$, corresponding to the $T_2$ theory. The four expected BPS hypermultiplets (thick blue) appear at various critical phases $\vartheta_{c,i}$ between $0$ and $\pi$.}
    \label{T2BPSstates}
\end{figure}

In Figure~\ref{T2BPSstates}, we apply this method to study the BPS spectrum of the so-called $T_2$ theory, that is the $A_1$ theory associated with a sphere with three regular punctures~\cite{Gaiotto:2009we}.
This theory describes a free hypermultiplet in the tri-fundamental of $SU(2)^{ 3}$ (4 BPS particles and 4 BPS anti-particles), with a trivial Coulomb branch, and a charge lattice $\Gamma$ of rank $f=3$.
As we vary the phase $\vartheta$ between $0$ and $\pi$, we observe four finite webs, corresponding to the four expected BPS states, with charges $\gamma_1,\gamma_2,\gamma_3, \gamma_4$. 
Any choice of three BPS states form an integral basis for $\Gamma$,
but there is a unique choice that leads to a \emph{positive} integral basis for $\Gamma$. Taking the basis $\{\gamma_1,\gamma_2, \gamma_4\}$, we can express the charge of the last BPS states as a positive linear combination: $\gamma_3 = \gamma_1 + \gamma_2 + \gamma_4$. 

An advantage of studying BPS states using spectral networks is that this method is systematic and algorithmic. 
In principle, we can study the whole BPS spectrum for any theory of class~$\CS$, anywhere on the Coulomb branch, by numerically plotting spectral networks at several phases
(for example with the softwares~\cite{swn-plotter} and~\cite{Longhi:2016rjt, loom}).
However, this procedure becomes quickly impractical for Riemann surfaces~$\CC$ with high genus or many punctures, and for higher-rank $A_{N-1}$ theories with $N>2$, because of the large number of BPS states and the complexity of the spectral networks involved.

\section{BPS graphs} \label{BPSgraphs}

We define the concept of a \emph{BPS graph} on a punctured Riemann surface~$\CC$ associated with an $A_{N-1}$ theory of class~$\CS$. BPS graphs arise from maximally degenerate spectral networks that appear on a very special locus of the Coulomb branch $\cB$. A key property of a BPS graph is that it provides a positive integral basis for the lattice $\Gamma$ of gauge and flavor charges of BPS states. A given surface~$\CC$ admits many topologically inequivalent BPS graphs, related by elementary moves.

\subsection{Definition}\label{def-bps-graph}

A generic spectral network undergoes several distinct topological jumps as the phase~$\vartheta$ varies between $0$ and $\pi$. At each critical phase~$\vartheta_{c,i}$, some part of the spectral network degenerates into a finite web, which can swirl in complicated ways and involve many joints.
In view of the difficulty of identifying all these finite webs, it is desirable to get a handle on the usually random distribution of the critical phases~$\vartheta_{c,i}$.
Our approach is to focus on a very special type of spectral network, for which there is only one topological jump at a \emph{single} critical phase $\vartheta_c$. 

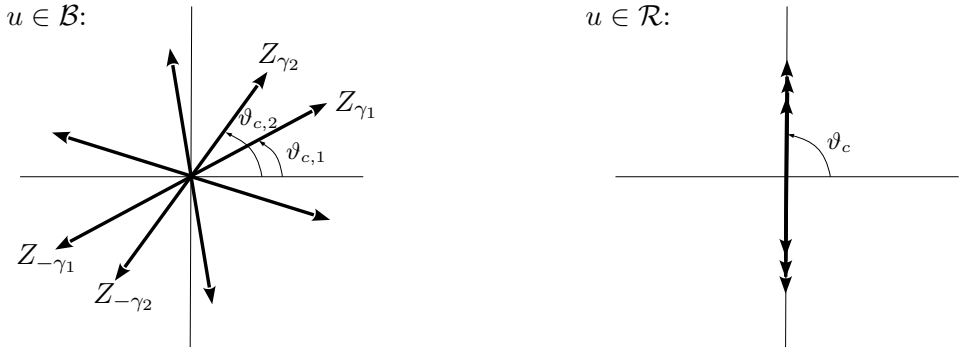
\begin{figure}[tbh]
    \centering
        \begin{overpic}[width=0.84\textwidth]{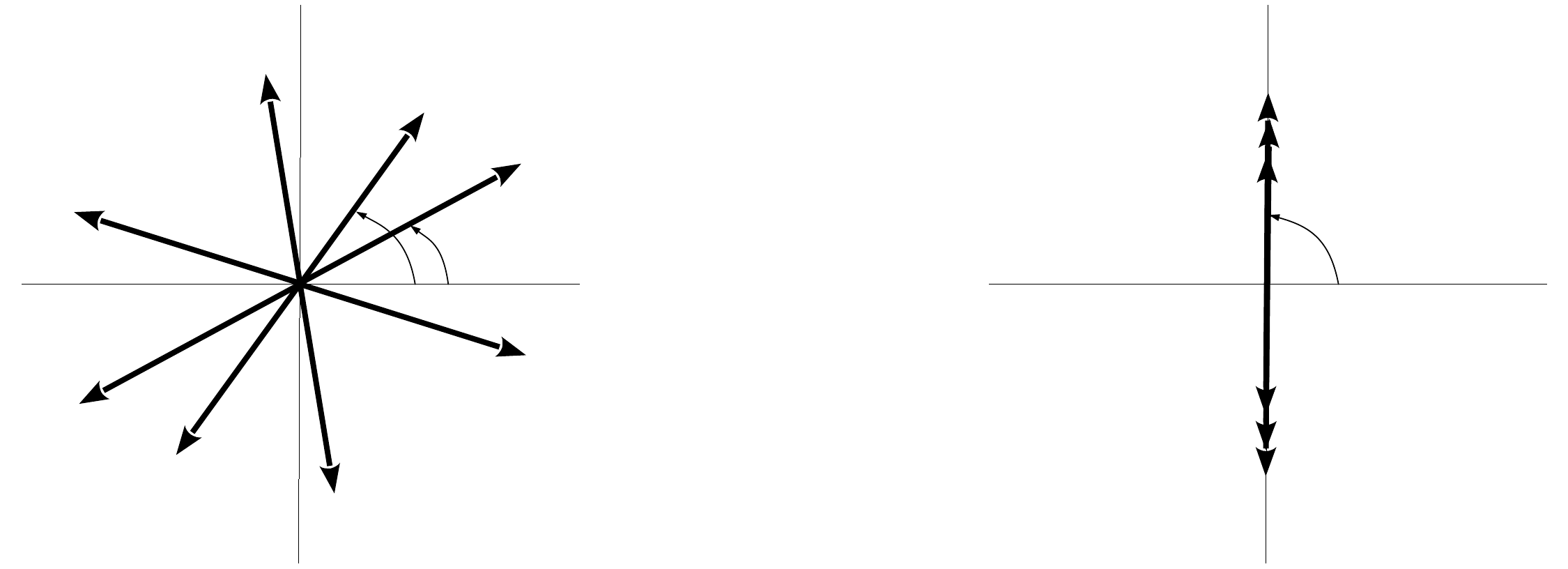}
 	\put (0, 34) {$u\in\cB$:}
	\put (60, 34) {$u\in\cR$:}
	\put (34, 25) {$Z_{\gamma_1}$}
	\put (26, 30) {$Z_{\gamma_2}$}
	\put (1, 9) {$Z_{-\gamma_1}$}
	\put (9, 5) {$Z_{-\gamma_2}$}
	\put (29, 20) {\footnotesize $\vartheta_{c,1}$}
	\put (24, 23.5) {\footnotesize $\vartheta_{c,2}$}
	\put (85, 21) {\footnotesize$\vartheta_c$}
\end{overpic}
    \caption{\emph{Left}: At a generic point of $\cB$, the central charges $Z_{\gamma_i}$ of the BPS (anti-)particles have phases $\vartheta_{c,i}$ distributed between $0$ and $\pi$ ($\pi$ and $2\pi$). \emph{Right}: On the Roman locus $\cR$, all the central charges have the same phase (here $\vartheta_c = \pi/2$).
    }
    \label{Zplane}
\end{figure}

In such a maximally degenerate spectral network, all the finite webs appear simultaneously, since the central charges $Z_{\gamma_i}$ of the corresponding BPS states all have the same phase~$\vartheta_c$ (Figure~\ref{Zplane}).%
\footnote{This is reminiscent of the situation in~\cite{Aganagic:2009kf}, where the analysis was simplified by moving to the locus where all the phases of central charges of the D2-D0 particles align.}
This implies that we are on a very special locus of the Coulomb branch $\cB$ that corresponds to a \emph{maximal intersection of walls of marginal stability}.%
\footnote{We are in fact considering the enlarged moduli space of Coulomb branch parameters and UV parameters (masses and gauge couplings), to which we extend the notion of walls of marginal stability.}
Because all roads lead to Rome, we call this maximal intersection the \emph{Roman locus} and denote it by~$\cR$. 
Since all the BPS bound states are marginally stable on~$\cR$, the BPS spectrum is maximally ambiguous. We would have to move away from~$\cR$ into a well-defined chamber, which amounts to choosing a side for each of the walls of marginal stability, in order to properly study the stable BPS spectrum.

Now comes the crucial point.
Among the finite webs in a maximally degenerate spectral network at~$\vartheta_c$, there is a unique set of $2r\!+\! f$ \emph{elementary webs} which provides a \emph{positive integral basis} for the lattice~$\Gamma$ of gauge and flavor charges of BPS states.
More precisely, the elementary webs lift to closed one-cycles $\gamma_i$ on the cover $\Sigma$, which form a positive integral basis for the half-lattice
\be\label{eq:half-lattice}
	\Gamma_+( u,\vartheta_c) := 
	\left\{\gamma\in\Gamma \,, 
	\ Z_\gamma(u)  \in {\rm e}^{\ii \vartheta_c}\mathbb{R}_+ \right\}\,,
\ee
with $u \in \cR$. We define the \emph{BPS graph} $\cG$ associated with the $N$-fold branched covering $\Sigma\to \CC$ as the set of these elementary webs.

From a purely \emph{topological} perspective,
a BPS graph $\cG$ associated with~$\Sigma$ is a trivalent undirected connected graph on $C$ made of the following data:
\begin{itemize} \parskip=-2pt
\item \emph{vertices} are branch points of~$\Sigma$,
\item \emph{edges} connect pairs of vertices,
\item \emph{webs} with joints connect three or more vertices (for $N>2$).
\end{itemize}
Note that in the terminology introduced above, edges and webs are all \emph{elementary webs}.
We showed an example of an edge between two branch points in Figure~\ref{DoubleWall}, and a web with one joint connecting three branch points in Figure~\ref{Yweb}.
Near a boundary component of~$\CC$, some vertices of $\cG$ may have one or two ``empty'' adjacent edges but we still consider them as trivalent.
Joints in webs are also typically trivalent, although they can have valence up to six in certain complicated situations (recall Figure~\ref{fig:network-intersections}).

The intersection form $\langle\cdot,\cdot \rangle$ on $\Gamma_+=H_1(\Sigma, \bZ)$%
\footnote{For theories of class~$\CS$, $\Gamma_+$ is in fact a certain sub-quotient of $H_1(\Sigma, \bZ)$, as mentioned in Section~\ref{SECspectralCurve}. If we considered $\mathfrak{gl}_{N}$ instead of $\mathfrak{sl}_{N}$, we would literally have $\Gamma_+ = H_1(\Sigma,\bZ)$.}
can easily be computed directly from the BPS graph $\cG$.
Given two edges (or webs) $p_1$ and $p_2$ in $\cG$ that lift to one-cycles $\gamma_1$ and $\gamma_2$ on~$\Sigma$, we can determine $\langle \gamma_1, \gamma_2 \rangle$ by looking at their common branch points (for simplicity we assume that $p_1$ and $p_2$ do not intersect on~$\CC$).
We define $\sigma(b;\gamma_1,\gamma_2)=\pm 1$ if $p_2$ ends on a branch point~$b$ on the slot immediately (counter-)clockwise of $p_1$.
The intersection form is then obtained by summing over all the common branch points:
\be\label{eq:pairing}
	\langle\gamma_1,\gamma_2\rangle = \sum_{ b}\sigma(b;\gamma_1,\gamma_2)\,.
\ee
There can be additional terms in cases where $p_1$ and $p_2$ have intersections on~$\CC$.

\subsection{Existence}\label{subsec:existence}

One may worry that for a general theory of class~$\CS$ the walls of marginal stability on the Coulomb branch $\cB$ would not have a maximal intersection, so that the Roman locus~$\cR$ would be empty. We do not have a full proof of the existence of a non-trivial $\cR$. It is clear however that~$\cR$ exists for all ``complete theories''~\cite{Cecotti:2011rv}, which include all $A_1$ theories of class~$\cS$. Indeed, by definition, the central charges of the BPS states in complete theories can be varied arbitrarily, and in particular we can choose them to all have the same phase~$\vartheta_c$. There is no such obvious proof for $A_{N-1}$ theories with $N>2$, but we have found explicit examples of BPS graphs for the sphere with three or four punctures (Sections~\ref{subsec.TN} and~\ref{subsec:4-puncture-An}). The best we can do is to show that the existence of $\cR$ is at least not ruled out by dimension counting: aligning all the BPS central charges imposes $2r+f-1$ real constraints on $r+f$ complex Coulomb and mass parameters, which would naively lead to a locus $\cR$ of real dimension $f+1$~\cite{Longhi:2016wtv}.%
\footnote{Note that this counting did not take into account the additional freedom of varying UV couplings, whose number depends on the topology of~$\CC$.}

We will see in examples that for a given theory of class~$\CS$ there is in general a whole family of BPS graphs on~$\CC$, parameterized by the multi-dimensional locus $\cR$. Moving along $\cR$ changes the shape of the corresponding BPS graph.
Even the topology of the BPS graph can change if we move from one component of~$\cR$ to another component through a singular interface where a BPS state becomes massless. We will describe these topological transformations in more detail in Section \ref{subsecMoves}.

\begin{figure}
    \centering
        \begin{overpic}[width=\textwidth]{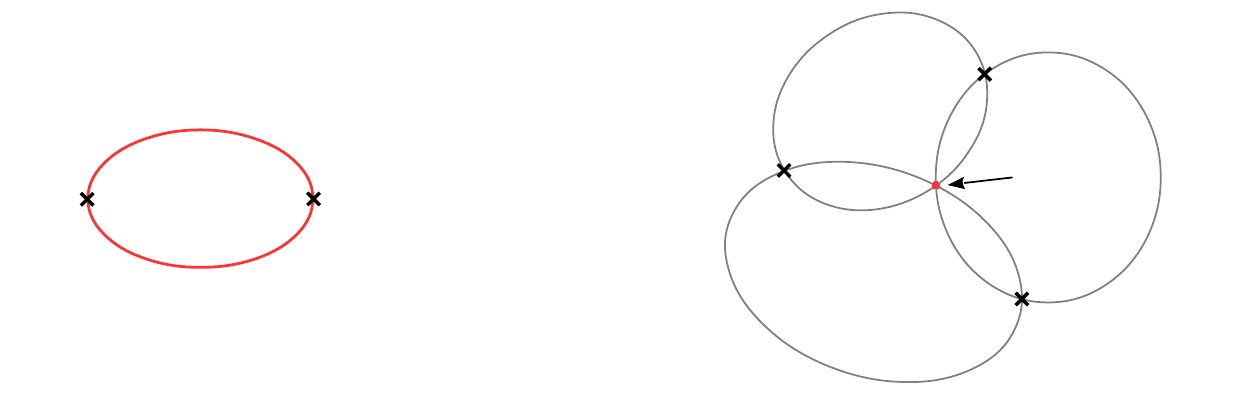}
        	\put (15, 23) {$\cR^{(1)}$}
	\put (15, 6.9) {$\cR^{(2)}$}
	\put (82, 17) {$\cR$}
\end{overpic}
    \caption{Schematic representations of the Roman locus $\cR$ (red). \emph{Left}: If there is a single wall of marginal stability, it coincides with $\cR$. Singularities where some BPS state becomes massless (black crosses) separate components. \emph{Right}: More generally, $\cR$ is the maximal intersection of walls of marginal stability.}
    \label{CBNf1}
\end{figure}

We sketch some simple examples of walls of marginal stability on the Coulomb branch in Figure~\ref{CBNf1}.
The case with a single wall of marginal stability could correspond to $SU(2)$ super Yang-Mills theory, or to the Argyres-Douglas theory AD$_3$ (see Section~\ref{secADtheories}).
The situation shown on the right could correspond to $SU(2)$ gauge theory with $N_f=1$ flavor (see Section~\ref{SU2Nf1}).%
\footnote{This cartoon is inspired by a plot obtained in unpublished joint work of some of the authors with Daniel S.~Park, in a study of ``$\mathcal{K}$-walls'' on the Coulomb branch.}
A detailed study of the Coulomb branches of $SU(2)$ theories with $N_f=2$ flavors was performed in~\cite{Bilal:1997st}, where it was found that multiple maximal intersections of walls of marginal stability occur, when the two flavor masses are set to be equal.

\subsection{Elementary moves}\label{subsecMoves}

As mentioned above, when we vary the point $u$ along the Roman locus $\cR$ for a given theory of class~$\CS$, the corresponding BPS graph on the surface~$\CC$ alters its shape and sometimes even its topology. 
We now illustrate two elementary local transformations which modify the topology of the BPS graph. 

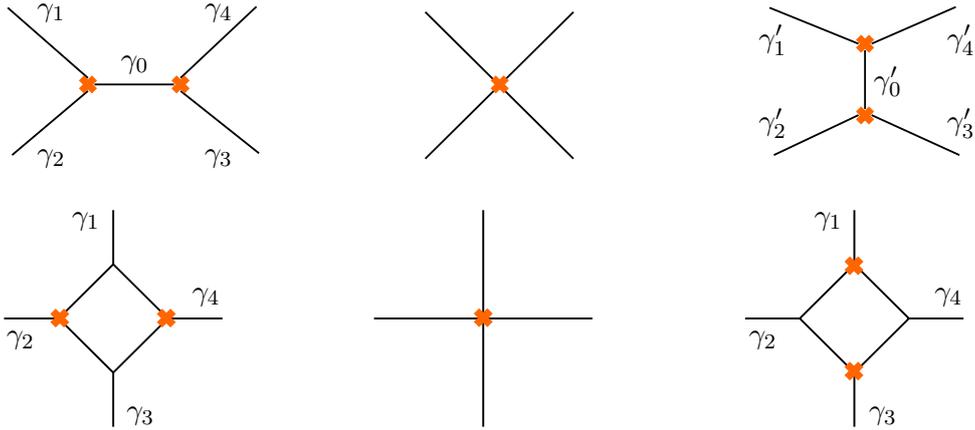
\begin{figure}
\begin{center}
        \begin{overpic}[width=0.91\textwidth]{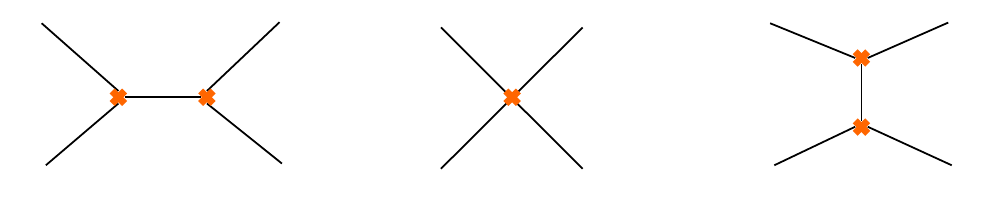}
 	\put (7, 17) {$\gamma_1$} \put (23, 17) {$\gamma_4$}
	\put (15, 12) {$\gamma_0$} 
	\put (7, 3) {$\gamma_2$} \put (23, 3) {$\gamma_3$}
	 \put (76, 14) {$\gamma_1'$} \put (94, 14) {$\gamma_4'$}
	\put (87, 10) {$\gamma_0'$} 
	\put (76, 6) {$\gamma_2'$} \put (94, 6) {$\gamma_3'$}
\end{overpic}\\ \vspace{0.1cm}
 \begin{overpic}[width=0.95\textwidth]{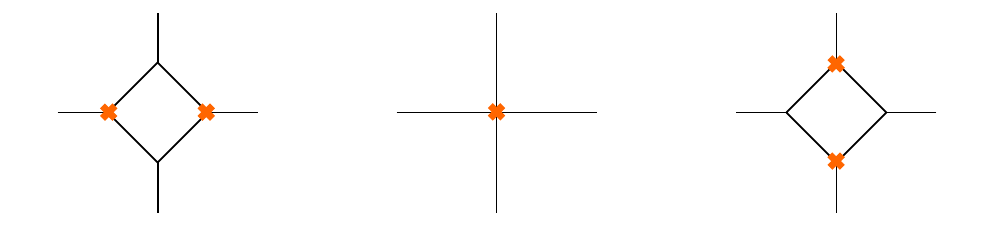}
 	\put (12, 20) {$\gamma_1$} \put (6, 9) {$\gamma_2$}\put (17, 2) {$\gamma_3$} \put (23, 13) {$\gamma_4$}
	\put (80, 20) {$\gamma_1$} \put (74, 9) {$\gamma_2$}\put (85, 2) {$\gamma_3$} \put (91, 13) {$\gamma_4$}
\end{overpic}
\caption{\emph{Top}: Flip move. \emph{Bottom}: Cootie move.}
\label{fig:flip}
\end{center}
\end{figure}

The first transformation is the \emph{flip move} shown on the top of Figure~\ref{fig:flip}.
It describes the complete shrinking of an edge $\gamma_0$ until its two branch points collide, followed by the transverse expansion of a new edge $\gamma_0'$ (beware that we label an edge on~$\CC$ by its lifted one-cycle~$\gamma_0$ on~$\Sigma$).
Note that right at the degeneration we do not really have a well-defined BPS graph because one generator of the charge lattice~$\Gamma$ is missing (that was also the reason to disallow vertices of valence higher than three). We are thus moving from one component of $\cR$ to another.
This requires passing through a complex codimension-one singularity on the Coulomb branch~$\cB$ where the BPS state associated with the shrunken edge becomes massless.
There is a well-known monodromy of the charge lattice around such singularities. In order to compare the lattice~$\Gamma$ before the flip with the lattice~$\Gamma'$ after the flip, we must therefore specify a choice of the path for the parallel transport in the transverse space (which has complex dimension one).
By convention, we choose the path defined by the following relations between the homology cycles $\gamma\in \Gamma$ and $\gamma'\in \Gamma'$:
\be\label{eq:mutation}
	\gamma_0' = -\gamma_0 \;, \qquad \gamma_1'=\gamma_1+\gamma_0  \;,\qquad \gamma_2'=\gamma_2 \;,\qquad \gamma_3'=\gamma_3+\gamma_0 \;,\qquad \gamma_4'=\gamma_4\;.
\ee

The second transformation is shown on the bottom of Figure~\ref{fig:flip}.
We call it the \emph{cootie move}, in reference to the origami known as the ``cootie catcher'' (or also ``fortune teller'').%
\footnote{
\textit{``Circle, circle, dot, dot: Now you've got the cootie shot!
Circle, circle, square, square:
Now you have it everywhere!
Circle, circle, knife, knife:
Now you've got it all your life!
Circle, circle, line, line:
Now protected all the time!''} \hfil Anon.}
It again describes the collision and transverse separation of two branch points, but this time without any edge shrinking. The net result is simply the exchange of the positions of the two branch points with those of two nearby joints, which modifies the topologies of the four elementary webs involved. 
Since no singularity is encountered, no ambiguity arises in identifying the charge lattices before and after the cootie move. 
For practical purposes, the charges of four elementary webs of the final BPS graph are simply identified with those of the initial BPS graph in the obvious way.
Note that the intersection form~\eqref{eq:pairing} of the corresponding homology cycles is the same before and after the cootie move.


\section{\texorpdfstring{BPS graphs for $A_1$ theories}{BPS graphs for A1 theories}}
\label{sec:A1-graphs}

In this section we present several examples of BPS graphs on punctured Riemann surfaces~$\CC$ associated with $A_1$ theories of class~$\CS$. The spectral curve $\Sigma$ is just given by
\bea\label{eq:T2-diff}
\lambda^2 + \phi_2 =0 \;,
\eea
so our task consists in finding quadratic differentials $\phi_2$ corresponding to the Roman locus~$\cR$ on the Coulomb branch. Although possible for simple enough surfaces, it is too hard in general. We will then rely on a relation to ideal triangulations to identify topological representatives of BPS graphs for all $A_1$ theories of class~$\CS$.

\subsection{\texorpdfstring{$T_2$ theory}{T2 theory}}
\label{sec:T2-example}

The case where the surface~$\CC$ is a sphere with three regular punctures corresponds to the $T_2$ theory, which describes a hypermultiplet in the tri-fundamental representation of $SU(2)^3$ (recall Section~\ref{sec:bps-states}). 
The quadratic differential~\eqref{phik} is given by
\bea  \label{phi2T2}
\phi_2 = \frac{M_2^az_{ab}z_{ac}(z-z_b)(z-z_c)  +\text{cyclic} } {(z-z_a)^2(z-z_b)^2(z-z_c)^2  } (\dd z)^2 \;.
\eea
Since the Coulomb branch is trivial, the central charges are linear combinations of the UV mass parameters $m_{a,b,c}$.
Therefore, we simply have to choose $M_2^{a,b,c}= m_{a,b,c}^2$ with the same phase $2\vartheta_c$ to guarantee that all the BPS states appear simultaneously at the single critical phase $\vartheta_c$. This leaves the three moduli $|M_2^{a,b,c}|$, which together with the phase $\vartheta_c$ give a locus $\cR$ of dimension $f+1=4$, as expected from the naive argument in Section~\ref{subsec:existence}.

\begin{figure}
    \centering
    \begin{subfigure}[b]{0.3\textwidth}
        \includegraphics[width=\textwidth]{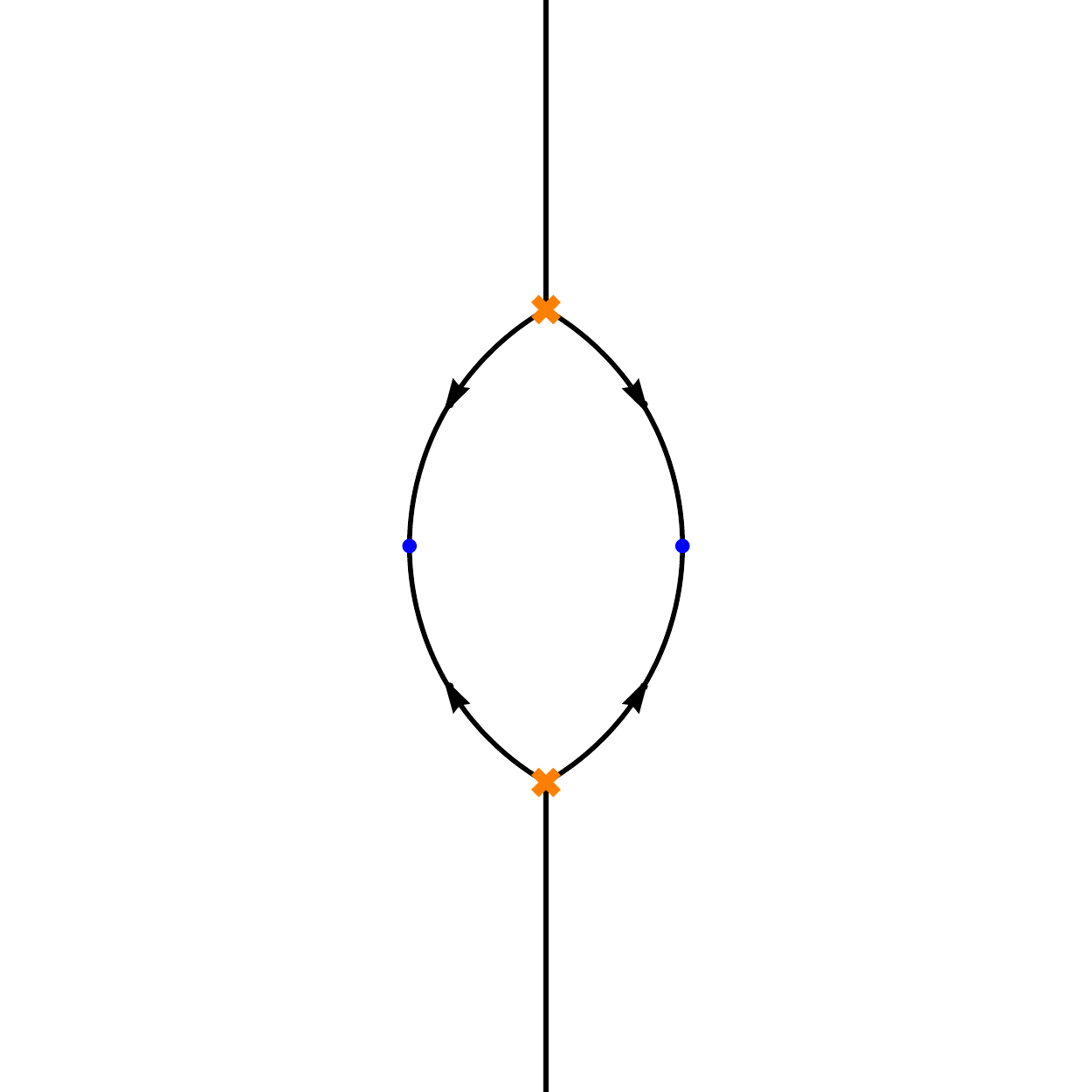}
        \caption{$\vartheta=\vartheta_c-\pi/2$}
        \label{T2graph0}
    \end{subfigure}
   \hfill
    \begin{subfigure}[b]{0.3\textwidth}
        \includegraphics[width=\textwidth]{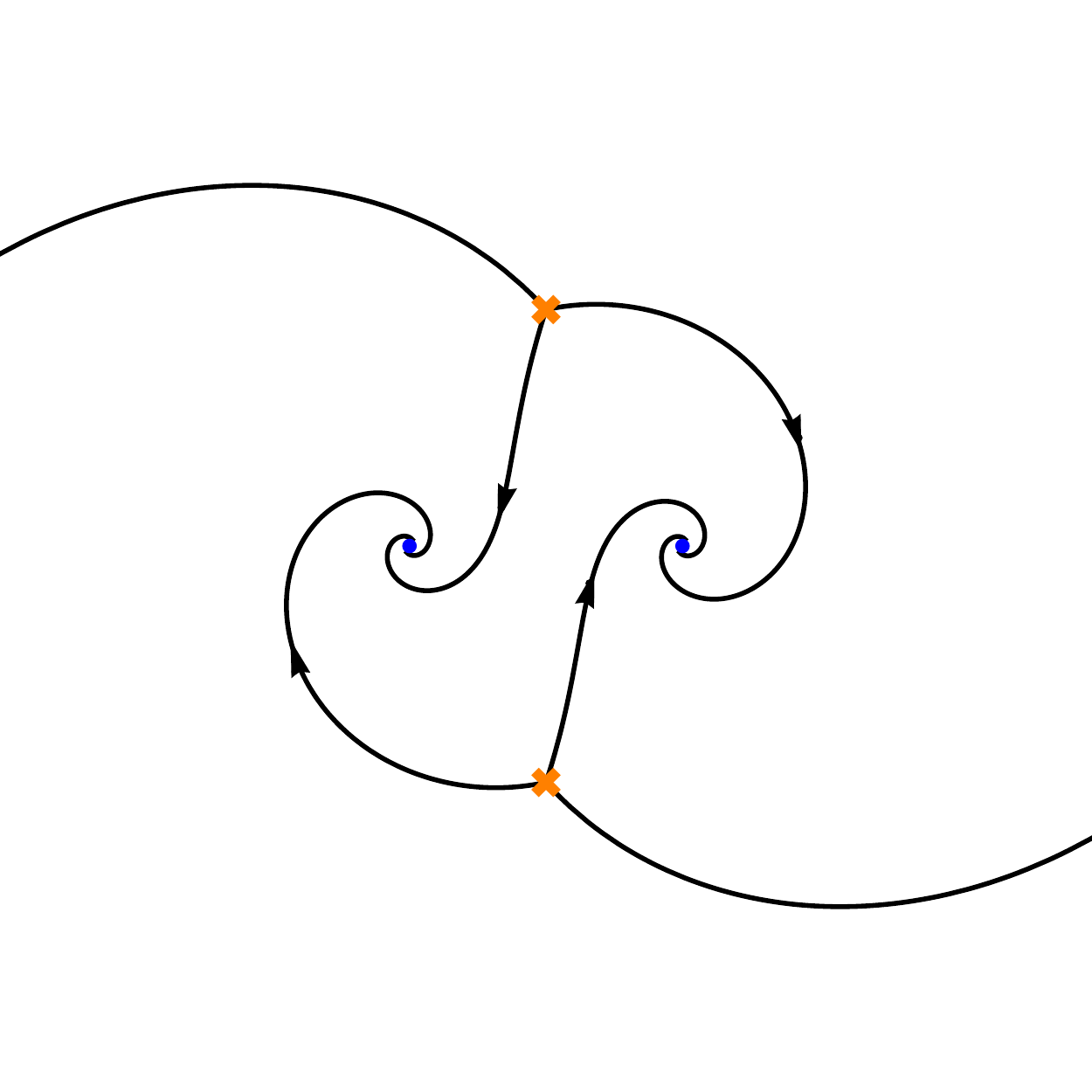}
        \caption{$\vartheta<\vartheta_c$}
        \label{T2graph1}
    \end{subfigure}
   \hfill
    \begin{subfigure}[b]{0.3\textwidth}
        \includegraphics[width=\textwidth]{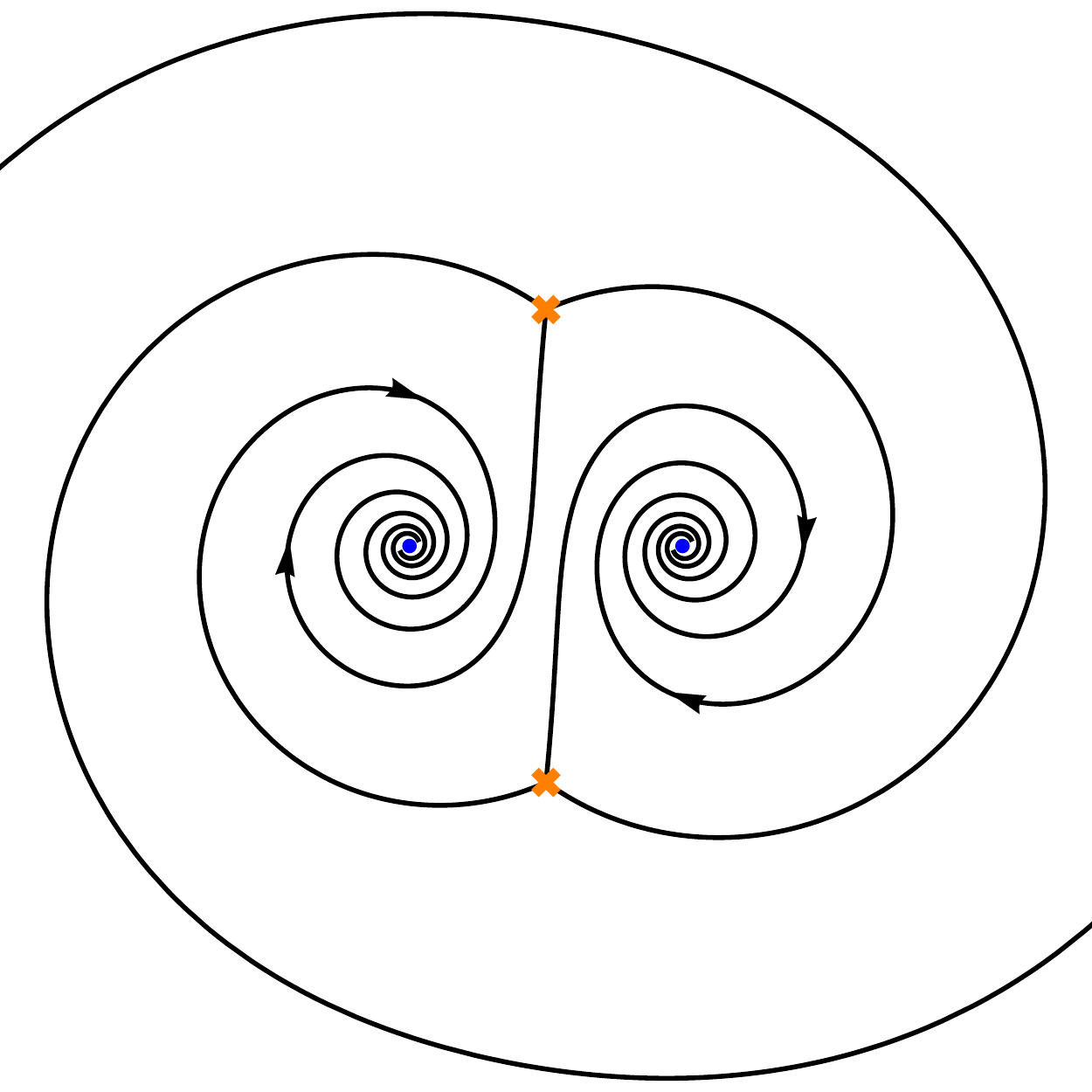}
        \caption{$\vartheta\lesssim\vartheta_c$}
        \label{T2graph2}
    \end{subfigure}
    
    \vspace{0.3cm}
    
    \begin{subfigure}[b]{0.35\textwidth}
        \includegraphics[width=\textwidth]{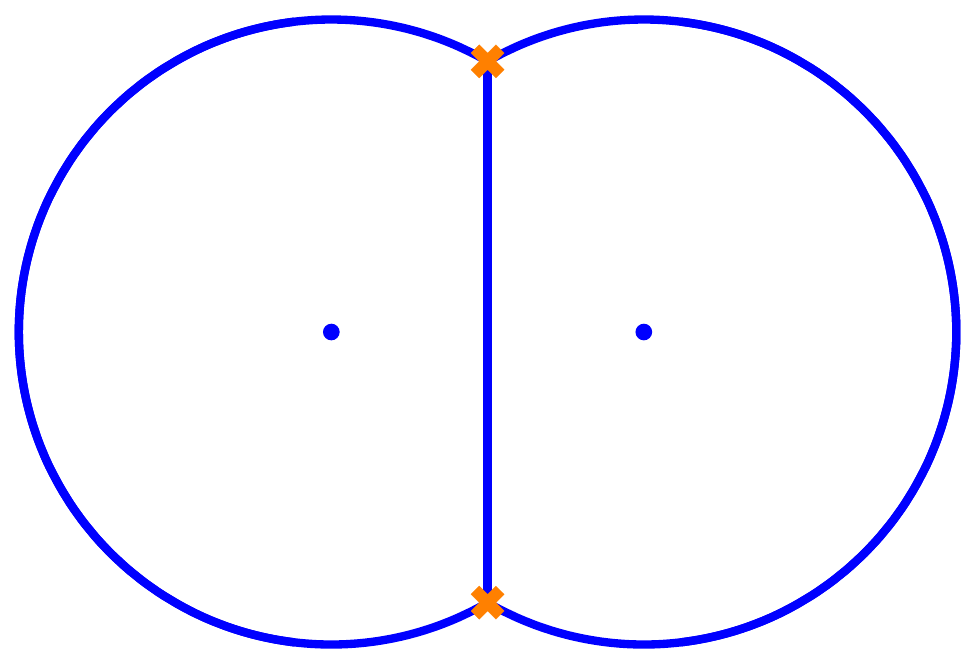}
        \caption{$\vartheta=\vartheta_c$}
        \label{T2graph3}
    \end{subfigure}
    \caption{Spectral networks on the 3-punctured sphere (with punctures at $\{\infty,1,-1\}$) for the $T_2$ theory. As the phase reaches its critical value~$\vartheta_c$, three double walls appear simultaneously and form the BPS graph, shown in~(d). Compare with Figure~\ref{T2BPSstates}.}
    \label{T2graph}
\end{figure}

In Figure~\ref{T2graph}, we choose $M_2^a=M_2^b=M_2^c=-1$ and we see three double walls appear at the same critical phase~$\vartheta_c=\pi/2$ (we obtained  a puncture at infinity by first setting $z_a=0$ in~\eqref{phi2T2} and then inverting $z$). 
This gives a BPS graph with three edges connecting the two branch points of the cover $\Sigma$.
Note that these three elementary edges can be identified with the three BPS states that appeared in Figures~\ref{T2BPS1}, \ref{T2BPS2}, \ref{T2BPS4}. They provide a positive integral basis for the charge lattice $\Gamma_+$, as discussed at the end of Section~\ref{sec:bps-states}.
In contrast, the BPS bound state which zig-zags between the punctures in Figure~\ref{T2BPS3} does not appear directly in the BPS graph, but would correspond to a combination of the three elementary edges.

\begin{figure}
\centering
\begin{minipage}{.34\textwidth}
\centering
\includegraphics[width=\linewidth]{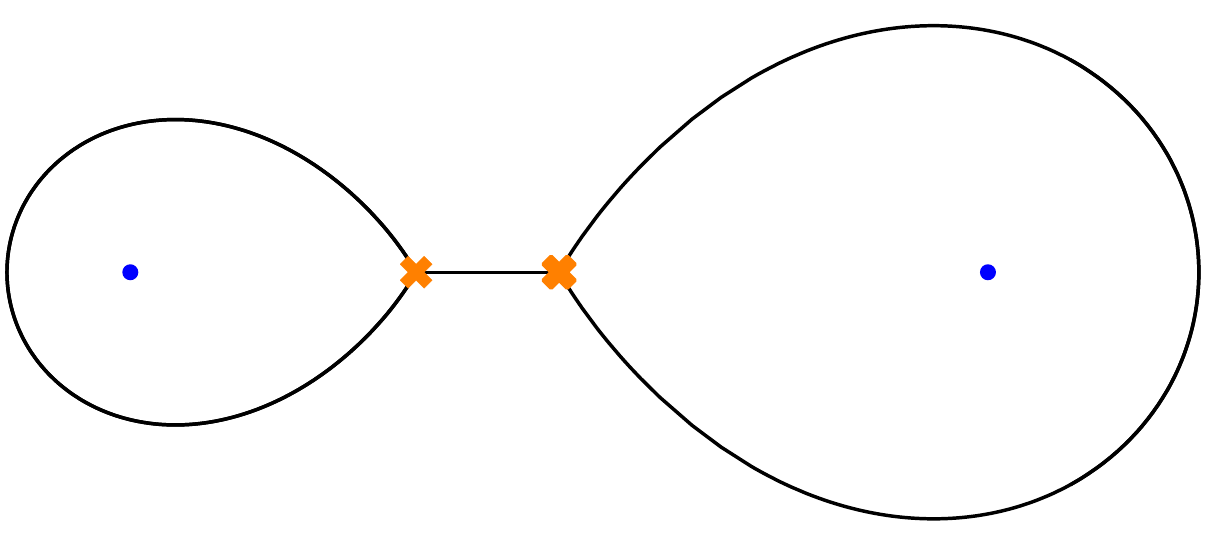}
\end{minipage}\hfill
    \caption{Another topology for the BPS graph on the 3-punctured sphere.}
    \label{T2graphFlipped}
\end{figure}

At a different point of $\cR$ we can obtain a BPS graph with a different topology. For example, with $M_2^a=-3$, $M_2^b=-1$, $M_2^c=-1/2$ we obtain the BPS graph in Figure~\ref{T2graphFlipped}.
These two topologies correspond to two components of $\cR$ that are separated by a singular interface on which two branch points collide and a BPS state becomes massless.

\subsection{\texorpdfstring{$SU(2)$ gauge theory with four flavors}{SU(2) with four flavors}}
\label{subsec:4-puncture-A1}

\begin{figure}
\centering
\begin{minipage}{.3\textwidth}
\centering
\includegraphics[width=\linewidth, trim={0 0 0 0}]{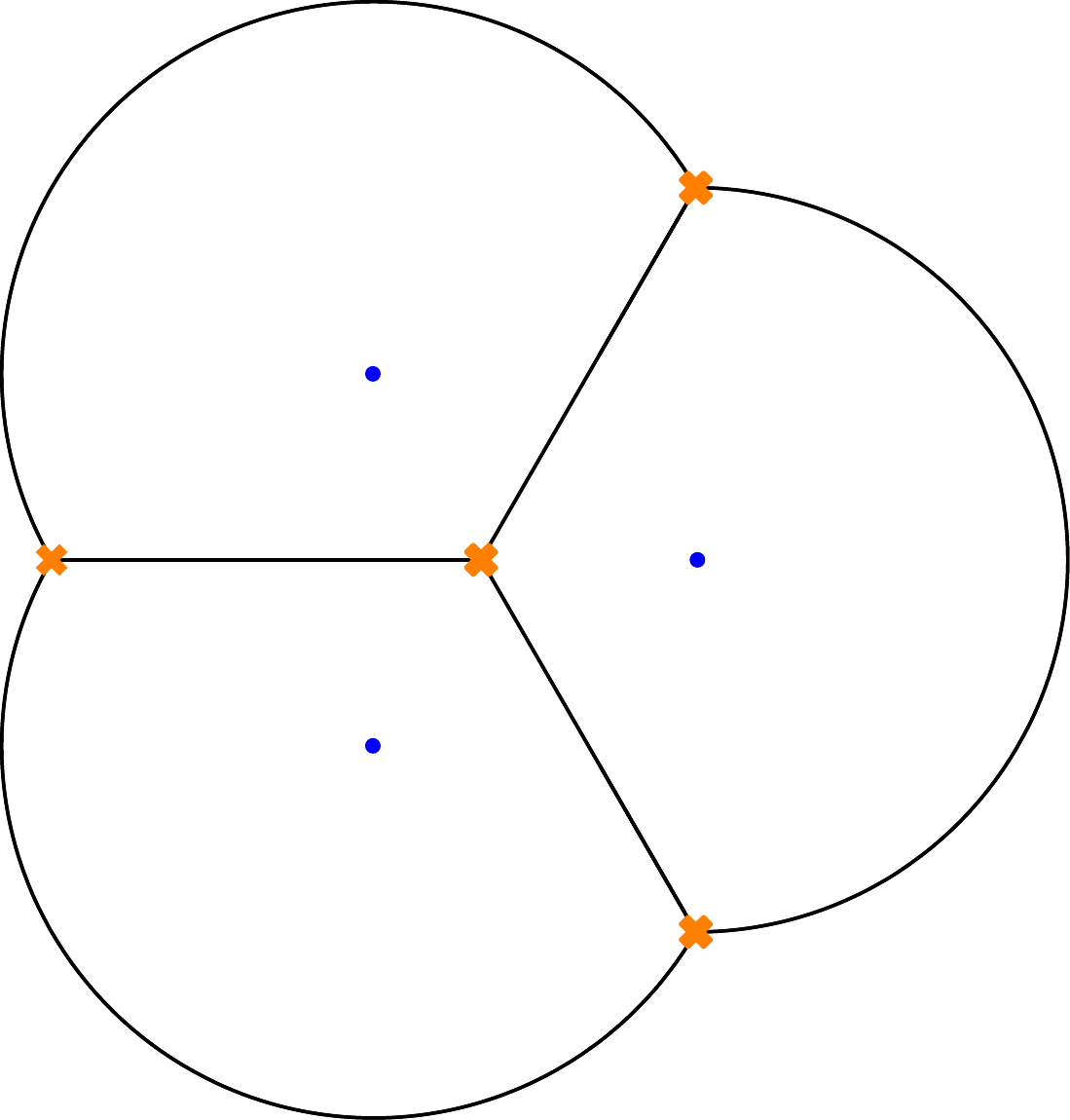}
\end{minipage}\hfill
\begin{minipage}{.34\textwidth}
\centering
\includegraphics[width=\linewidth, trim={0 0 0 0}]{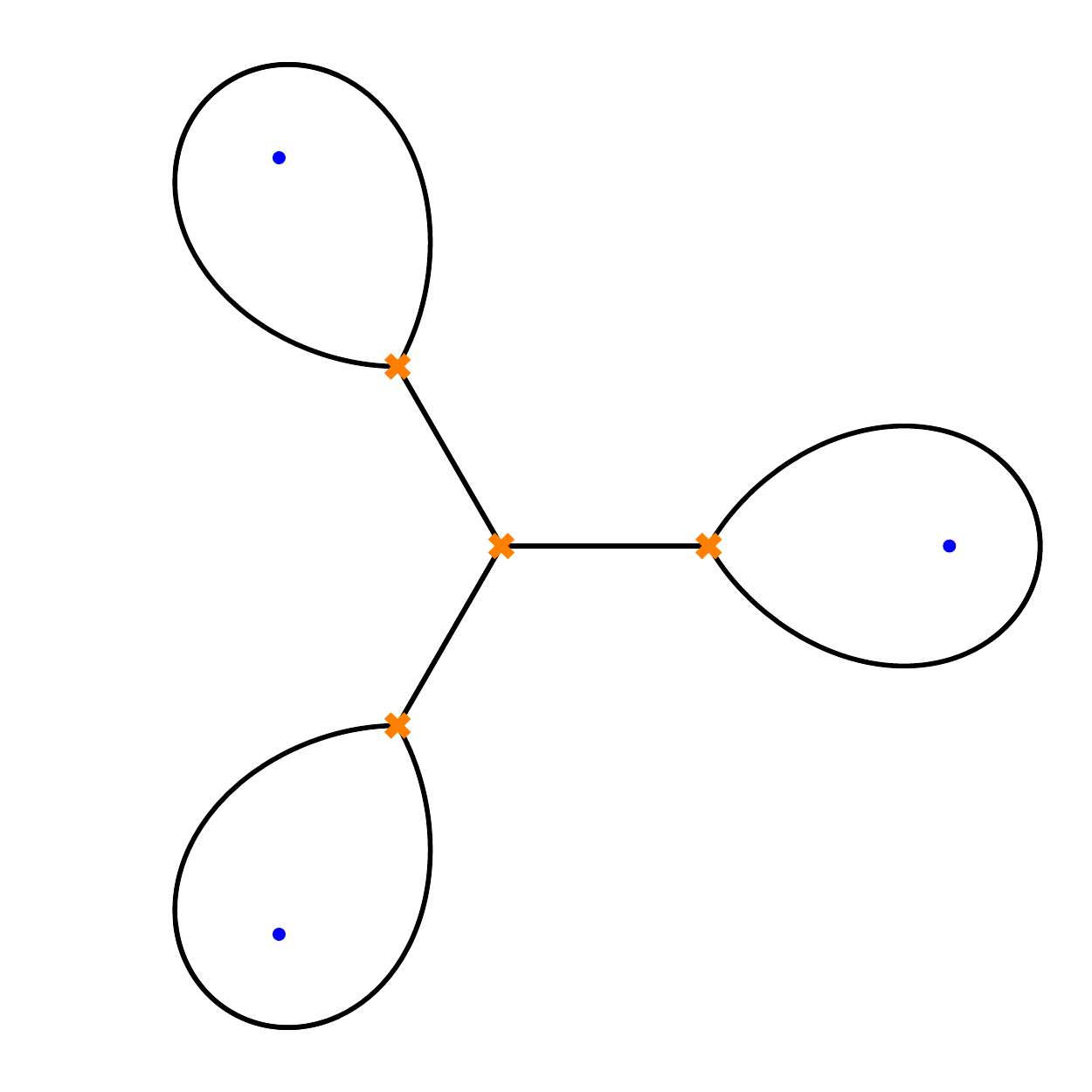}
\end{minipage}
    \caption{Two BPS graphs on the 4-punctured sphere, with different topologies.}
    \label{4punctures-A1}
\end{figure}

Our next example is the sphere with four regular punctures, which corresponds to $SU(2)$ super-QCD with  $N_f=4$ flavors.
The quadratic differential is
\be \label{phi24P}
\phi_2 = \frac{M_2^az_{ab} z_{ac} z_{ad} (z\! -\!z_b)(z\!-\!z_c)(z\!-\!z_d)  +\text{cyclic} +u \, (z\! -\! z_a) (z \!- \!z_b) (z\! -\! z_c)(z\! -\! z_d)} {(z-z_a)^2(z-z_b)^2(z-z_c)^2(z-z_d)^2  } (\dd z)^2 \;, 
\ee
with four mass parameters $M_2^{a,b,c,d}$ and one Coulomb branch parameter $u$.
To obtain the BPS graph, it is convenient to fix the complex structure of~$\CC$ by placing the punctures at $z_a=\infty$ and $z_{b,c,d} = 1, \ex^{\pm 2\ii\pi/3}$. We can then find a sublocus of $\cR$ by setting $u=0$ and identifying the phases of the mass parameters. We must also set $M_2^{b}=M_2^{c}=M_2^{d}$ in order to get a connected graph. BPS graphs for the choices $M_2^{a,b,c,d}=-3$, and $M_2^{a}=-10$, $M_2^{b,c,d}=-1$ are plotted in Figure~\ref{4punctures-A1}.
Each one has six elementary edges connecting the four branch points of $\Sigma$. This matches the rank $2r+f$ of the charge lattice $\Gamma$ given that $r=1$ and $f=4$.

\subsection{\texorpdfstring{Argyres-Douglas theories}{Argyres-Douglas theories}}\label{secADtheories}

\begin{figure}
\centering
\begin{minipage}{.31\textwidth}
\centering
\includegraphics[width=\linewidth]{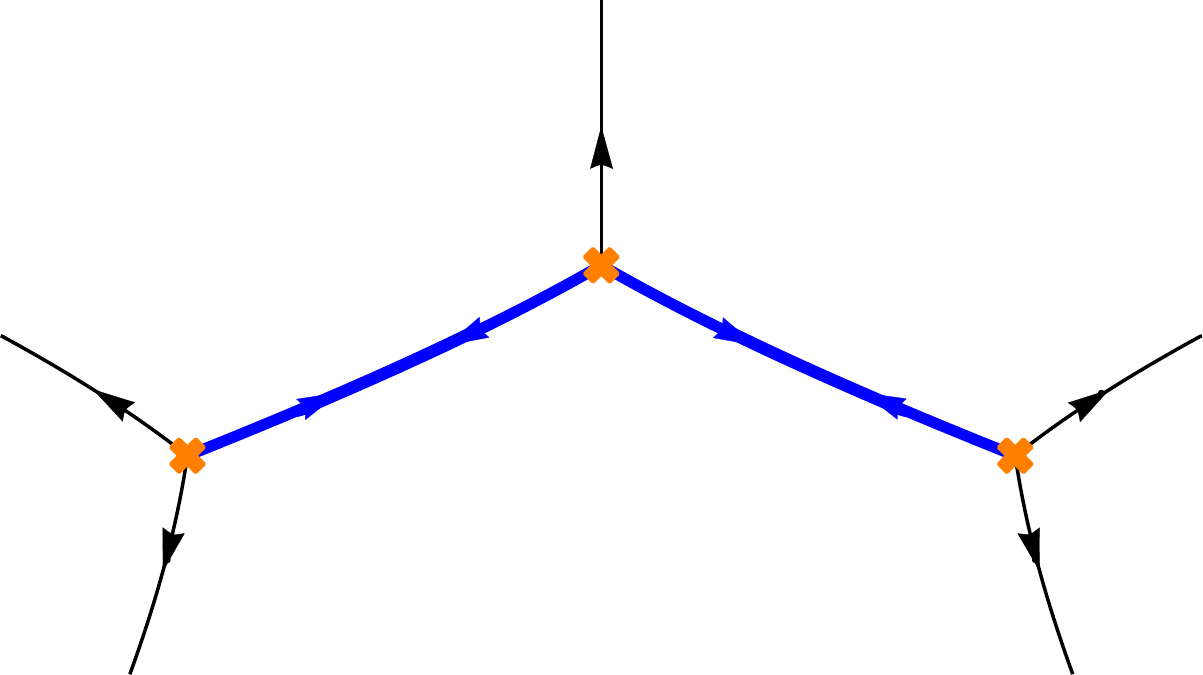}
\end{minipage}\hfill\vline\hfill
\begin{minipage}{.33\textwidth}
\centering
\includegraphics[width=\linewidth]{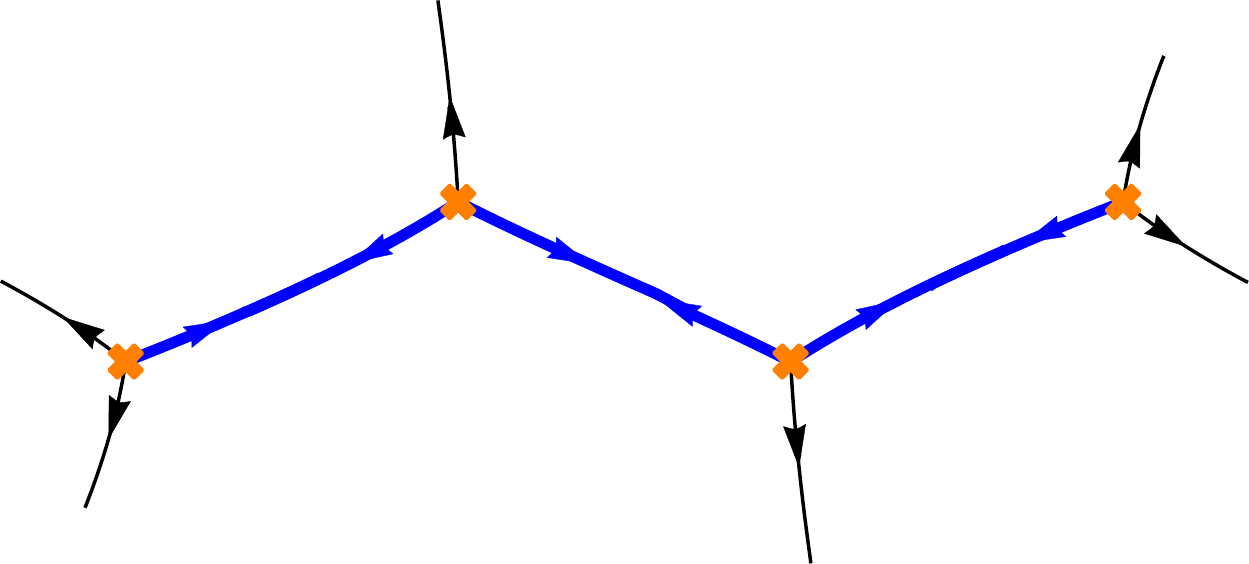}
\end{minipage}\hfill\vline\hfill
\begin{minipage}{.34\textwidth}
\centering
\includegraphics[width=\linewidth]{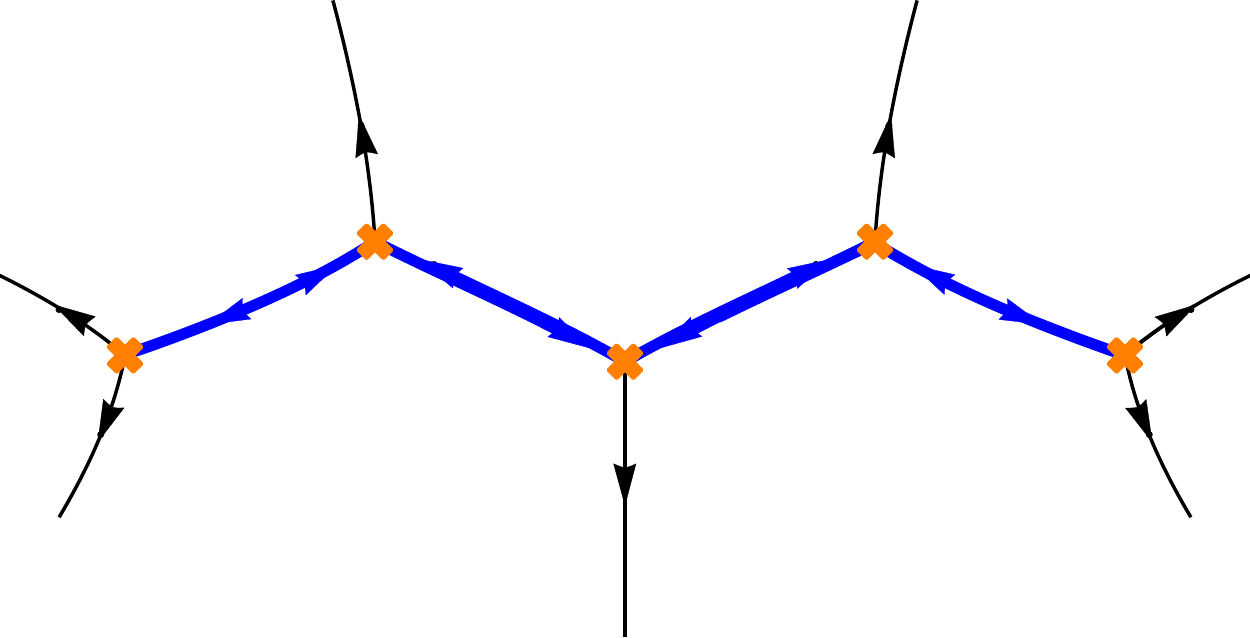}
\end{minipage}
    \caption{BPS graphs on the sphere with one irregular puncture at $z=\infty$ for the Argyres-Douglas theories AD$_n$ with $n=3,4,5$.}
    \label{ADgraphs}
\end{figure}

A simple example involving an irregular puncture is provided by the series of $A_1$ theories known as Argyres-Douglas theories AD$_{n}$ (or $(A_1, A_{n-1})$ theories)~\cite{Gaiotto:2009hg, Xie:2012hs, Gaiotto:2012db}.
In this case, the surface~$\CC$ is a sphere with one irregular puncture (say at $z=\infty$), which we can represent as a disk with $n+2$ marked points on the boundary, and the Seiberg-Witten curve $\Sigma$ at the superconformal point is given by 
\be
\lambda^2 = z^{n} (\dd z)^2 .
\ee
We can turn on some Coulomb branch and mass deformations, and the charge lattice $\Gamma$ is of rank
\be
\text{rank} \ \Gamma = n-1.
\ee
To find the BPS graph we choose the following quadratic differential:
\be
\phi_2 = -(z-z_1)(z-z_2)\cdots (z-z_n) (\dd z)^2,
\ee
where $z_1,z_2,\ldots, z_n$ are the locations of the branch points of $\Sigma$. We place them at unit intervals on the real axis and then adjust their positions on the imaginary axis in order to have all double walls appear at the same phase $\vartheta_c$. The prototypical double wall that we showed in Figure~\ref{DoubleWall} in fact corresponds to the BPS graph for the AD$_2$ theory. We show BPS graphs for the AD$_n$ theories with $n=3,4,5$ in Figure~\ref{ADgraphs}. The maximally degenerate spectral networks have $n-1$ double walls between pairs of branch points, and $n+2$ single walls that escape towards the irregular puncture at $z=\infty$. The BPS graphs are made principally of the $n-1$ double walls, but we also keep track of the one or two ``empty'' edges at each trivalent vertex.

\subsection{\texorpdfstring{$SU(2)$ gauge theory with one flavor}{SU(2) with one flavor}}\label{SU2Nf1}

\begin{figure}
\centering
\includegraphics[width=0.17\textwidth]{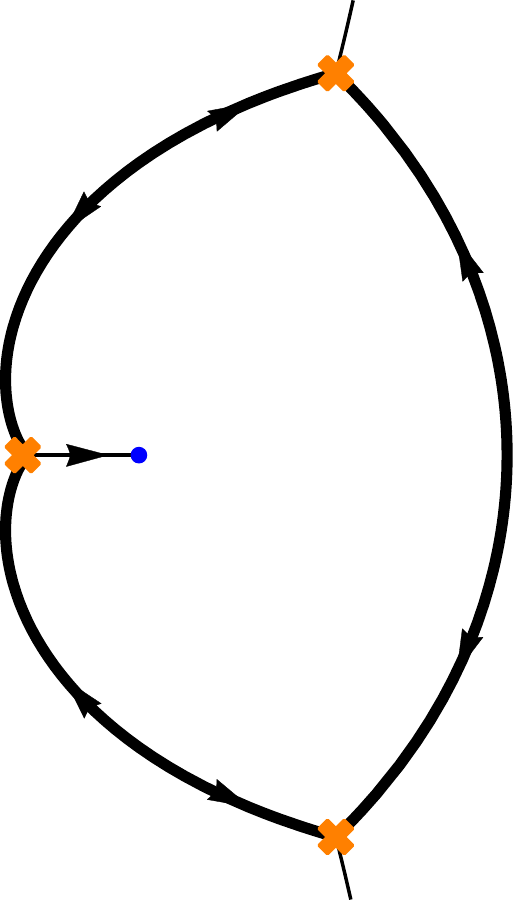}
    \caption{BPS graph on the sphere with two irregular punctures for $SU(2)$ gauge theory with $N_f=1$ flavor.}
    \label{Nf1graph}
\end{figure}

We can also consider $SU(2)$ gauge theory with $N_f=1$ flavor, corresponding to an annulus with one and two marked points on the boundary components (or equivalently a sphere with two irregular punctures at $z=0,\infty$).
The Seiberg-Witten curve has the following form~\cite{Gaiotto:2009hg}:
\bea
\lambda^2 = \left( \frac{\Lambda^2}{z^3} + \frac{3u}{z^2} + \frac{2 \Lambda m }{z} + \Lambda^2 \right) (\dd z)^2.
\eea
We find a BPS graph by setting for example $\Lambda=1$, $u=3/5$, and $m=-1/2$ (Figure~\ref{Nf1graph}).
The Coulomb branch looks like the sketch shown on the right of Figure~\ref{CBNf1}.

\subsection{\texorpdfstring{Strebel differentials and Fenchel-Nielsen networks}{Strebel differentials}}\label{StrebelFN}

BPS graphs for $A_1$ theories of class~$\CS$ are closely related to Strebel differentials~\cite{Strebel} and to the ``Fenchel-Nielsen networks'' introduced in~\cite{Hollands:2013qza}.

For an $A_1$ theory, the spectral curve $\Sigma$ given in~\eqref{SigmaNfold} is determined by a meromorphic quadratic differential of the form $\phi_2 = \phi(z) \dd z^2$ on the surface~$\CC$.
A curve $\xi(t)$ on~$\CC$ is called \emph{horizontal} if it satisfies 
\be\label{eq:horizontal-leaves}
	\phi\left(\xi(t)\right) \left(\xi'(t)\right)^2 \ \in \ \bR_+ \;.
\ee
A \emph{Strebel differential} is such that the set of its open horizontal curves is of measure zero~\cite{Strebel}.
For the case of a surface~$\CC$ with $n$ punctures (without boundary), a Strebel differential induces a metric that makes it look like $n$ cylinders glued along a \emph{critical graph}.
Strebel's theorem guarantees the existence of such a differential, which is moreover unique if the residues at the poles are in $\bR_+$ (corresponding to the radii of the cylinders).

Comparing the horizontality condition~\eqref{eq:horizontal-leaves} with the differential equation~\eqref{eq:S-wall} that characterizes the walls of a spectral network, we see that imposing that $\ex^{-2\ii\vartheta} \phi_2$ be a Strebel differential implies that the resulting spectral network defines a foliation of~$\CC$ with only compact leaves (in the absence of irregular punctures). For the 3-punctured sphere, the differential~\eqref{phi2T2} is indeed well-known to be Strebel when the parameters $M_2^{a,b,c}$ have the same phase, and the corresponding critical graph is precisely the BPS graph. 

However, for more general surfaces~$\CC$, the critical graph of a Strebel differential typically consists of disconnected copies of critical graphs for the 3-punctured sphere, like the ones in Figures~\ref{T2graph3} and~\ref{T2graphFlipped}. This defines a pair-of-pants decomposition of~$\CC$, and the related spectral networks were used in~\cite{Hollands:2013qza} to construct complexified Fenchel-Nielsen coordinates on the moduli space of flat $SL(2,\bC)$-connections on~$\CC$.

In contrast, it follows from the definition of a BPS graph that it only has one connected component (otherwise some BPS states would appear away from the critical phase $\vartheta_c$, see the example in Figure~\ref{4Ppants}).
We thus conclude that $A_1$ BPS graphs arise from a special type of Strebel differentials whose critical graphs are connected. In the terminology of~\cite{Hollands:2013qza}, they agree with fully \emph{contracted} Fenchel-Nielsen networks, which are obtained from generic Fenchel-Nielsen networks by merging all pants graphs together. 
It leads to a decomposition of~$\CC$ into punctured disks, at the intersection of many pants decompositions.

\subsection{\texorpdfstring{BPS graphs from ideal triangulations}{BPS graphs from triangulations}}\label{secTriangulations}

Even though we know from Section~\ref{subsec:existence} that BPS graphs exist for all $A_1$ theories of class~$\CS$, it is often hard to find the explicit form of the quadratic differentials $\phi_2$ that produce them. Luckily, for most applications, we are only interested in the topology of a BPS graph, and we can resort to a relation to ideal triangulations to obtain \emph{topological} BPS graphs.

A BPS graph for an $A_1$ theory is indeed naturally dual to an ideal triangulation of the associated punctured surface~$\CC$. Trivalent vertices (branch points) are dual to triangles ending on punctures, and edges of the BPS graph are transverse to edges of the triangulation. 

\begin{figure}[h!]
\centering
\begin{minipage}{.37\textwidth}
\centering
\includegraphics[width=\linewidth ]{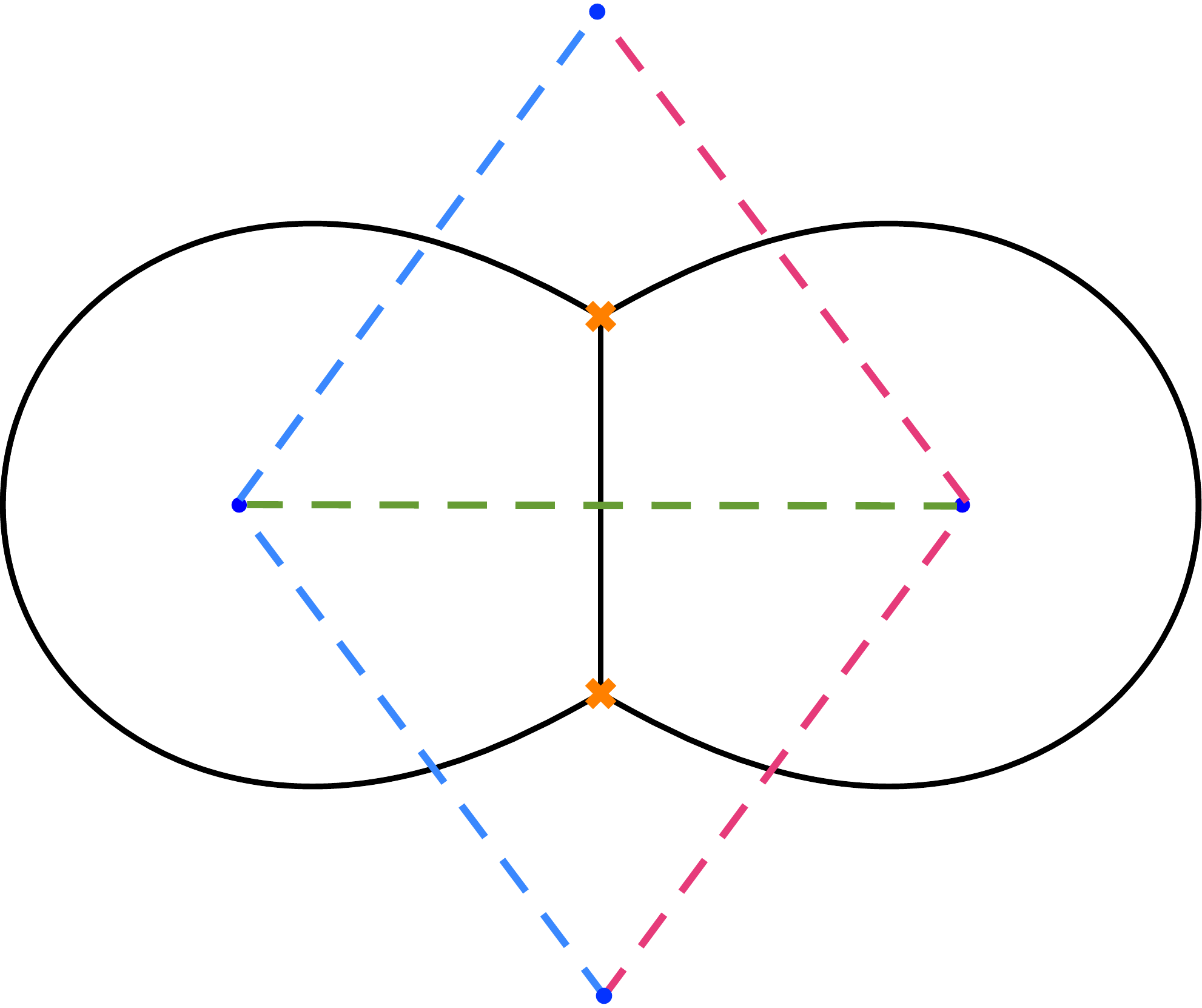}
\end{minipage}\hfill
\begin{minipage}{.3\textwidth}
\centering
\includegraphics[width=\linewidth ]{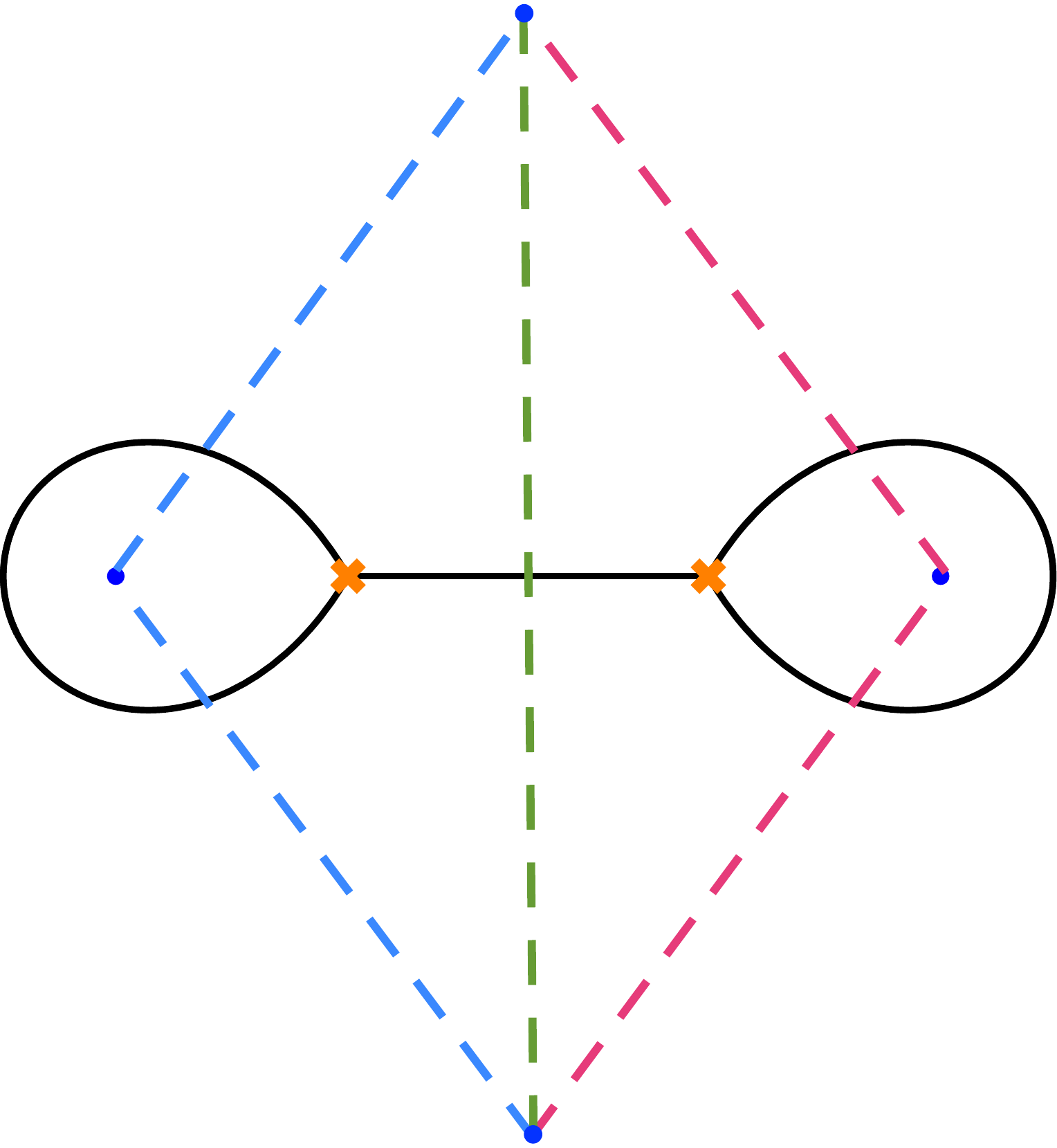} 
\end{minipage}
    \caption{Ideal triangulations (dashed) dual to two BPS graphs for the $T_2$ theory (the top and bottom punctures represent the same puncture at $\infty$, and triangle sides are identified accordingly).}
    \label{figT2molec12}
\end{figure}

\begin{figure}
\centering
\begin{minipage}{.29\textwidth}
\centering
\includegraphics[width=\linewidth ]{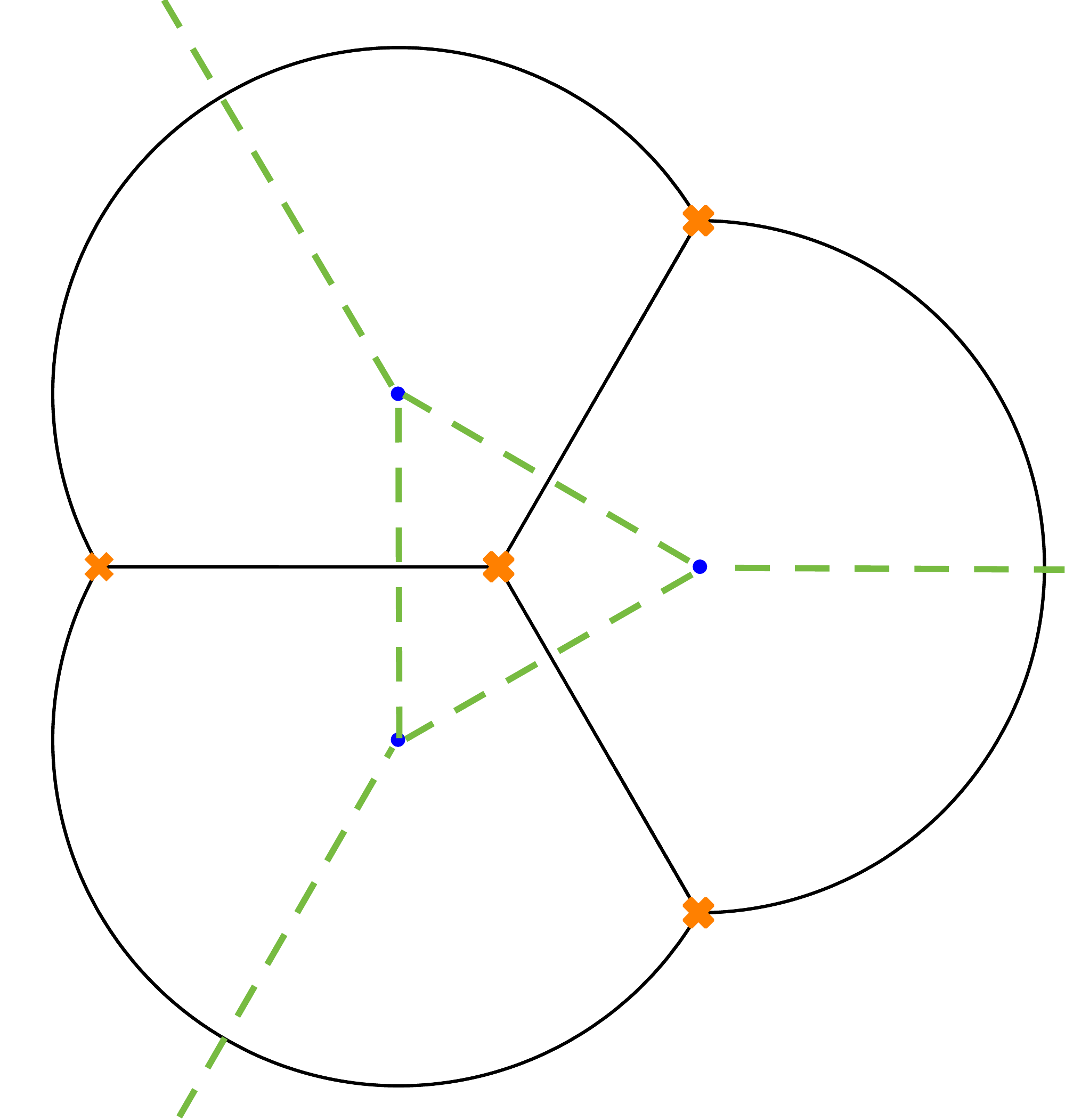}
\end{minipage}\hfill
\begin{minipage}{.29\textwidth}
\centering
\includegraphics[width=\linewidth ]{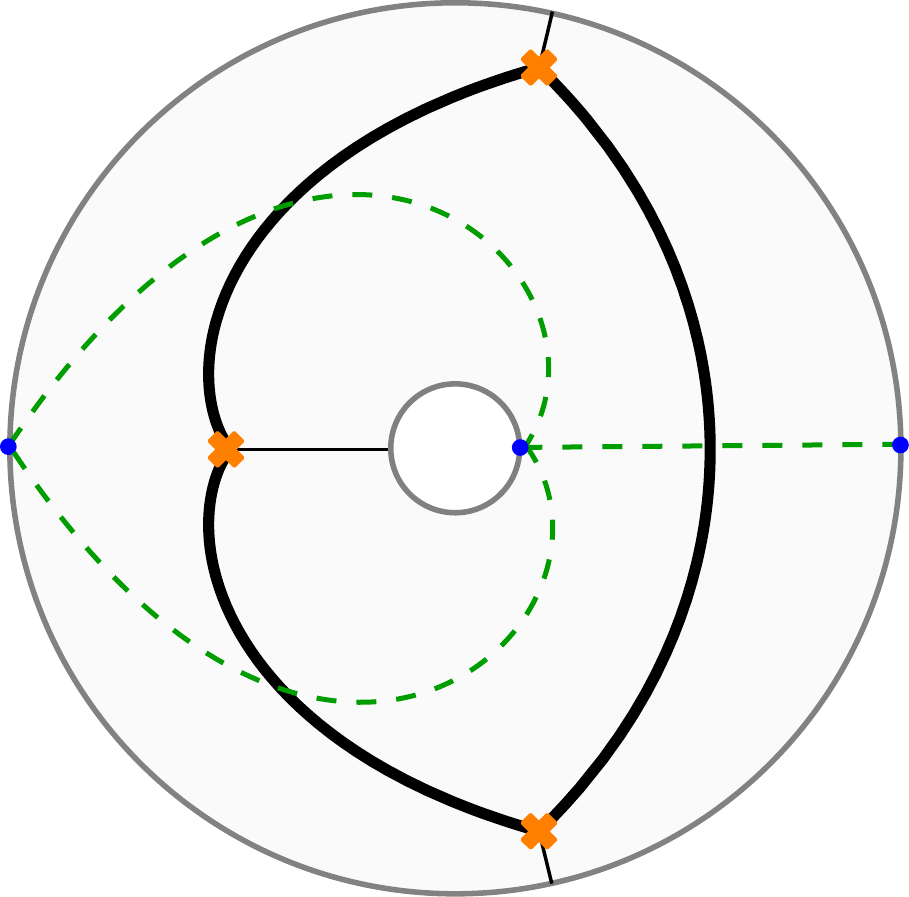} 
\end{minipage}
\hfill
\begin{minipage}{.31\textwidth}
\centering
\includegraphics[width=\linewidth ]{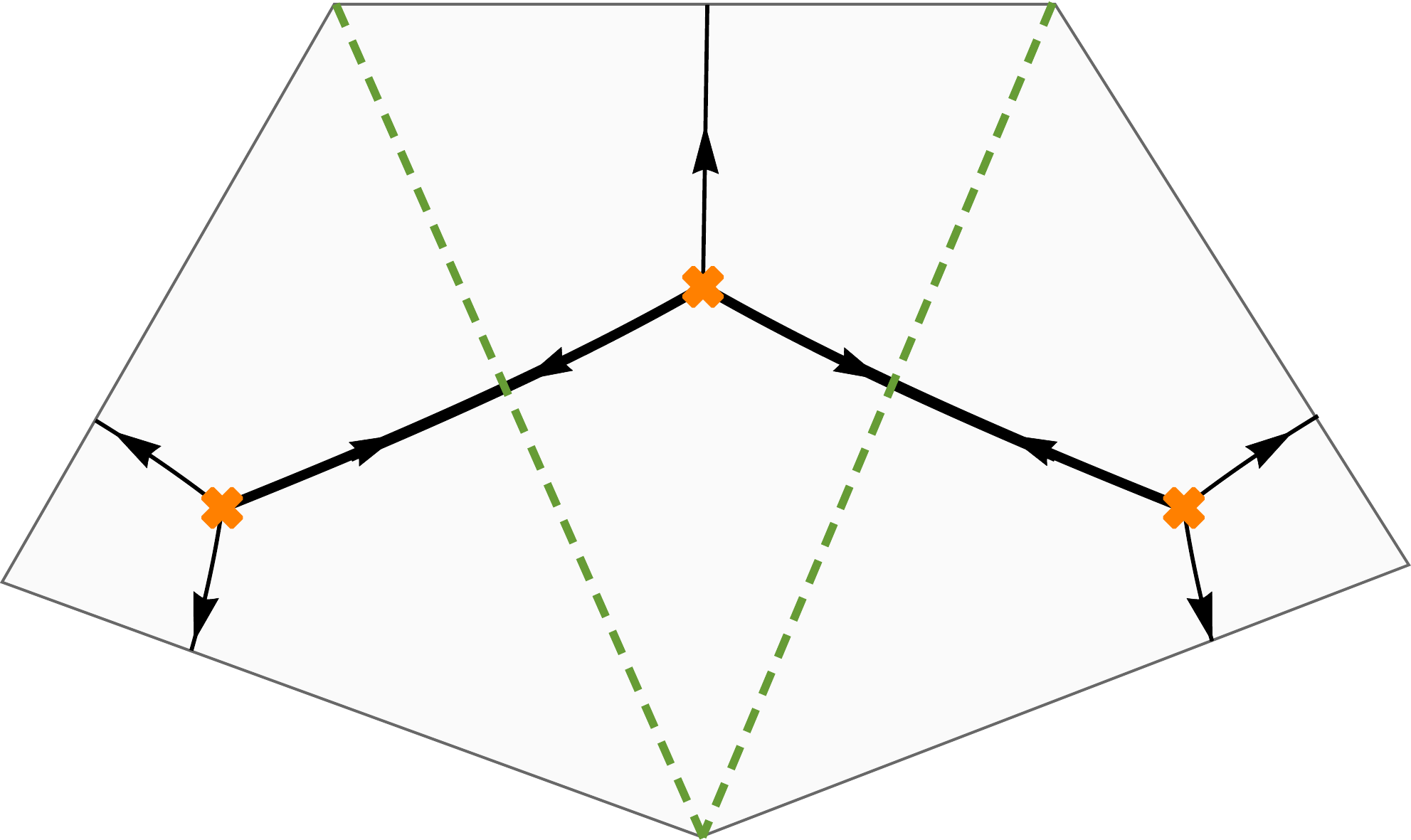} 
\end{minipage}
    \caption{Ideal triangulations (dashed) dual to BPS graphs for various surfaces~$\CC$. \emph{Left}: 4-punctured sphere, dual to $SU(2)$ gauge theory with $N_f=4$ (the three outer edges meet at the puncture at infinity). \emph{Middle}: Annulus with one marked point on the inner boundary and two on the outer, dual to $SU(2)$ gauge theory with $N_f=1$. \emph{Right}: Disk with five marked points on the boundary, dual to the Argyres-Douglas AD$_3$ theory.}
    \label{4pAD3triangulations}
\end{figure}

For example, the BPS graphs for the $T_2$ theory in Figures~\ref{T2graph3} and~\ref{T2graphFlipped} are dual to the ideal triangulations shown in Figure \ref{figT2molec12}. We see that the topological transformation between the two BPS graphs on different components of $\cR$ corresponds to a \emph{flip} between the dual triangulations. Note that a vertex with a loop in the BPS graph is dual to a self-folded triangle. In Figure~\ref{4pAD3triangulations}, we show the ideal triangulations dual to some of the other examples that we have encountered so far. 

As a remark, recall that the relation between spectral networks and ``WKB triangulations'' described in~\cite{Gaiotto:2009hg} was such that the appearance of a double wall at $\vartheta_c$ signaled the transition point between two well-defined triangulations. From this perspective, since BPS graphs are entirely made of double walls, they wouldn't seem to correspond to any specific WKB triangulation. However, given a BPS graph at $\vartheta_c$, we can find the associated WKB triangulation by considering a related spectral network at any $\vartheta\neq \vartheta_c $, which by definition does not contain any double wall. This is particularly easy to do at $\vartheta= \vartheta_c \pm \pi/2$, as can be seen for the $T_2$ theory in Figure~\ref{T2graph0}.

\begin{figure}
\centering
\includegraphics[width=\textwidth]{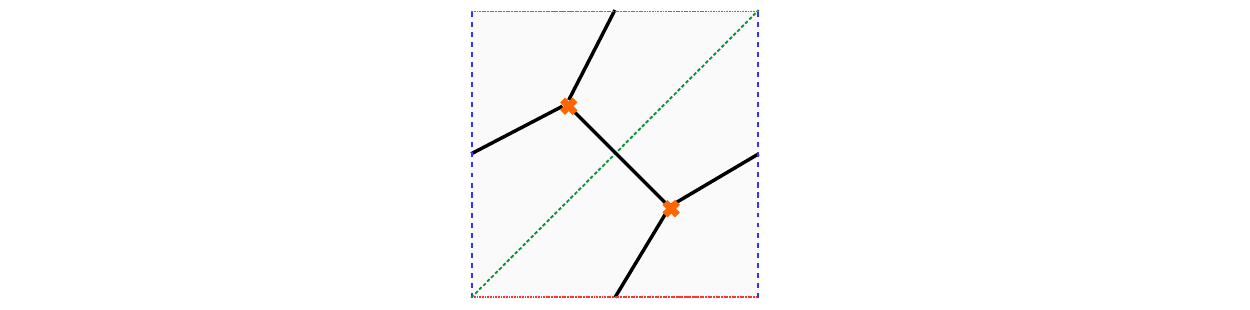}
    \caption{Topological BPS graph on the torus with one puncture for the $\cN=2^*$ $SU(2)$ gauge theory (opposite sides of the triangulation are identified).}
    \label{torusGraph}
\end{figure}

Given the trivalent property of BPS graphs, we expect that their relation to ideal triangulations extends to arbitrary Riemann surfaces~$\CC$ with any genus and any number of punctures or boundary components.
This provides us with a powerful method to find a topological representative of a BPS graph for any $A_1$ theory of class~$\CS$: we simply choose an ideal triangulation of~$\CC$ and take its dual graph. Different choices of triangulations should correspond to different components of $\cR$. As an example, we show in Figure~\ref{torusGraph} a BPS graph on the torus with one puncture, associated with the $\cN=2^*$ $SU(2)$ gauge theory. 

\section{\texorpdfstring{BPS graphs for $A_{N-1}$ theories}{BPS graphs for A(N-1) theories}}
\label{sec:N-lift-graphs}

We now turn to BPS graphs for higher-rank $A_{N-1}$ theories of class~$\CS$ (with \emph{full} punctures). 
We obtain several explicit examples of $A_{N-1}$ BPS graphs by using the ``level-$N$ lift'' technique introduced in~\cite{Gaiotto:2012db}. 
We then formulate a general conjecture based on a relation to the $N$-triangulations featured in Fock and Goncharov's work on higher Teichm\"uller theory~\cite{FockGoncharovHigher}, and to the ``ideal bipartite graphs'' recently introduced by Goncharov~\cite{2016arXiv160705228G}.

\subsection{\texorpdfstring{$N$-lifts}{N-lifts}} \label{Lifts}

While general spectral networks of $A_{N-1}$ theories with $N>2$ can be very complicated, an interesting family of tractable spectral networks was identified in~\cite{Gaiotto:2012db}.
Their advantage is that they are closely related to  $A_1$ spectral networks.
The idea is to start with the spectral curve~$\Sigma$ for an $A_1$ theory,
\be \label{SigmaCurve_Phi}
\det (\lambda  - \varphi )=
\lambda^2 + \phi_2 = 0 \;,
\ee
where $\varphi$ is valued in $\mathfrak{sl}_{2}$,
and apply to it a homomorphism $\rho: SU(2) \to SU(N)$ given by the $N$-dimensional irreducible representation of $SU(2)$ to obtain the \emph{level-$N$ lift} of $\Sigma$:
\be \label{SigmaCurve_phi}
\det \left(\lambda  - \rho(\varphi) \right)=
\lambda^N +\sum_{k=2}^N  \lambda^{N-k} \phi_k = 0 \;.
\ee
The resulting curve is reducible, but can be made irreducible by turning on generic small perturbations~$\delta \phi_k$.  
This has the effect of splitting each branch point of the 2-fold cover into $\frac12 N(N-1)$ nearby branch points of the $N$-fold cover. 
Similarly, the associated $A_{N-1}$ spectral network first consists of $\frac12 N(N-1)$ superimposed copies of the original $A_1$ spectral network, which are then resolved by the perturbations.
Far away from branch points, the walls of the $N$-lifted spectral network align closely with the walls of the $A_1$ spectral network.
For generic $\delta\phi_k$, regular punctures in an $A_1$ theory give \emph{full} punctures in the $N$-lifted theory, because the homomorphism $\rho$ produces a residue matrix~\eqref{eq:regular-puncture} with distinct eigenvalues.

\begin{figure}
    \centering
    \begin{subfigure}[b]{0.4\textwidth}
        \includegraphics[width=\textwidth]{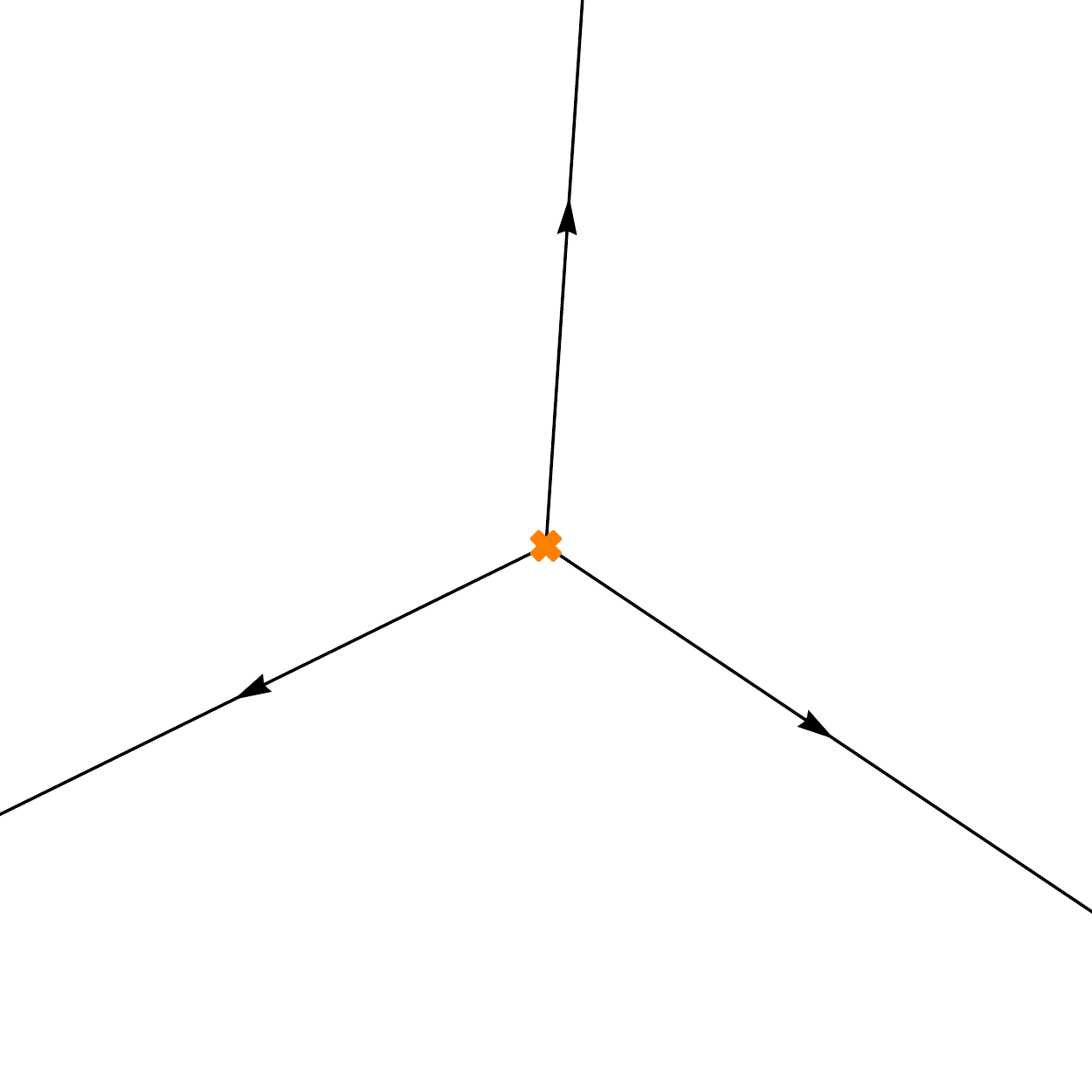}
        \caption{$N=2$}
        \label{AD1N2}
    \end{subfigure}
    \hfill
    \begin{subfigure}[b]{0.4\textwidth}
        \includegraphics[width=\textwidth]{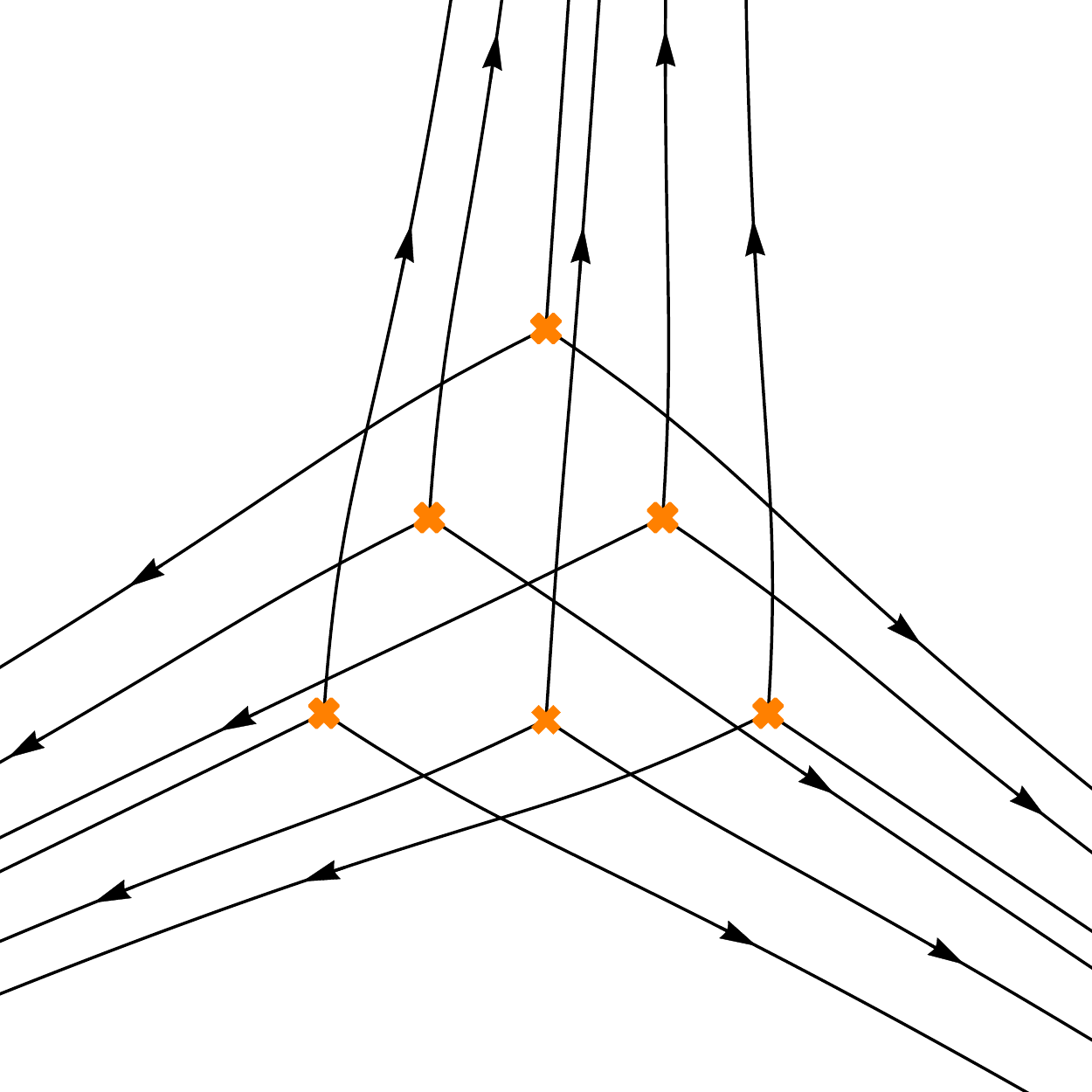}
        \caption{$N=4$}
        \label{level4lift2}
    \end{subfigure}
    \caption{The effect of a 4-lift on a branch point is to split it into six nearby branch points. The walls of the lifted $A_3$ spectral network asymptote to those of the $A_1$ spectral network far away from the branch points. 
    }
    \label{level4lift}
\end{figure}

For example, for $N=3,4$, this lifting procedure yields the following curves:
\bea\label{eq:K-lifted-curves}
\lambda^3 + (4\phi_2 + \delta \phi_2) \lambda + \delta \phi_3 &=& 0 \;,\\
\lambda^4 + (10\phi_2 + \delta \phi_2) \lambda^2 + \delta \phi_3 \lambda + ( 9 \phi_2^2 + \delta\phi_4) &=& 0 \;.  
\label{eq:4-lifted-curve}
\eea
Figure~\ref{level4lift} shows the 4-lift of a single branch point, which results in six slightly separated branch points.

We want to use this lifting procedure to obtain $A_{N-1}$ BPS graphs from $A_1$ BPS graphs. 
For this purpose, it is crucial that the perturbations $\delta\phi_k$ that we turn on preserve the alignment of the central charges of the BPS states. It is relatively easy to guess such perturbations for symmetric configurations of punctures, but we do not have a general method to identify them.

\subsection{\texorpdfstring{$T_N$ theories}{TN theories}} \label{subsec.TN}

We can obtain $N$-lifted BPS graphs on the sphere with three full punctures for the $T_N$ theories starting from the BPS graphs for the $T_2$ theory presented in Section~\ref{sec:T2-example}.

\begin{figure}
\centering
\begin{subfigure}[b]{0.45\textwidth}
\centering
\includegraphics[width=\linewidth]{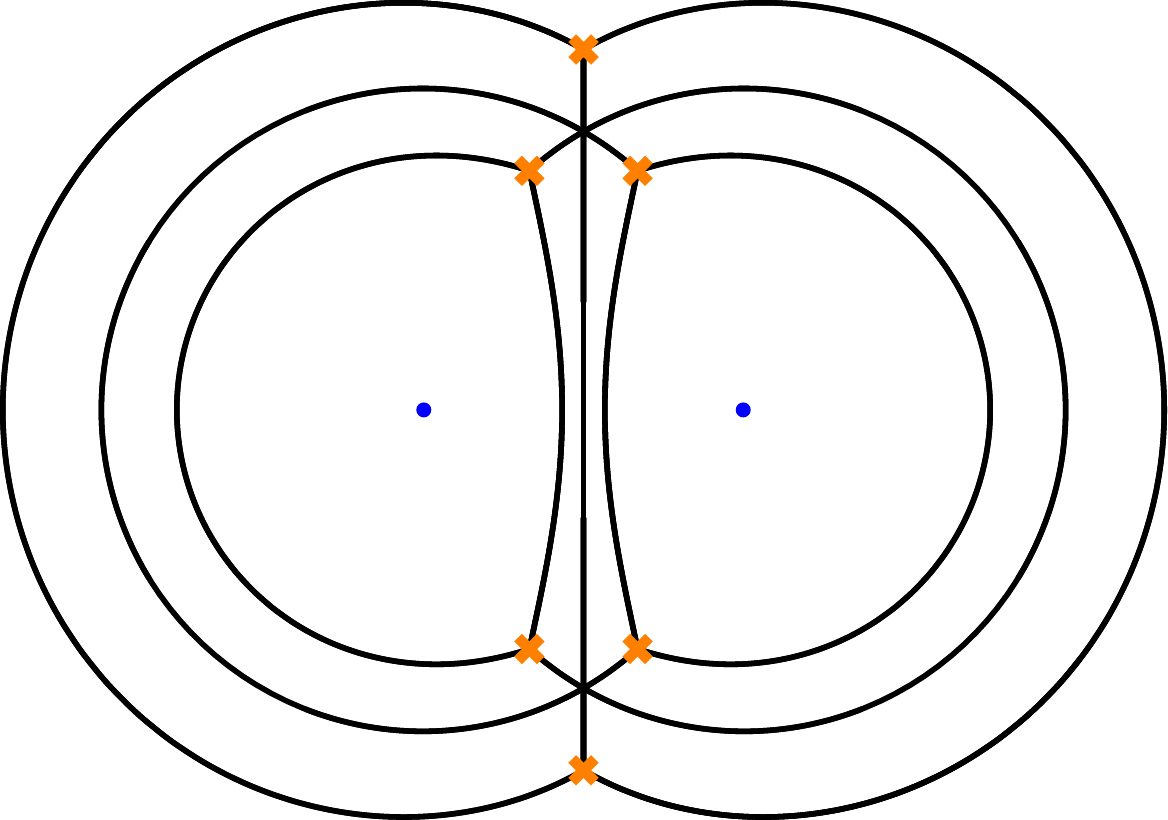}
\caption{}
\label{T3graphAll}
\end{subfigure}\hfill
\begin{subfigure}[b]{0.45\textwidth}
\centering
\includegraphics[width=\linewidth]{images/T3graphY}
\caption{}
\label{T3graphWebs}
\end{subfigure}\hfill
    \caption{\emph{Left}: Maximally degenerate 3-lifted spectral network on the 3-punctured sphere. \emph{Right}: $A_2$ BPS graph for the $T_3$ theory, with three pairs of edges and two Y-webs.}
    \label{T3graph}
\end{figure}

For $N=3$, the curve can be taken as in (\ref{eq:K-lifted-curves}) with $\phi_2$ and $\delta\phi_3$ as in~\eqref{phik}. We found the following alignment-preserving perturbations:
\be
	M_2^{a,b,c}=1\, ,\qqq  \delta M_3^{a,b,c}=-\frac{\ii}{10}\,.
\ee
This produces a maximally degenerate spectral network that looks like three nearby copies of the BPS graph for $T_2$, plotted on the left of Figure~\ref{T3graph}.
It contains a total of 12 finite webs: three triplets of edges, two Y-webs, and one web with two joints shaped like an horizontal $\theta$ (see Appendix~\ref{AppWebs} for a systematic identification of these finite webs). 
In order to find the BPS graph, we need to identify the $\text{rank}\,\Gamma = 8$ elementary webs which provide a positive integral basis for $\Gamma_+$. 
Direct examination reveals that among all the finite webs that go through the intersections, only the Y-webs are elementary.
We conclude that the BPS graph for the $T_3$ theory is made of three pairs of edges and two Y-webs, as shown on the right of Figure~\ref{T3graph}.

\begin{figure}
    \centering
        \includegraphics[width=0.5\textwidth]{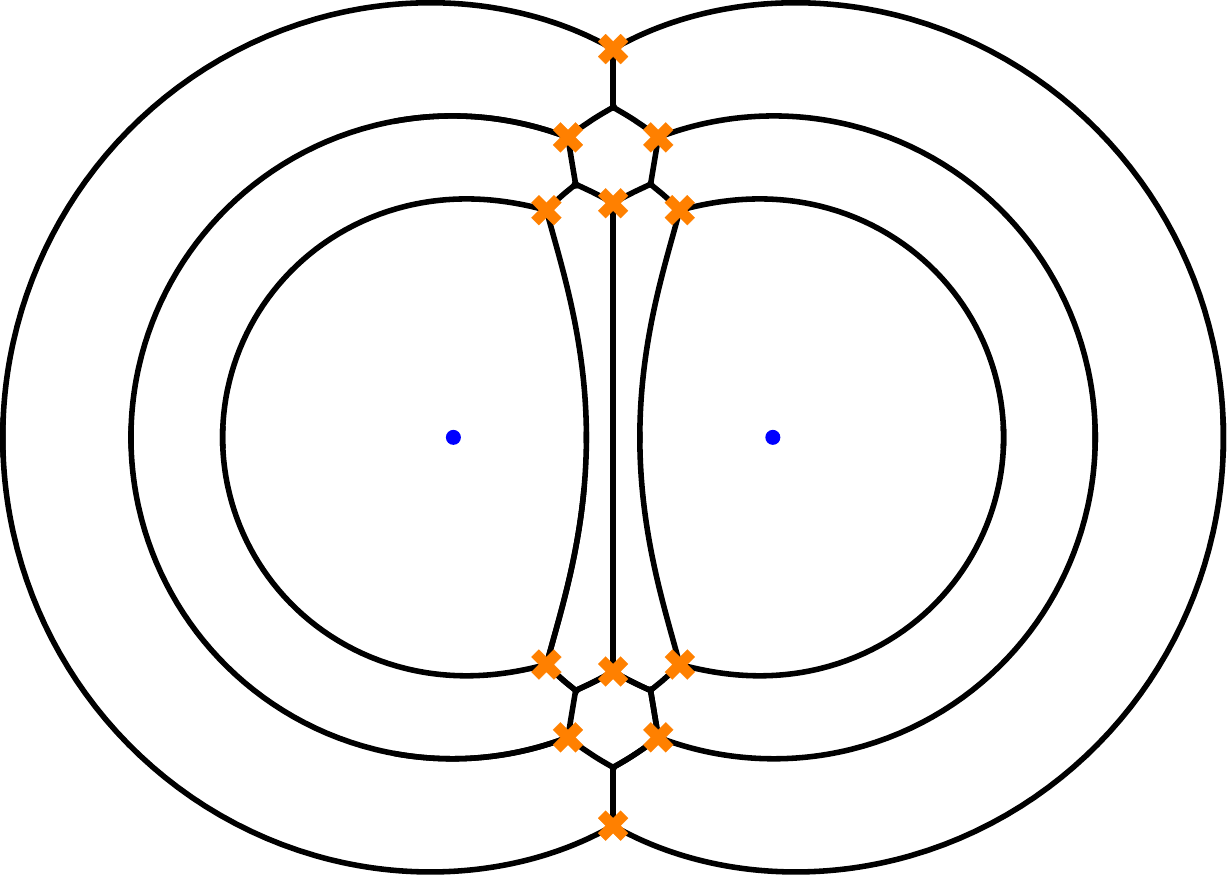}
    \caption{$A_3$ BPS graph for the $T_4$ theory, with 9 edges and 6 Y-webs.}
    \label{T4graph}
\end{figure}

For $N=4$, we set $M_2^{a,b,c}=1$ and take the alignment-preserving perturbations $\delta M_3^{a,b,c}=-\ii$ in~\eqref{eq:4-lifted-curve}. After eliminating redundant edges and $\theta$-webs, we find the BPS graph shown in Figure~\ref{T4graph}, with 9 edges and 6 Y-webs.

\subsection{\texorpdfstring{$A_{N-1}$ theories with four full punctures}{Four full punctures}}
\label{subsec:4-puncture-An}

We can also study the $N$-lifts of $A_1$ BPS graphs on the 4-punctured sphere obtained in Section~\ref{subsec:4-puncture-A1}, which correspond to non-Lagrangian $A_{N-1}$ theories with four full punctures.

\begin{figure}
\centering
\begin{minipage}{.34\textwidth}
\centering
\includegraphics[width=\linewidth]{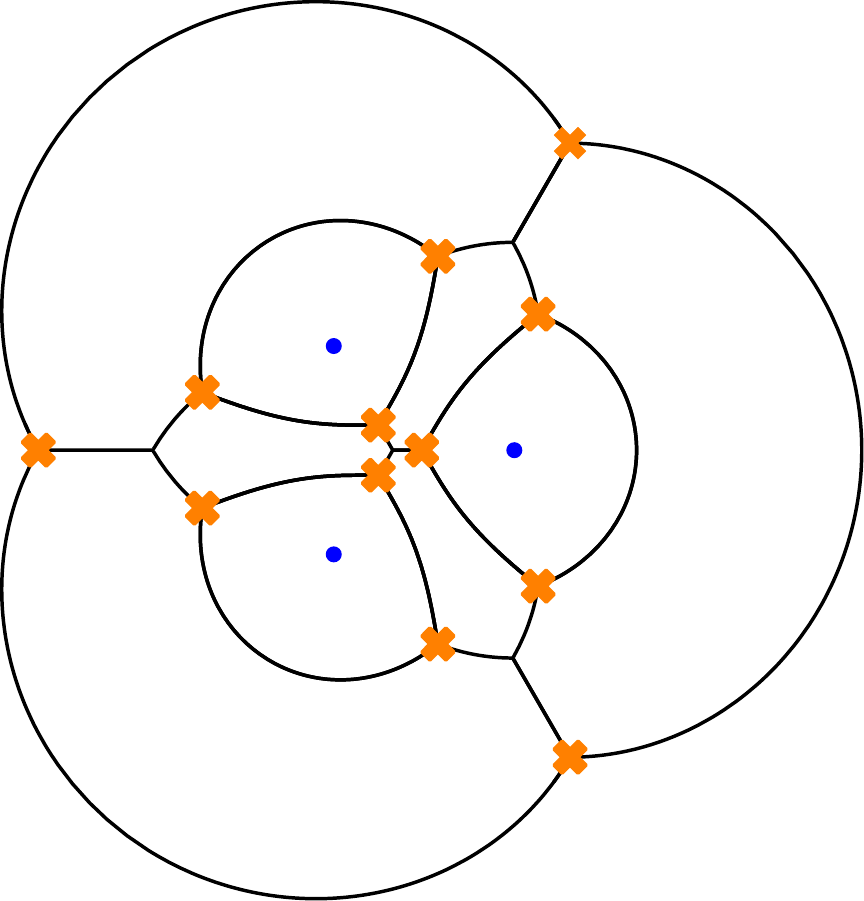}
\end{minipage}\hfill
\begin{minipage}{.36\textwidth}
\centering
\includegraphics[width=\linewidth]{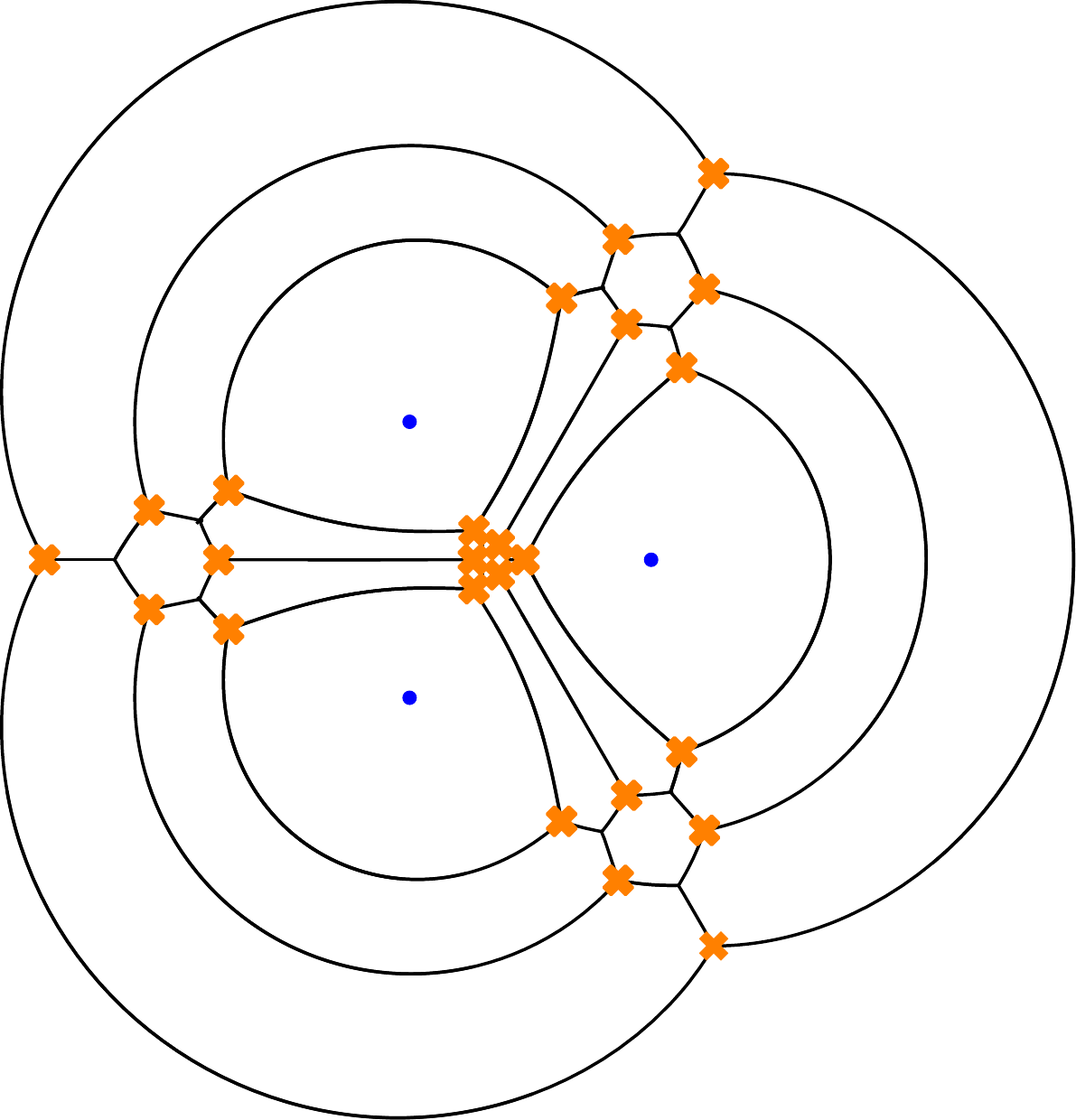}
\end{minipage}
    \caption{$A_2$ and $A_3$ BPS graphs on the sphere with four full punctures.}
    \label{4punctures}
\end{figure}

For $N=3$, we set $M_2^{a,b,c,d}=1$ and $u=0$ in the expression~\eqref{phi24P} for the quadratic differential and turn on the following perturbations:
\be
\delta \phi_3 = \frac{\delta M_3^a z_{ab}^2 z_{ac}^2 z_{ad}^2(z\! -\!z_b)(z\!-\!z_c)(z\!-\!z_d)  +\text{cyclic} +\delta P_2 (z\! -\! z_a) (z \!- \!z_b) (z\! -\! z_c)(z\!-\!z_d)} {(z-z_a)^3(z-z_b)^3(z-z_c)^3(z-z_d)^3  } (\dd z)^3 \, ,
\ee
with $\delta M_3^{a,b,c,d}=\ii$ and $\delta P_2 = 8\ii z^2$. 
For $N=4$, we also set $M_2^{a,b,c,d}=1$ and $u=0$, and turn on the perturbations $\delta M_3^{a,b,c,d}=2\ii$ and $\delta P_2 = 16\ii z^2$. 
The resulting $A_2$ and $A_3$ BPS graphs are plotted in Figure~\ref{4punctures}.

\subsection{\texorpdfstring{BPS graphs from $N$-triangulations}{BPS graphs from N-triangulations}}\label{subsec:N-triangulations}

\begin{figure}
\centering
\begin{minipage}{.3\textwidth}
\centering
\includegraphics[width=\linewidth]{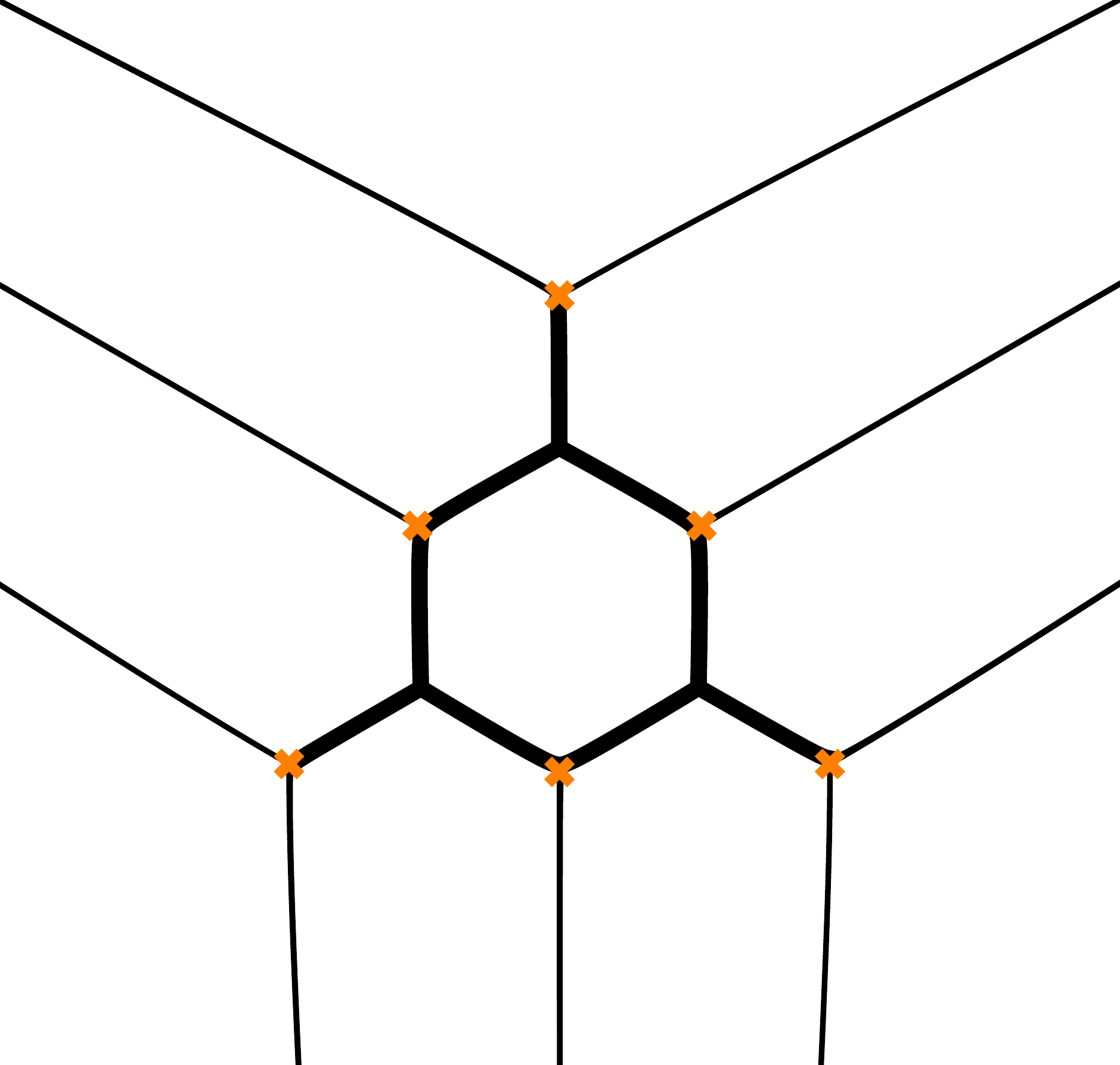}
\end{minipage}\hfill
\begin{minipage}{.34\textwidth}
\centering
\includegraphics[width=\linewidth]{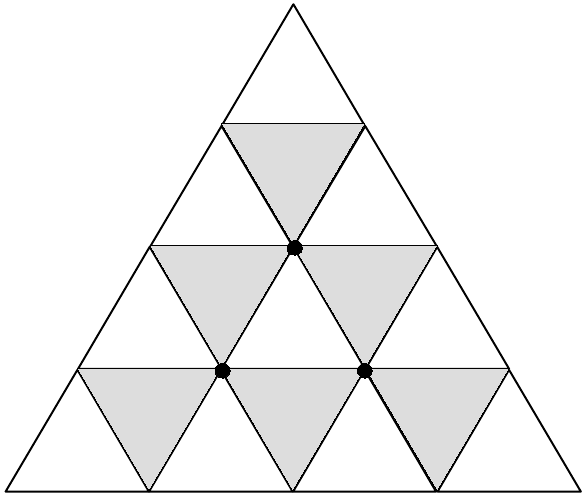}
\end{minipage}
    \caption{\emph{Left}: 4-lift of a branch point, with three Y-webs and three triplets of parallel edges. \emph{Right}: 4-triangulation of a triangle. Branch points correspond to black triangles, Y-webs to internal vertices, and edges to peripheral vertices.}
    \label{level4Ywebs}
\end{figure}

The examples of 3-lifted and 4-lifted BPS graphs in the previous sections present a clear combinatorial pattern that can easily be extended to higher $N$.
Moreover, in the spirit of the relation between topological $A_1$ BPS graphs and ideal triangulations of~$\CC$ discussed in Section~\ref{secTriangulations}, we can propose natural candidate $A_{N-1}$ BPS graphs for all higher-rank theories of class~$\CS$
with full punctures, without having to find suitable $k$-differentials $\phi_k$ explicitly. 
We simply start with an $A_1$ BPS graph and replace each branch point by a triangular array of $\binom{N}{2}$ branch points supporting $\binom{N-1}{2}$ Y-webs, and each edge by $(N-1)$ edges between different pairs of branch points (Figure~\ref{level4Ywebs}). 
In the case of a surface~$\CC$ with genus $g$ and $n$ full punctures, the ideal triangulation has $-2\chi$ triangles and $-3\chi$ edges, where $\chi = 2-2g-n$ is the Euler characteristic. 
The $N$-lifted BPS graph then contains a total of $-\chi(N^2-1)$ edges and Y-webs, which agrees with $\text{rank}\, \Gamma=2r+f$ using~\eqref{eq:flavor-rank} and~\eqref{eq:gauge-rank}.
It was indeed already pointed out in Section 4.3 of~\cite{Gaiotto:2012db} that this set of cycles forms a basis for $\Gamma$. 

We conjecture that the Roman locus $\cR$ exists for $A_{N-1}$ theories associated with Riemann surfaces with full punctures, and that it includes at least a point where the \emph{physical} BPS graph (as obtained from spectral networks) is topologically equivalent to the candidate BPS graph obtained by this construction.

\begin{figure}
\centering
\begin{minipage}{.3\textwidth}
\centering
\includegraphics[width=\linewidth]{images/T4graph}
\end{minipage}\hfill
\begin{minipage}{.2\textwidth}
\centering
\includegraphics[width=\linewidth]{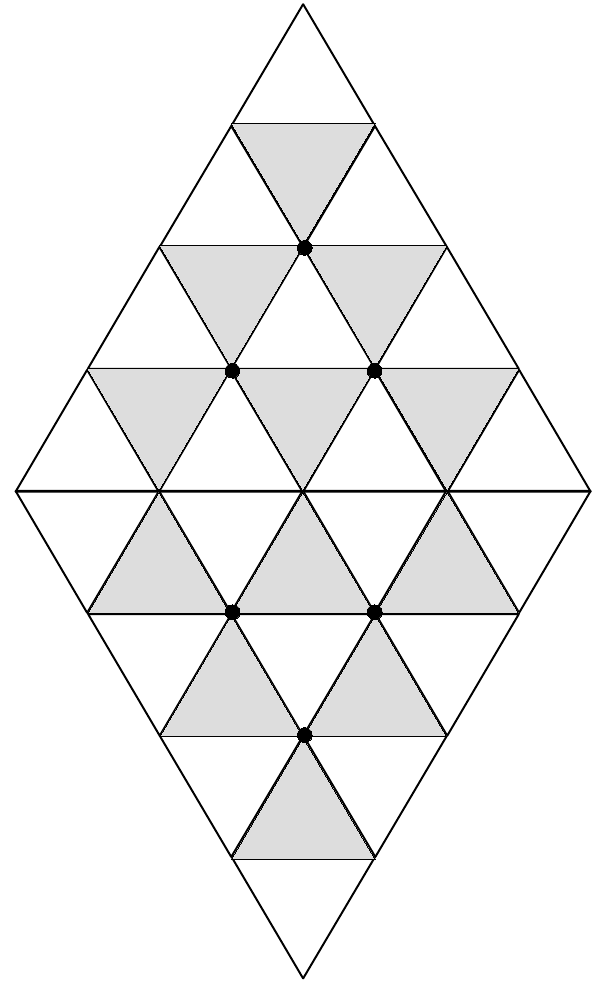}
\end{minipage}\ \vline \,
\begin{minipage}{.23\textwidth}
\centering
\includegraphics[width=\linewidth]{images/4pN3}
\end{minipage}\hfill
\begin{minipage}{.24\textwidth}
\centering
\includegraphics[width=\linewidth]{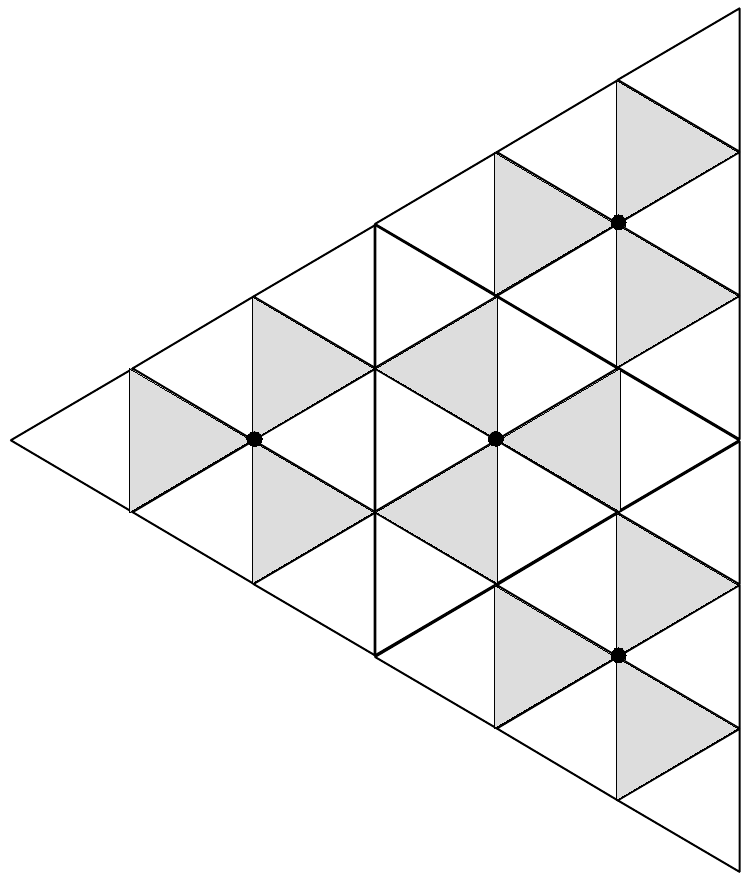}
\end{minipage}
    \caption{BPS graph and dual 4-triangulation of the 3-punctured sphere and vice versa.}
    \label{figT4triangles}
\end{figure}

This method naturally produces topological $A_{N-1}$ BPS graphs that are dual to the \emph{$N$-triangulations} used by Fock and Goncharov to define cluster coordinate systems on the moduli space of flat connections in~\cite{FockGoncharovHigher}, and related to spectral networks in~\cite{Gaiotto:2012db}. 
An $N$-triangulation is obtained as a refinement of an ideal triangulation of~$\CC$ by further dividing each ideal triangle into smaller triangles. 
For each original ideal triangle, there are $\binom{N+1}{2}$ ``upright'' and $\binom{N}{2}$ ``upside-down'' small triangles, as shown on the right of Figure~\ref{level4Ywebs}. 
The branch points of the BPS graph are dual to small upside-down triangles of the $N$-triangulation. Similarly, edges and Y-webs of the BPS graph are dual to vertices that are respectively on the edges and faces of the $N$-triangulation.
Examples of $N$-triangulations dual to BPS graphs on spheres with three and four full punctures are shown in Figure~\ref{figT4triangles}.

It was shown in~\cite{Gaiotto:2012db} that $N$-triangulations are related to a particular class of spectral networks, known as ``minimal Yin--Yang'' networks.
Explicit examples could be found up to $N=5$, but there is no general proof of the existence of a Yin--Yang locus on the Coulomb branch.
We found evidence that the two conjectural Roman and Yin--Yang loci generically intersect.
More precisely, given a BPS graph dual to an $N$-triangulation, we expect to find a Yin or Yang network at the phase $\vartheta=\vartheta_c \pm\pi/2$ (see for example Figure~15 in~\cite{Gaiotto:2012db}). However, as we will discuss momentarily, in general BPS graphs are not dual to $N$-triangulations, and hence not related to minimal Yin--Yang networks.

\subsection{\texorpdfstring{$N$-flips}{N-flips}}\label{subsec:N-flips}

An interesting consequence of the relation between BPS graphs and $N$-triangulations is that it leads to a concrete geometric realization of the combinatorics of higher mutations studied by Fock and Goncharov (see Section~10 of~\cite{FockGoncharovHigher}). The flip of an edge in the $N$-triangulation, or \emph{$N$-flip}, indeed corresponds to a sequence of the elementary local transformations of BPS graphs described in Section~\ref{subsecMoves}. 
More explicitly, an $N$-flip involves a total of $\binom{N+1}{3}$ elementary flip moves (intertwined with cootie moves). Note that thinking about flipping an edge in terms of gluing a tetrahedron, we see that the number of elementary flips matches nicely the number of octahedra in the $N$-decomposition of a tetrahedron~\cite{Dimofte:2013iv}. 

\begin{figure}\hrule
\centering\vline
\begin{minipage}{.25\textwidth}
\centering
\includegraphics[width=\linewidth]{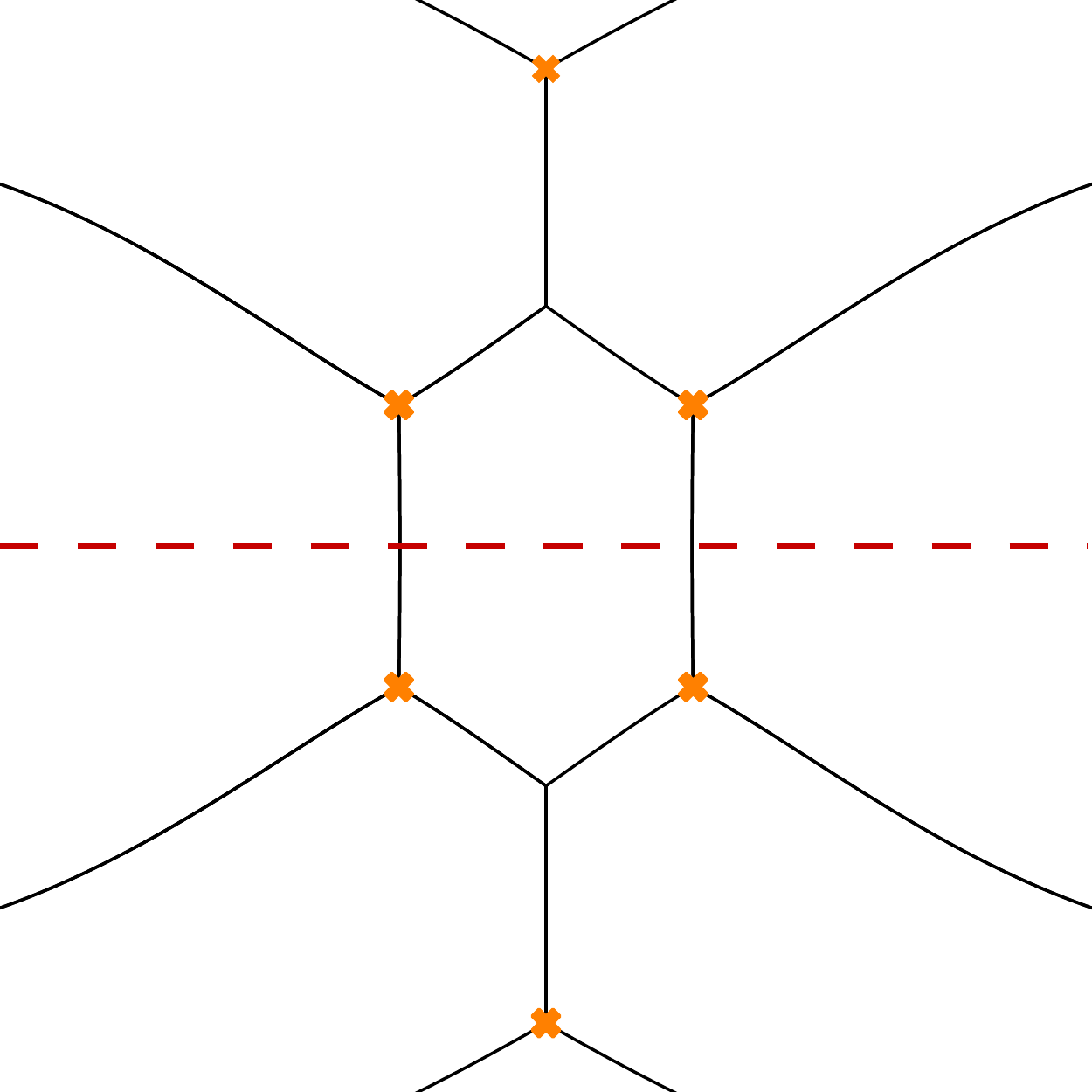}
\end{minipage}\vline
\begin{minipage}{.25\textwidth}
\centering
\includegraphics[width=\linewidth]{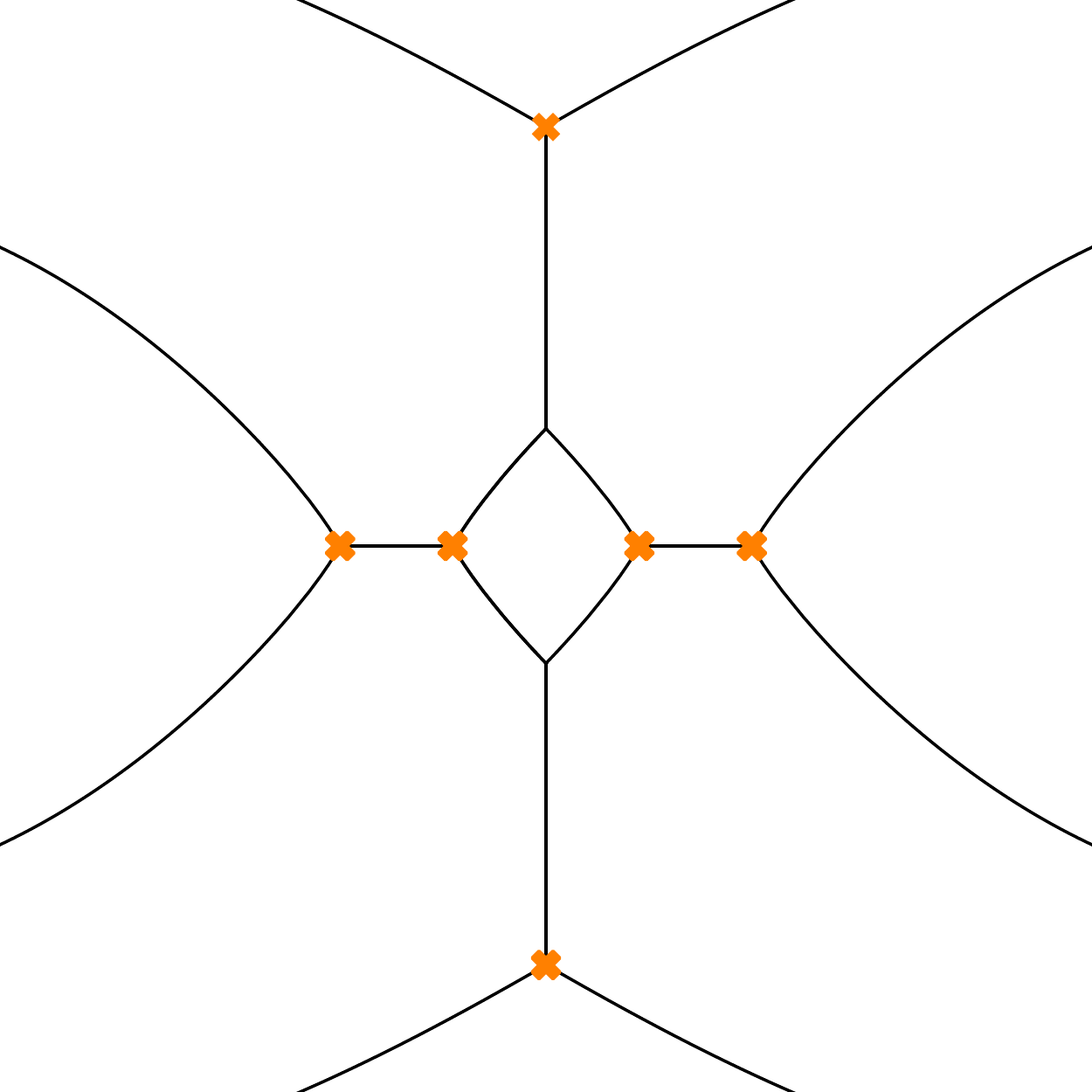}
\end{minipage}\vline
\begin{minipage}{.25\textwidth}
\centering
\includegraphics[width=\linewidth]{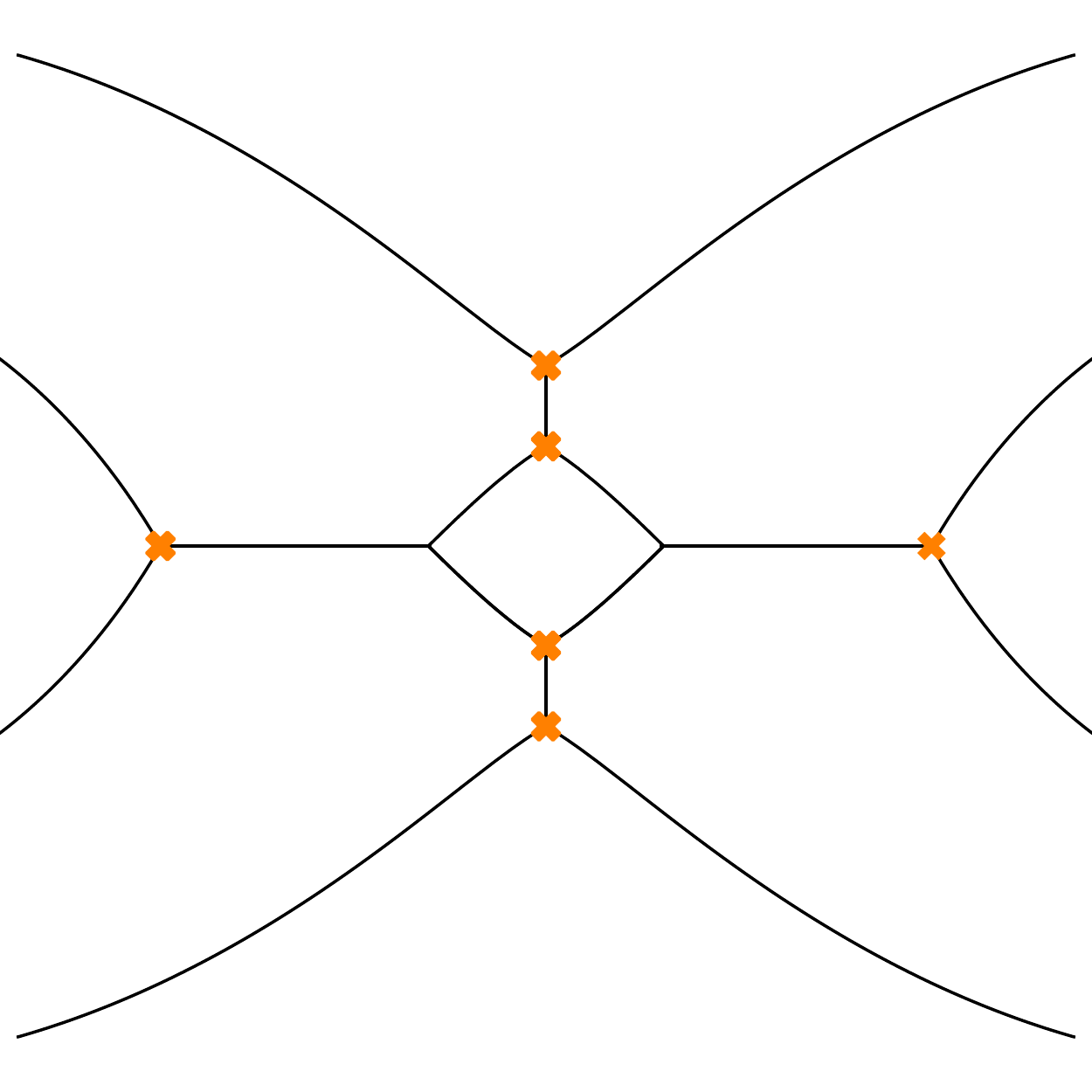}
\end{minipage}\vline
\begin{minipage}{.25\textwidth}
\centering
\includegraphics[width=\linewidth]{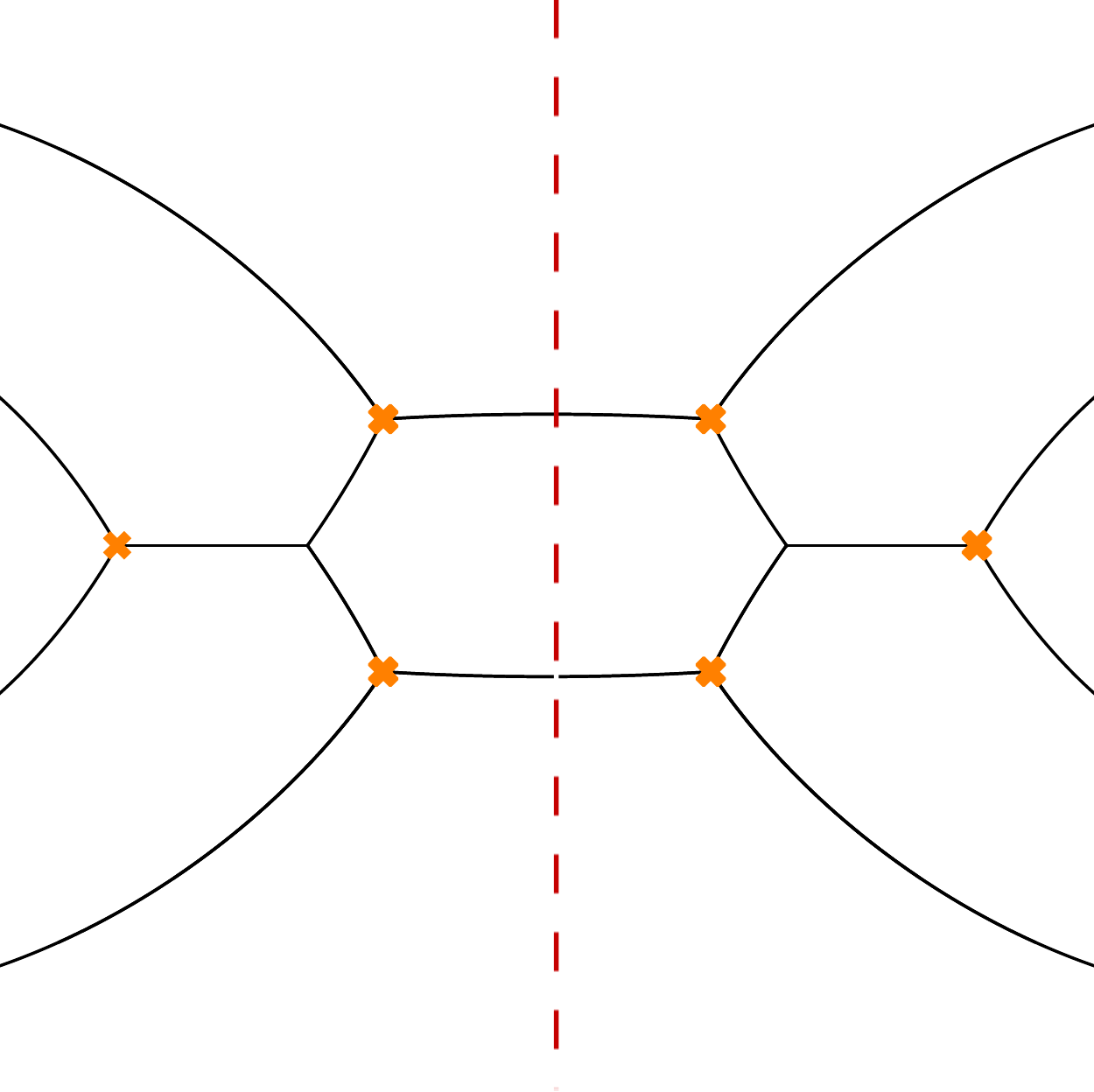}
\end{minipage}\vline \hrule\
    \caption{The decomposition of a 3-flip in terms of elementary moves: two flips, one cootie, two flips.}
    \label{mutationN3}
\end{figure}

As an example, we plotted in Figure~\ref{mutationN3} the sequence of elementary transformations of the BPS graph for the $T_3$ theory corresponding to a $3$-flip. 
In the initial configuration, there are two distinct sets of three branch points, located inside the faces of two triangles separated by a horizontal edge. 
As we vary the mass parameter $M_2$ for the puncture at infinity, the two pairs of branch points closest to the edge collide pairwise vertically and separate horizontally, in a synchronized pair of flip moves. 
This is followed by a cootie move involving a pair of branch points. Finally, after another pair of synchronized flip moves, we arrive at a BPS graph with two sets of three branch points, now separated by a vertical edge.

We stress that the BPS graphs appearing at intermediate steps of an $N$-flip do not correspond to any specific $N$-triangulation.
Starting from an $N$-lifted BPS graph, we can indeed apply arbitrary sequences of flip and cootie moves to it, and obtain BPS graphs that are much more general than $N$-triangulations!
This should be related to the conjecture made in~\cite{Gaiotto:2012db} that via the ``nonabelianization map'' spectral networks provide the \emph{cluster altas}, that is the set of all cluster coordinate systems on the moduli space of flat rank-$N$ connections over~$\CC$.

\subsection{\texorpdfstring{Goncharov's ideal bipartite graphs}{Ideal bipartite graphs}}\label{subsec:ideal-webs}

\begin{figure}
    \centering
        \includegraphics[width=\textwidth]{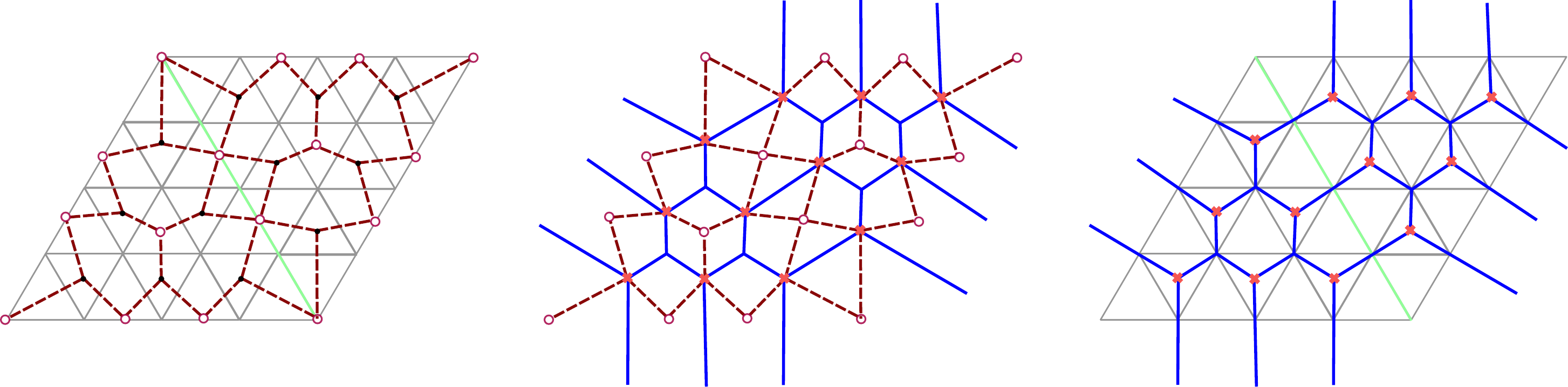}
    \caption{4-triangulation of two glued ideal triangles. \emph{Left}: Ideal bipartite $A_3$ graph. \emph{Center}: Duality. \emph{Right}: $A_3$ BPS graph.}
    \label{fig:BPS-bipartite}
\end{figure}

Recently, Goncharov introduced in~\cite{2016arXiv160705228G}, the notion of an ``ideal bipartite $A_{N-1}$ graph'' on a Riemann surface~$\CC$ with marked points, and used it to construct cluster coordinate systems on moduli spaces of local $A_{N-1}$ systems over~$\CC$.
These cluster coordinate systems contain those defined by Fock and Goncharov using $N$-triangulations of~$\CC$~\cite{FockGoncharovHigher}, but also generalizes them considerably.
Given an ideal triangulation $T$ of~$\CC$, 
the corresponding ideal bipartite $A_{N-1}$ graph~$\Gamma_{A_{N-1}}(T)$ is defined by placing black vertices $\bullet$ in small upside-down triangles of the related $N$-triangulation, and white vertices $\circ$ in small upright triangles (see the example on the left of Figure~\ref{fig:BPS-bipartite}).
More general ideal bipartite $A_{N-1}$ graphs can then be obtained by applying ``two-by-two moves`` and ``shrink-expand moves.''

The relation between ideal bipartite $A_{N-1}$ graphs and $A_{N-1}$ BPS graphs should now be obvious. For a given $N$-triangulation of~$\CC$, the ideal bipartite graph~$\Gamma_{A_{N-1}}(T)$ is simply dual to the corresponding $N$-lifted BPS graph (constructed as in Section~\ref{subsec:N-triangulations}): black vertices~$\bullet$ are dual to branch points, while square and hexagonal faces are dual to edges and Y-webs, respectively (Figure~\ref{fig:BPS-bipartite}).
Moreover, the ``two-by-two move'' and the ``shrink-expand move'' on ideal bipartite graphs correspond precisely to the flip and cootie moves on BPS graphs described in Section~\ref{subsecMoves}.
We summarize the relation between ideal bipartite graphs and BPS graphs in the following table:
\be\label{eq:bipartite-dictionary}
\begin{array}{|c|c|}
\hline
\text{\qqq Ideal bipartite graph\qqq } & \text{BPS graph} \\
\hline
\hline
\bullet \text{ vertex} & \text{branch point}\\
\hline
\circ \text{ vertex} &\text{face}\\
\hline
\text{$(4+2k)$-gon face} & \text{\qquad elementary web with $k$ joints\qquad }\\
\hline
\hline
\text{two-by-two move} & \text{flip move}\\
\hline
\text{shrink-expand move} & \text{cootie move}\\
\hline
\end{array}
\ee
Note that we have allowed for a generalization to faces with more than four or six sides, which would correspond to elementary webs with $k\ge2$ joints. We will indeed see in Section~\ref{PartialP} that such multi-joint webs do appear in BPS graphs on surfaces~$\CC$ with partial punctures (decorated with partial flags).

Another point of contact between BPS graphs and ideal bipartite $A_{N-1}$ graphs is their relation to quivers (which we will discuss in Section~\ref{BPSquivers}). 
For surfaces without boundaries, the quiver associated with an ideal bipartite graph is defined by a set of nodes $\{e_G\}$ and a bilinear skew-symmetric form $\langle \cdot,\cdot \rangle$. 
The generators $\{e_G\}$ are identified with the internal faces of the ideal bipartite graph. According to table~\ref{eq:bipartite-dictionary}, these correspond to the elementary webs of BPS graphs, in agreement with our proposal to identify them with quiver nodes (see Section~\ref{subsecGraphs2Quivers}). 
The intersection form $\langle \cdot,\cdot \rangle$ is given in Lemma 2.26 of~\cite{2016arXiv160705228G} by a combinatorial formula, which is easily seen to agree with our prescription~\eqref{eq:pairing} to compute the intersection form of elementary webs. The quivers associated with ideal bipartite $A_{N-1}$ graphs thus exactly coincide with the BPS quivers associated with BPS graphs.

\section{Partial punctures} \label{PartialP}

In this section we study BPS graphs for $A_{N-1}$ theories with partial punctures (see for example~\cite{Chacaltana:2012zy, Xie:2012dw,FominP,Gang:2015wya}).
We describe the topological transformation of the BPS graph as a full puncture becomes a partial puncture. 
We restrict to the case of partial punctures of type $[k,1, \dots, 1]$, leaving a more general analysis to future work.

\subsection{\texorpdfstring{Puncture degeneration and web fusion}{Degeneration and web fusion}}

At a partial puncture, some of the eigenvalues of the residue matrix~\eqref{eq:regular-puncture} of the Higgs field~$\varphi$ are degenerate:
\bea 
\varphi(z) \sim \frac 1 z \begin{pmatrix} m_1 \bI_{k_1} & &  &\\ & m_2 \bI_{k_2} & & \\ & & \ddots & \\ & & & m_\ell \bI_{k_\ell} \end{pmatrix} \dd z + \cdots \;,
\eea
where $[k_1, k_2, \ldots, k_\ell]$ is a partition of $N$, and $\sum_\alpha k_\alpha m_\alpha = 0$. 

In order to degenerate a full puncture $[1,1,\ldots,1]$ into a partial puncture $[k_1, k_2, \ldots, k_\ell]$, we need not only to take the limit where some of the eigenvalues coincide, but also to select the correct $SL(N)$-orbit of the residue matrix~\cite{Gukov:2006jk}.\footnote{As a simple example, 
for $N=2$ the matrices 
$\left(\begin{smallmatrix} m_1 & 0 \\ 0 & m_2 \end{smallmatrix}\right)$ and 
$\left(\begin{smallmatrix} m_1 & 1 \\ 0 & m_2 \end{smallmatrix}\right)$
are conjugate for $m_1 \neq m_2$, but not for $m_1 = m_2$.}
This is achieved by imposing a condition on the order of the pole of the discriminant $\Delta$ of the spectral curve~$\Sigma$~\eqref{SigmaNfold} at the puncture~\cite{Gaiotto:2009we}. 
Geometrically, this condition implies that some zeroes of~$\Delta$, which correspond to branch points of the covering $\Sigma\to C$, end up coinciding with the puncture and lower the order of its pole. 
This amounts to the \emph{absorption of pairs of branch points} by the puncture, which changes the topology of the spectral curve~$\Sigma$. 
As two eigenvalues $m_i, m_j$ are tuned to coincide, the homology cycle $\gamma_f$ whose period is $Z_{\gamma_f}=m_i-m_j$ shrinks, and eventually disappears when the two branch points supporting it are absorbed by the puncture.

The corresponding BPS graph transforms in an interesting way. After a pair of branch points is absorbed by the puncture, the elementary webs that were originally attached to them are \emph{fused} together. Starting with an $N$-lifted BPS graph with only edges and Y-webs, we can thus obtain BPS graphs with new types of webs involving several joints on a surface~$\CC$ decorated by partial punctures.

\subsection{\texorpdfstring{Puncture $[2,1]$}{Puncture [2,1]}}

As a simple example, we consider the BPS graph on the sphere with three full punctures for the $T_3$ theory, discussed in Section~\ref{subsec.TN} (in this section we put the punctures at $z_a=0$, $z_{b,c}=\pm 1$). The differentials $\phi_2$ and $\phi_3$ are taken as in~\eqref{phik}, where $P_0=u$ parameterizes the Coulomb branch~$\cB$. 

\begin{figure}
\centering
\begin{minipage}{.32\textwidth}
\centering
\includegraphics[width=\linewidth]{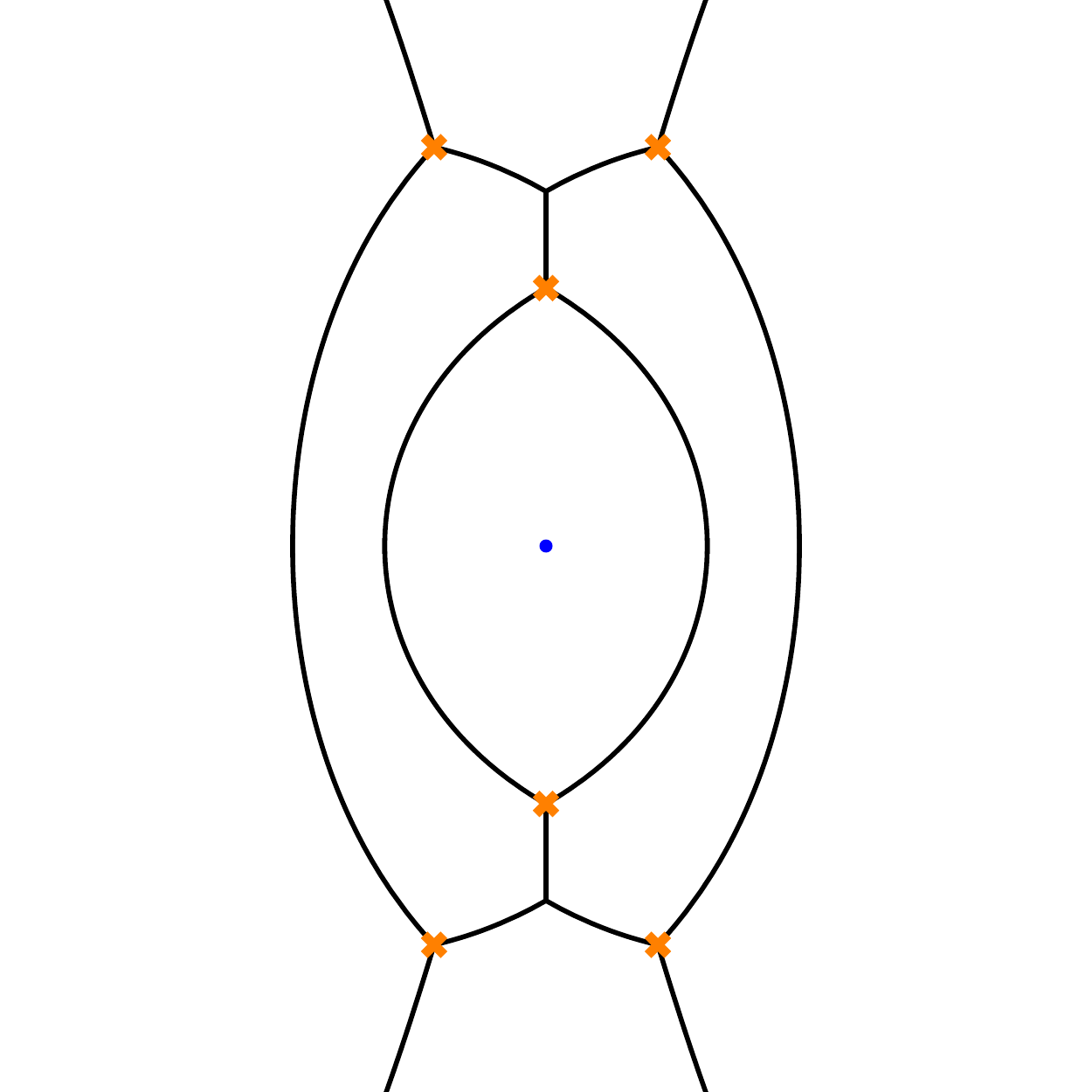}
\end{minipage}\hfill
\begin{minipage}{.32\textwidth}
\centering
\includegraphics[width=\linewidth]{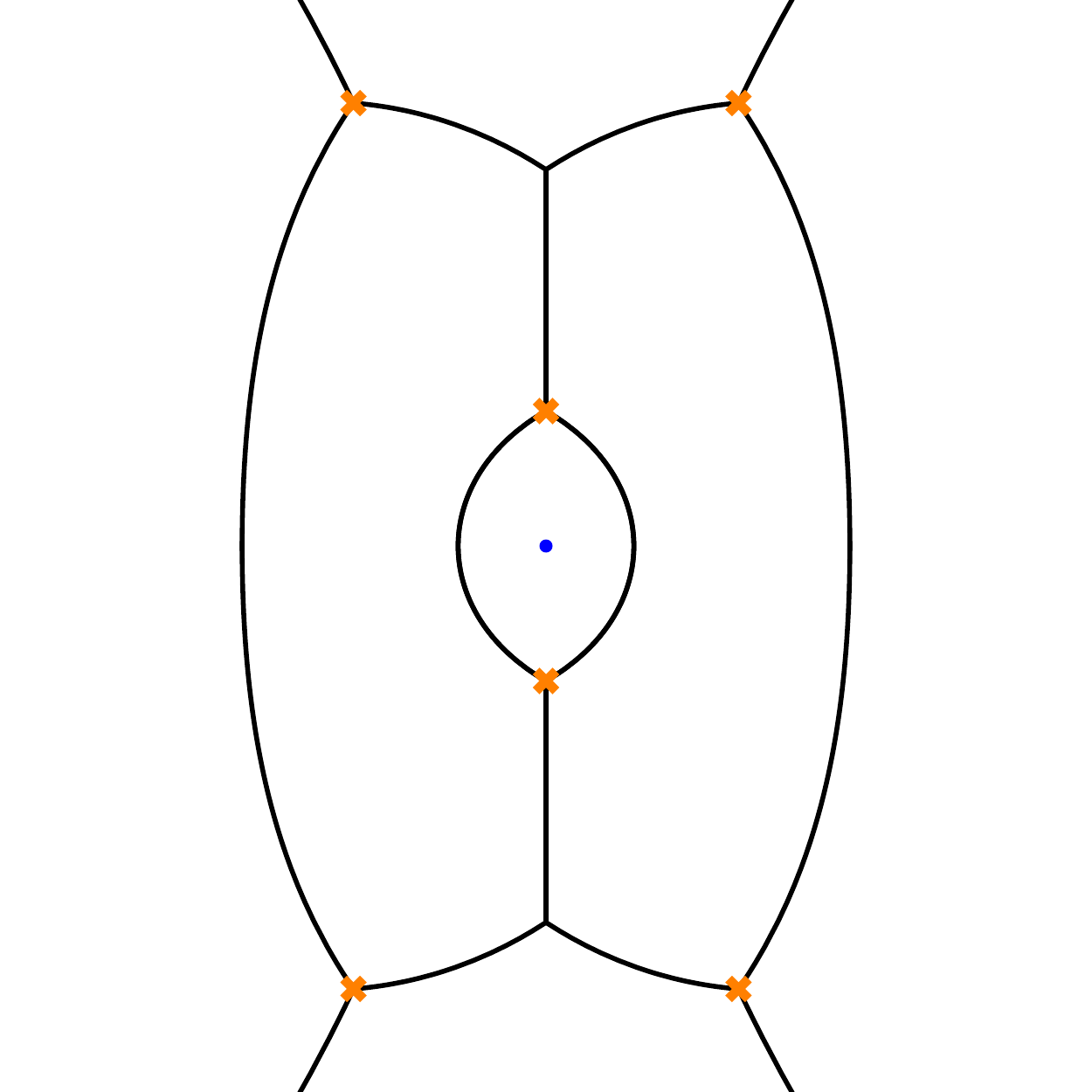}
\end{minipage}\hfill
\begin{minipage}{.32\textwidth}
\centering
\includegraphics[width=\linewidth]{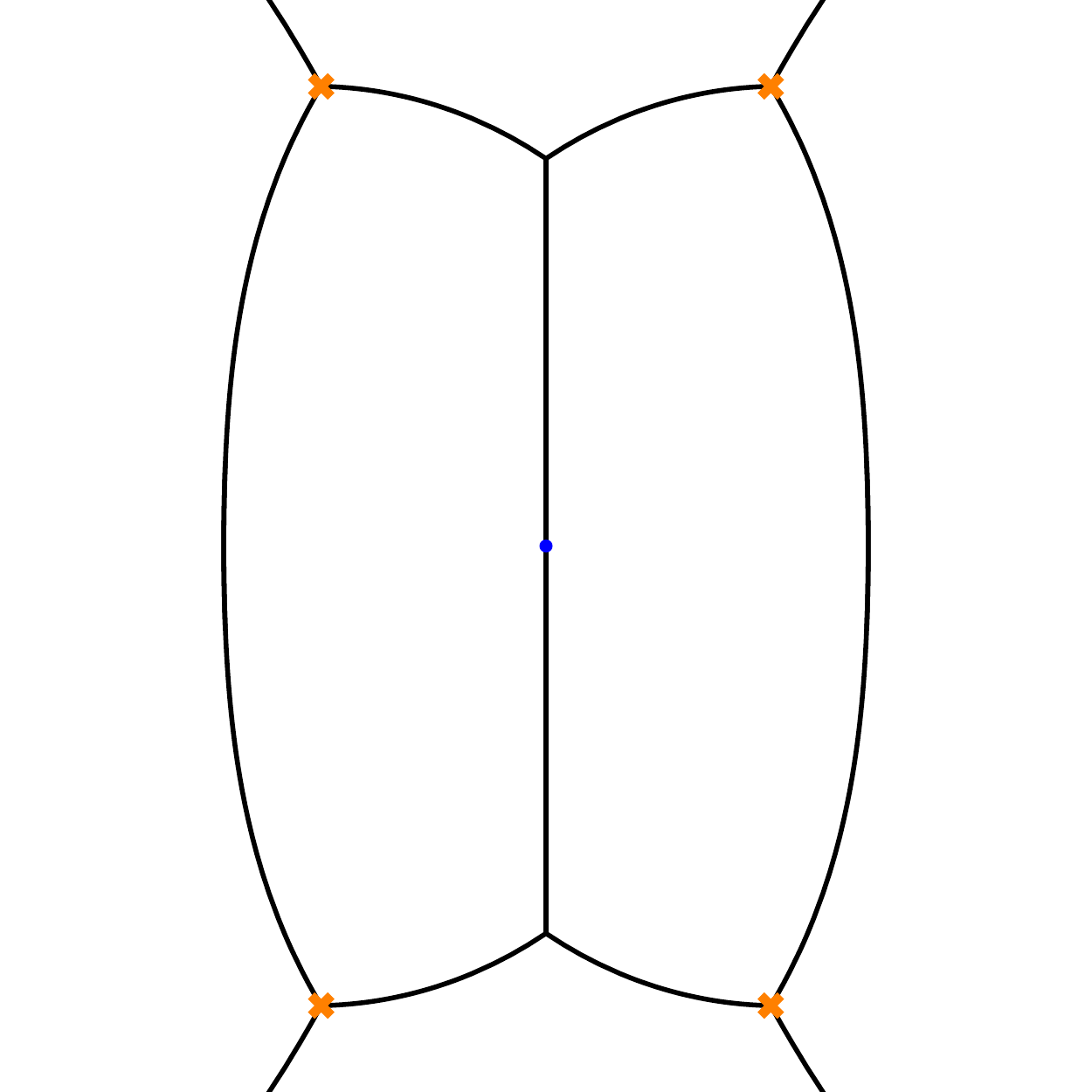}
\end{minipage}
    \caption{Degeneration from a full puncture $[1,1 ,1]$ to a simple puncture $[2, 1]$ on the 3-punctured sphere. A pair of branch points is absorbed by the puncture, and two Y-webs fuse into an H-web.}
    \label{21puncture}
\end{figure}

In order to degenerate a full puncture $[1,1,1]$ into a partial puncture $[2,1]$,  we first take the limit $m_1^a \to m_2^a \equiv m$, which gives $M_2^a = -3m^2$ and $M_3^a = 2 m^3$. 
In addition, we must ensure that the order of the pole at $z=z_a$ of the discriminant $\Delta = -4 \phi_2^3 -27 \phi_3^2$ drops from 6 to 4. This fixes the value of the Coulomb branch parameter:
\be\label{uT3}
u = -2m(M_2^b - M_2^c) + 4 ( M_3^b - M_3^c) \;.
\ee
Starting with the parameters $M_2^{a,b,c}=3$, $M_3^{a,b,c} = -\ii/4$ for the case of a full puncture, we vary $M_3^a\to- 2 \ii$ along the imaginary axis until we satisfy $m_1^a = m_2^a=\ii$ and $u=0$ in~\eqref{uT3}.
We plotted the corresponding degeneration of the BPS graph in Figure~\ref{21puncture}.
We see that the two closest branch points collide with the puncture, and the two connected edges shrink completely. 
The resulting BPS graph involves an H-shaped elementary web, or \emph{H-web}, arising from the fusion of the two Y-webs originally connected to the absorbed branch points. 
Note that the fact that the newly formed H-web passes through the puncture is not an issue. Since the branch points have been absorbed, the distance between the two sheets of $\Sigma$ to which the H-web lifts remains finite along the fiber of $T^*\CC$.\footnote{By studying the behavior of roots of (\ref{SigmaNfold}) near a puncture at $z_0$, we indeed find that $\lim_{z\to z_0}\lambda_i-\lambda_j$ is a finite constant.}
This ensures that the central charge of the H-web is in fact finite.
We can moreover easily deform the H-web away from the puncture.

\begin{figure}
\centering
\begin{minipage}{.25\textwidth}
\centering
\includegraphics[width=\linewidth]{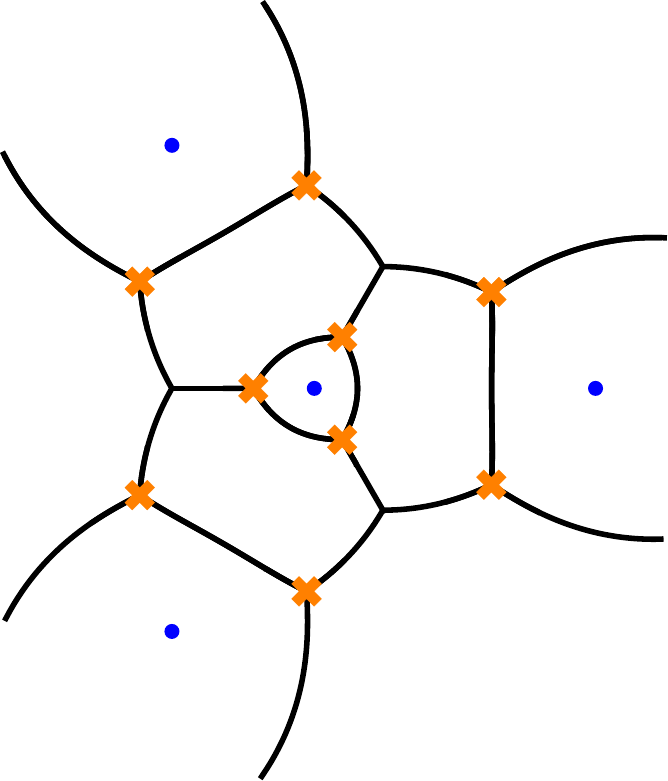}
\end{minipage}\hspace{5cm}
\begin{minipage}{.25\textwidth}
\centering
\includegraphics[width=\linewidth]{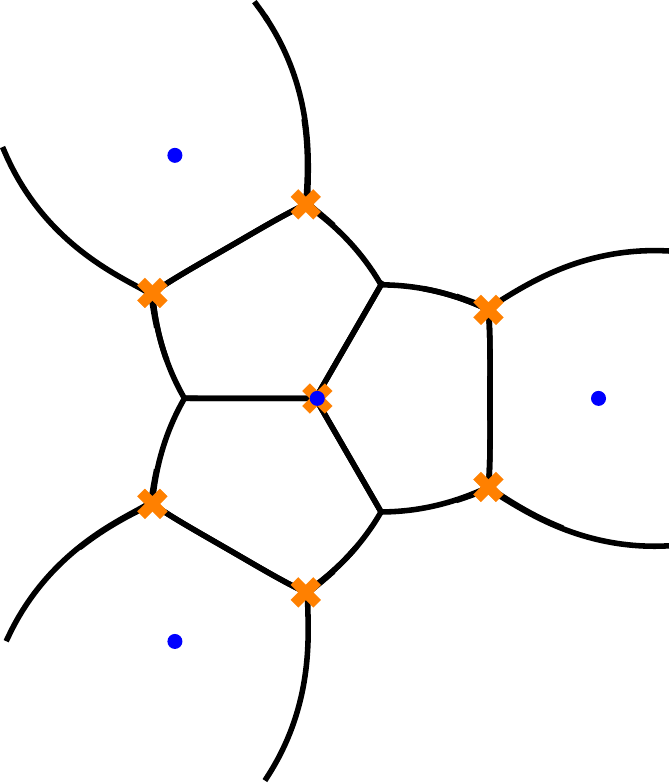}
\end{minipage}
    \caption{Degeneration from a full puncture $[1,1 ,1]$ to a simple puncture $[2, 1]$ with three incident triangles on the 4-punctured sphere.}
    \label{4p21}
\end{figure}

In Figure~\ref{4p21}, we show the degeneration to a puncture $[2,1]$ for a BPS graph on the 4-punctured sphere. In this case, there are three triangles in the underlying ideal triangulation that are incident at the degenerating puncture. Because of the $\bZ_3$ symmetry of the configuration, we see three branch points converging towards the puncture. However, only two of them are absorbed by the puncture. The three edges originally connected to them disappear, but the Y-webs connected to the third branch point remain distinct.

\subsection{\texorpdfstring{Puncture $[3,1]$}{Puncture [3,1]}}

\begin{figure}
\centering
\begin{minipage}{.17\textwidth}
\centering
\includegraphics[width=\linewidth]{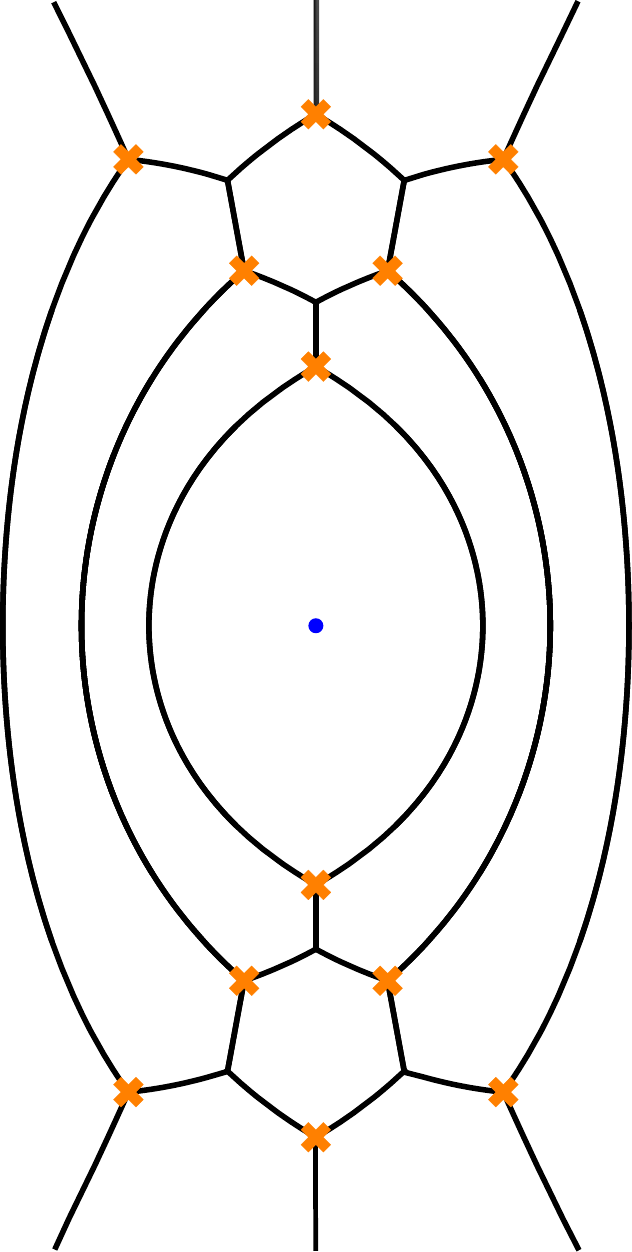}
\end{minipage}\qqq\qqq
\begin{minipage}{.34\textwidth}
\centering
\includegraphics[width=\linewidth]{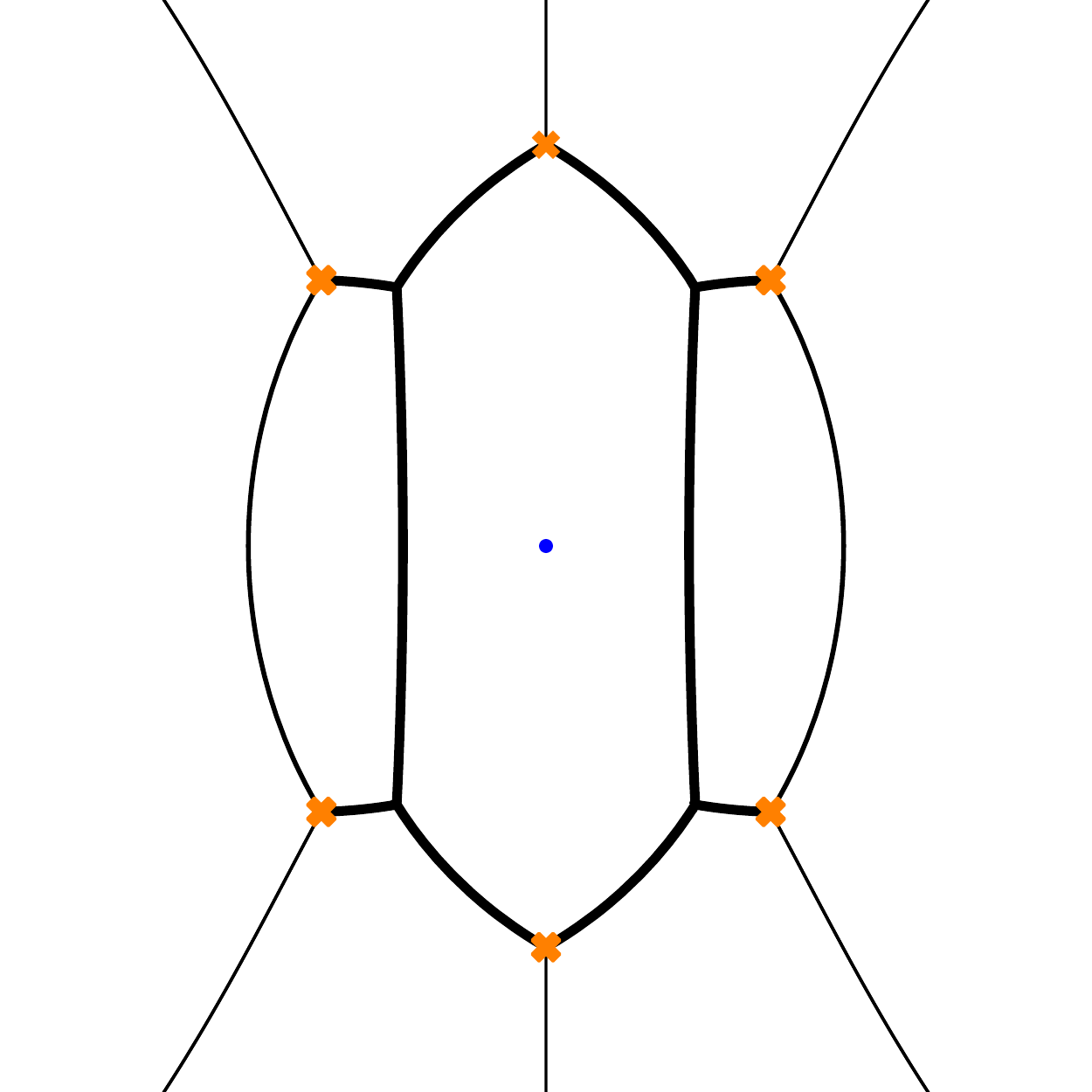}
\end{minipage}
    \caption{Degeneration from a full puncture $[1,1,1,1]$ to a partial puncture of type $[3,1]$ on the 3-punctured sphere. Two pairs of Y-webs fuse into two H-webs.}
    \label{31p}
\end{figure}

We can also consider the degeneration of a full puncture $[1,1,1, 1]$ into a partial puncture~$[3, 1]$ for the $T_4$ theory.
The $k$-differentials take the form (recall~\eqref{eq:4-lifted-curve})
\bea
\phi_2 &=&   \frac{(10M_2^a + \delta M_2^a) z_{ab}z_{ac}(z\! -\!z_b)(z\!-\!z_c) +\text{cyclic} } {(z-z_a)^2(z-z_b)^2(z-z_c)^2  } (\dd z)^2 \;, \nn
\phi_3 &=&   \frac{\delta M_3^az_{ab}^{2}z_{ac}^{2}(z\! -\!z_b)(z\!-\!z_c)  +\text{cyclic} +P_{0}\, (z\! -\! z_a) (z \!- \!z_b) (z\! -\! z_c)} {(z-z_a)^3(z-z_b)^3(z-z_c)^3  } (\dd z)^3 \;, \nn
\phi_4 &=&  \frac{9 \left[ M_2^a z_{ab}z_{ac}(z\! -\!z_b)(z\!-\!z_c) +\text{cyclic}\right]^2 + \delta M_4^az_{ab}^{3}z_{ac}^{3}(z\! -\!z_b)(z\!-\!z_c)  +\text{cyclic} } {(z-z_a)^4(z-z_b)^4(z-z_c)^4  } (\dd z)^4  \nn
&& + \frac{P_{1}\, (z\! -\! z_a) (z \!- \!z_b) (z\! -\! z_c)} {(z-z_a)^4(z-z_b)^4(z-z_c)^4  } (\dd z)^4 \;, 
\eea
with $P_0=u_0$ and $P_1 = u_1 + u_2 z$. 
To obtain a $[3,1]$ puncture, we first take the limit $m_1^a \to m_2^a\to m_3^a \equiv m$, which gives $10M_2^a + \delta M_2^a = -6m^2$, $\delta M_3^a = 8 m^3$, and $9 (M_2^a)^2 + \delta M_4^a = -3m^4$.
We must then impose that the discriminant~$\Delta$ has a pole of order 6, which fixes the Coulomb branch parameters:
\bea
u_0 &=&  -40 m (M_2^b-M_2^c) -4m (\delta M_2^b-\delta M_2^c)+ 4 (\delta M_3^b-\delta M_3^c) \;,  \nn
u_1 &=&- 4 (5 m^2 - 9 M_2^a) (M_2^b - M_2^c) -2  m^2 (\delta M_2^b-\delta M_2^c )  + 
 8 (\delta M_4^b - \delta M_4^c)  \;, \nn
u_2 &=&   m^4 - 9 (M_2^a)^2 + (20 m^2+36 M_2^a) (M_2^b+M_2^c)   +  36 (M_2^b-M_2^c)^2  \nn
&&  +2m^2( \delta M_2^b + \delta M_2^c )  - 4 m (\delta M_3^b +\delta M_3^c) + 8 (\delta M_4^b + \delta M_4^c)\;.
\eea
We plotted the degeneration of the BPS graph in Figure~\ref{31p}, with the following choice of parameters: $m=-\ii$, $M_2^{a,b,c}=3$, $\delta M_3^{a,b,c} = 8\ii$, $\delta M_4^a = -84$, $\delta M_2^a = -24$, $u_2 = 384$, and the others vanishing.
Three pairs of branch points are absorbed by the puncture, four edges and two Y-webs shrink to zero size, while two pairs of Y-webs fuse into two H-webs (deformed away from the puncture for clarity).

\subsection{\texorpdfstring{Punctures $[k,1,\dots,1]$}{Punctures [k,1,...,1]}}\label{k1p-2-triangles}

\begin{figure} 
    \centering
    \includegraphics[width=0.67\textwidth]{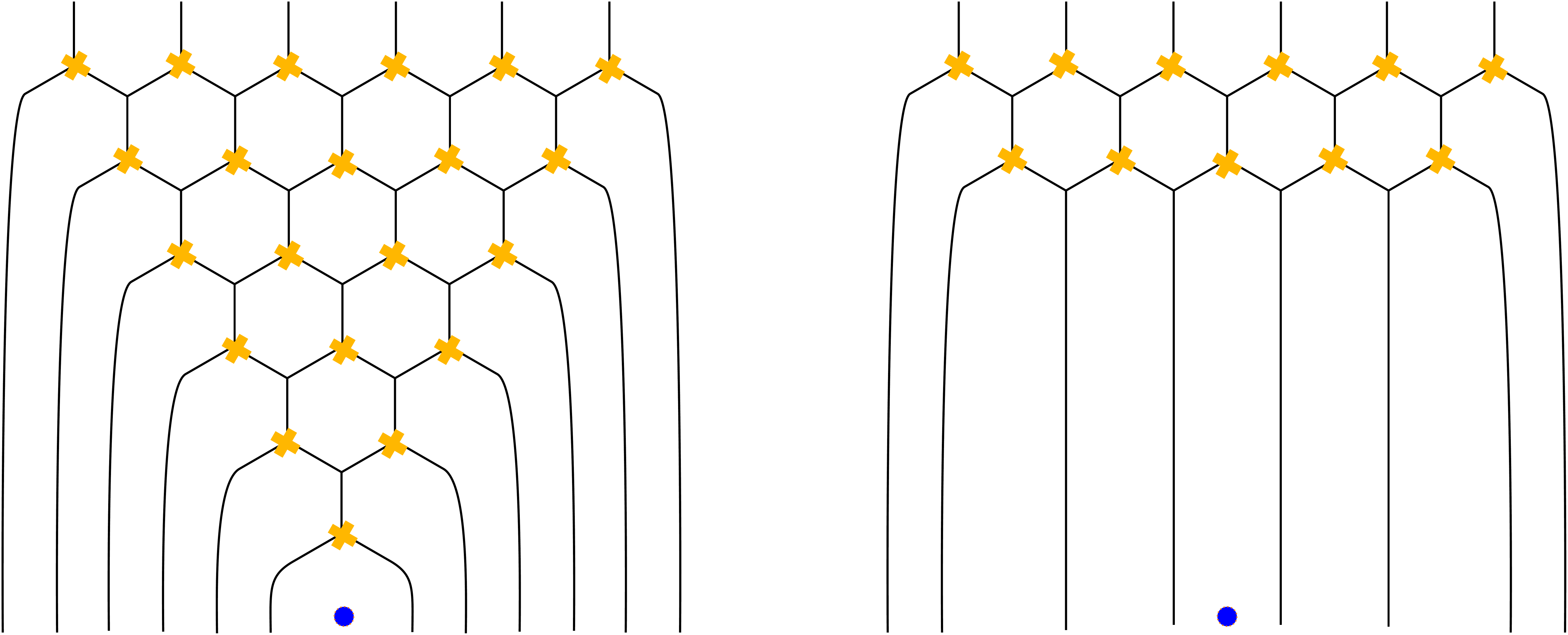}
    \caption{Schematic transformation of the BPS graph upon degeneration of a full puncture $[1,1,\ldots,1]$ into a partial puncture $[k,1\dots,1]$ (the identical half of the BPS graph in the lower triangle is left out).}
    \label{k1p}
\end{figure}

The examples discussed in the two previous sections admit a natural generalization to punctures of type $[k,1,\dots, 1]$. When two ideal triangles are incident at the puncture (like for the examples on the 3-punctured sphere),
this degeneration involves the absorption of $k(k-1)$ branch points by the puncture,
the disappearance of $2(k-1)$ edges and $k(k-1)$ Y-webs, and the appearance of $(k-1)$ H-webs.
The net change in the rank of the charge lattice~$\Gamma$ is that $k^2-1$ generators are lost, which agrees with the change in the genus of~$\Sigma$, $\Delta r=-\frac12 k(k-1)$,  and the shrinking of $\Delta f=(k-1)$ flavor cycles at the puncture.
The transformation of the BPS graph is depicted schematically in Figure~\ref{k1p} (for $k=5$).

We can also produce topological BPS graphs when there are more than two triangles incident at a partial puncture. 
The idea is to perform of incomplete $N$-flips (sequences of flips and cooties) such that the set of branch points that the puncture should absorb are in a configuration corresponding to two incident triangles.
We can then apply the general degeneration procedure just described.

\begin{figure} 
    \centering
    \includegraphics[width=\textwidth]{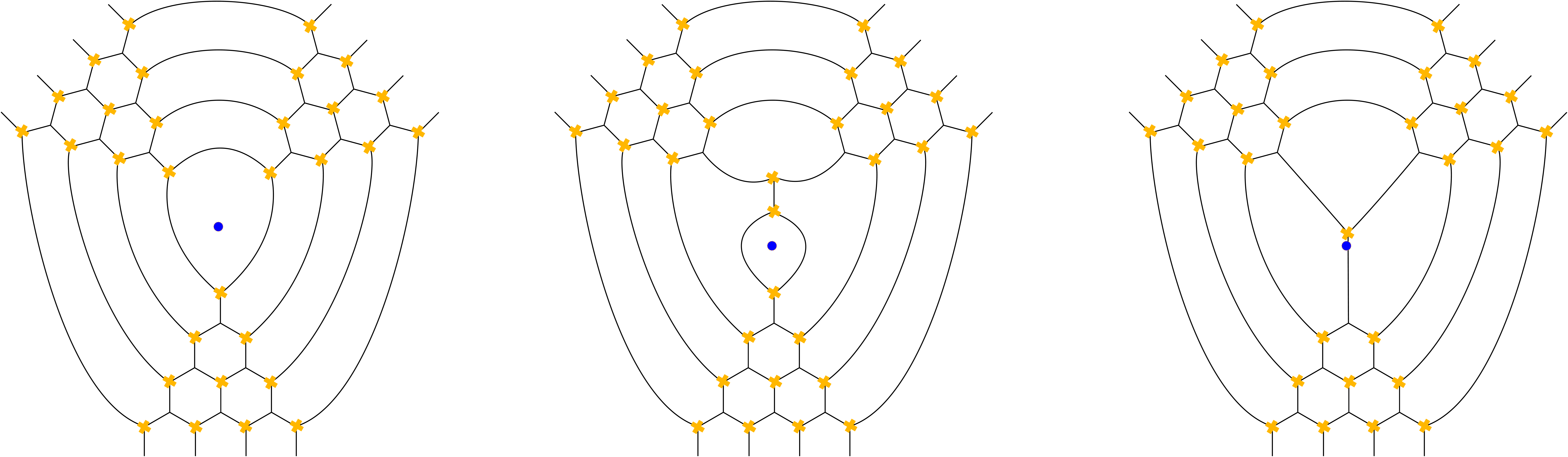}
    \caption{The degeneration of a full puncture with three incident ideal triangles into a partial puncture $[2,1,\dots,1]$ can be decomposed into an elementary flip move followed by the absorption of a pair of branch points by the puncture.}
    \label{21p-3-triangles}
\end{figure}

For example, let us consider the degeneration from a full puncture to a partial puncture $[2,1,\dots,1]$ when there are three incident triangles (Figure \ref{21p-3-triangles}).
In this case, we just need to perform an incomplete $N$-flip involving a single elementary flip move along the edge between the two upper ideal triangles.
The resulting BPS graph contains a pair of edges surrounding the puncture, which we can then shrink as in Figure~\ref{21puncture}. The final BPS graph is $\bZ_3$-symmetric and consistent with the degeneration on the 4-punctured sphere plotted in Figure~\ref{4p21}.

\begin{figure} 
    \centering
    \includegraphics[width=\textwidth]{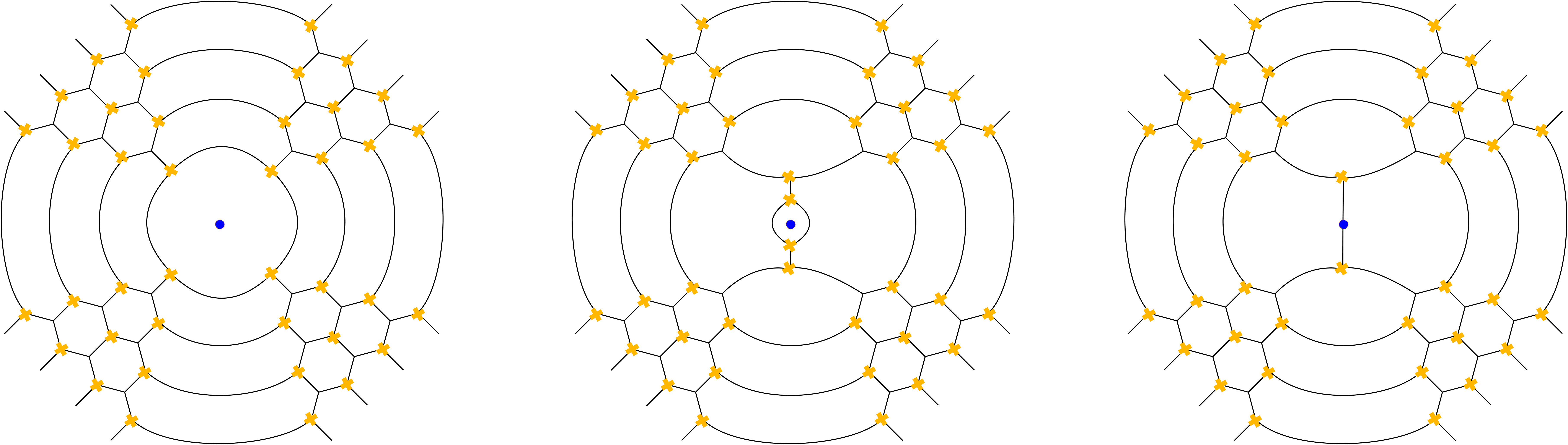}
    \caption{The degeneration of a full puncture with four incident ideal triangles into a partial puncture $[2,1,\dots,1]$, decomposed into two flip moves and the absorption of two branch points.}
    \label{21p-4-triangles}
\end{figure}

The degeneration to a partial puncture $[2,1,\dots,1]$ with four incident ideal triangles is depicted in Figure \ref{21p-4-triangles}.
Note that, once again, only two branch points are absorbed, as expected from the order of the pole of the discriminant at a puncture $[2,1,\dots,1]$.
If instead of applying elementary flip moves on the two vertical edges of the triangulation we had chosen the two horizontal ones, the resulting BPS graph would simply differ by a flip move on the central edge.

\subsection{\texorpdfstring{$N$-flips with partial punctures}{N-flips with partial punctures}}  \label{subsec:generalized-N-flips}

We now describe an approach to performing the analog of an $N$-flip on surfaces with partial punctures of type $[k,1,\dots,1]$, in terms of a sequence of elementary moves as in Section~\ref{subsec:N-flips}.

\begin{figure} 
    \centering
    \includegraphics[width=0.99\textwidth]{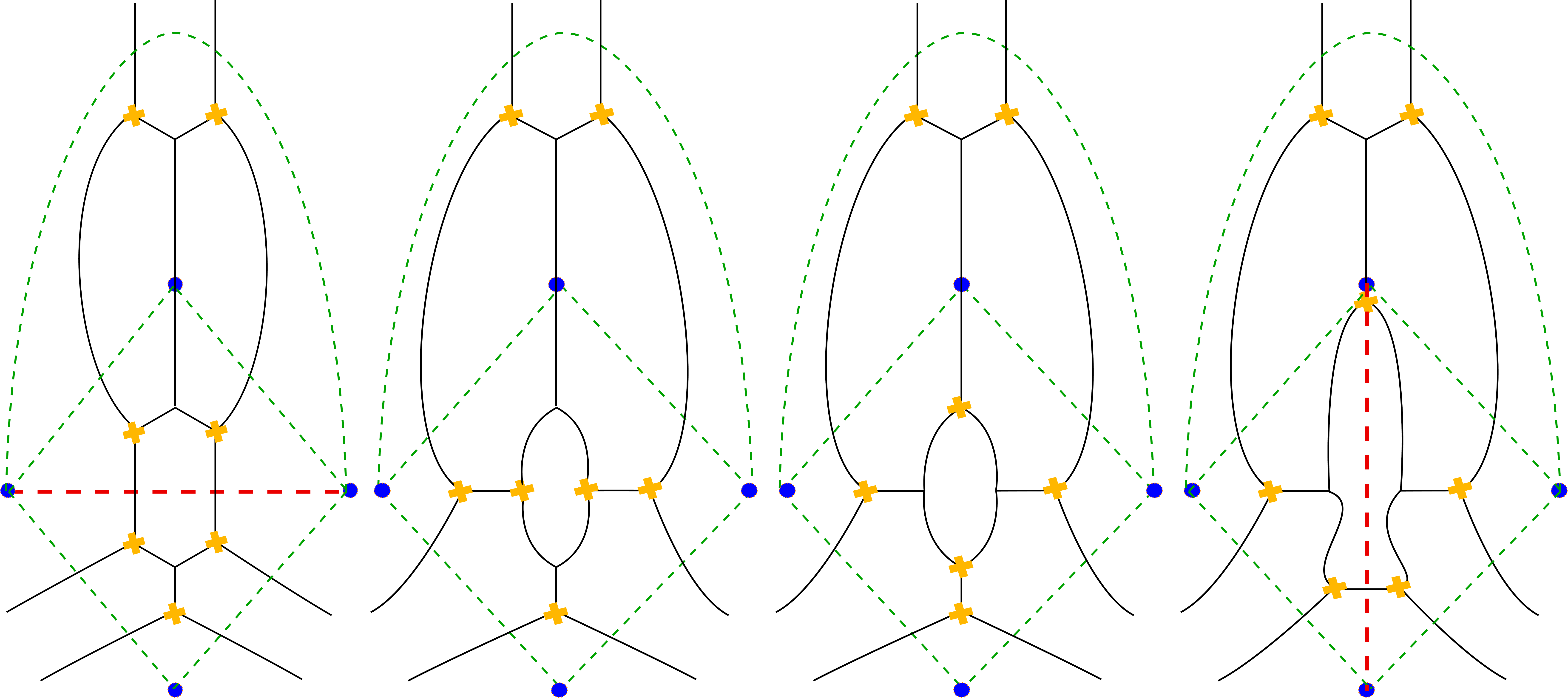}
    \caption{Sequence of elementary moves corresponding to the flip of an edge (red dashed) near a puncture~$[2,1]$ (top).}
    \label{fig:21-2-triangles-flip-a}
\end{figure}

As a first example, let us consider an $A_2$ theory with a puncture $[2,1]$ with two incident triangles, in which case the BPS graph looks locally like in Figure~\ref{fig:21-2-triangles-flip-a}.
By performing a sequence of moves, we can obtain the BPS graph dual to the flipped underlying ideal triangulation, where now the puncture $[2,1]$ has three incident triangles (compare with Figure \ref{21p-3-triangles}).
The sequence of moves is easily obtained as a subset of the $N$-flip sequence for the case with a full puncture: first two flips on the edges crossing the horizontal edge of the triangulation, then a cootie move in the center, and finally a \emph{single} flip for the lower edge.
The result is consistent, in the sense that this is the same BPS graph we would have obtained by first doing a $3$-flip of the BPS graph with full punctures, and then degenerating the upper puncture to $[2,1]$, as can be seen by comparing with Figure~\ref{21p-3-triangles}.
This process can of course be reversed, by reading Figure~\ref{fig:21-2-triangles-flip-a} from right to left, starting from a puncture~$[2,1]$ with three incident triangles.

\begin{figure} 
    \centering
    \includegraphics[width=\textwidth]{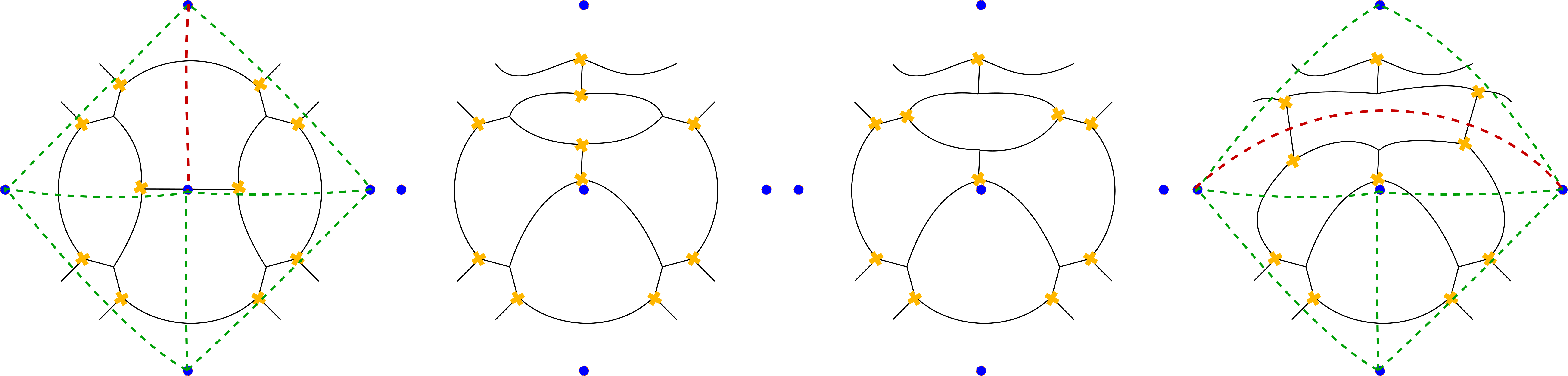}
    \caption{Ideal triangulation around a puncture $[2,1]$ with four (left) or three (right) incident triangles. The associated local BPS graphs are related by a sequence of elementary moves: two flips, one cootie, two flips.}
    \label{fig:21-4-triangles-flip}
\end{figure}

Another interesting example involves an $A_2$ theory with a puncture $[2,1]$ with four incident ideal triangles (Figure \ref{fig:21-4-triangles-flip}).
We can flip an edge by performing a sequence of moves and obtain a BPS graph dual to a triangulation where the puncture $[2,1]$ has three incident triangles.
The resulting BPS graph is again consistent, in the sense that this is the same BPS graph we would have obtained by first doing a $3$-flip of the BPS graph with full punctures, and then degenerating the central puncture to $[2,1]$.

\section{BPS quivers}\label{BPSquivers}

In this section, we explain how to obtain a quiver from the topology of a BPS graph, and identify it with the BPS quiver~\cite{Cecotti:2010fi,Cecotti:2011rv,Alim:2011ae,Alim:2011kw} of the corresponding $A_{N-1}$ theory of class~$\CS$. This thus provides a bridge between spectral networks and BPS quivers. 
The mutations that relate BPS quivers in the same equivalence class correspond to elementary transformations, such as flip and cootie moves, that relate BPS graphs on different components of the Roman locus~$\cR$. 
We also propose an expression for the quiver superpotential and perform several checks.

\subsection{Review of BPS quivers}\label{secReviewQuivers}

A BPS quiver consists of a set $Q_0$ of nodes and a set $Q_1$ of arrows, constructed as follows (for more details see e.g.~\cite{Alim:2011ae}). 
Given a generic point $u$ on the Coulomb branch of a 4d $\cN=2$ gauge theory and an angle~$\vartheta$, a positive half-lattice $\Gamma_+\subset\Gamma$ is defined as the set of charges which get mapped to the positive half-plane $\text{Re} (\ex^{-\ii\vartheta}Z) >0$ in the complex plane of central charges,
and it admits a unique positive integral basis. 
The generators $\{\gamma_i\}$ in this basis correspond to the nodes of the BPS quiver, and the electromagnetic pairing determines the arrows: $\langle\gamma_i,\gamma_j\rangle=-k$ implies that there are $k$ arrows from $\gamma_i$ to~$\gamma_j$.

The spectrum of BPS states can be determined by studying the representation theory of the quiver: each charge $\gamma= \sum_i n_i \gamma_i \in \Gamma$ corresponds to an $\cN=4$ quiver quantum mechanics theory.
The cohomology of the moduli space of solutions of this quiver quantum mechanics encodes the BPS index. The F-term equations are fixed by a choice of superpotential $W$, which can be constructed as a formal sum of cycles in the quiver path algebra. 

An important notion in the context of BPS quiver is the \emph{mutation}. This is the change in $(Q_0,Q_1,W)$ induced by a rotation of the positive half-plane such that one of the central charges exits.
This operation is also known as crossing a ``wall of the second kind''~\cite{Kontsevich:2008fj}.
By construction, the BPS state which exits the positive half-plane is part of the positive integral basis for $\Gamma_+$, and therefore a mutation is associated with a node $\gamma_0$ of the quiver. 
The new basis of elementary BPS states $\{\gamma_i'\}$ is given by
\be \label{mutationg0}
\gamma_0' = -\gamma_0 , \qqq
\gamma_j' =  \begin{cases}  \gamma_j + \langle\gamma_j,\gamma_0\rangle \gamma_0 & \quad \text{if } \langle\gamma_j,\gamma_0\rangle >0 , \\ \gamma_j &  \quad \text{if } \langle\gamma_j,\gamma_0\rangle \le 0 .
\end{cases} 
\ee
Mutations are ``equivalence relations'' in the sense that they leave the solutions of the quiver quantum mechanics, and hence the BPS spectrum, unchanged. Physically, they are a version of Seiberg duality.

\subsection{From BPS graphs to BPS quivers}\label{subsecGraphs2Quivers}

It is straightforward to construct a BPS quiver $(Q_0,Q_1)$ from a BPS graph $\cG$. 
The set of elementary webs of~$\cG$, whose lifted one-cycles $\{\gamma_i\}$ on~$\Sigma$ provide by definition a basis for the charge lattice $\Gamma$, corresponds to the set $Q_0$ of nodes.
The set $Q_1$ of arrows is then determined by identifying the electromagnetic pairing with the intersection form~\eqref{eq:pairing}. Focusing on one branch point $b$ at a time, we see that this means that two elementary webs ordered clockwise around $b$ correspond to two nodes with a negatively-oriented arrow between them. More explicitly, $\sigma(b;\gamma_1,\gamma_2)= +1$ corresponds to one arrow from node~$\gamma_2$ to node $\gamma_1$, and $\sigma(b;\gamma_1,\gamma_2)= -1$ to one arrow from $\gamma_1$ to $\gamma_2$. 

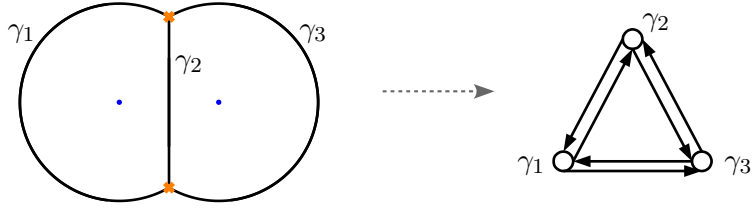
\begin{figure}
\centering
\begin{minipage}{.27\textwidth}
\centering
	\begin{overpic}[width=\textwidth]{images/T2graphB}
 		\put (-2,55) {$\gamma_1$}
		\put (52,45) {$\gamma_2$}
		\put (92,55) {$\gamma_3$}
	\end{overpic}
\end{minipage}\quad
\begin{minipage}{.34\textwidth}
\centering
	\begin{overpic}[width=\textwidth]{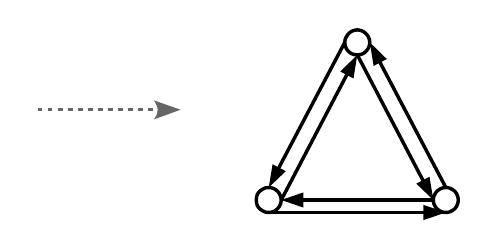}
 		\put (42,8) {$\gamma_1$}
		\put (74,45) {$\gamma_2$}
		\put (95,8) {$\gamma_3$}
	\end{overpic}
\end{minipage}
    \caption{From BPS graph to BPS quiver for the $T_2$ theory. }
    \label{T2quiver}
\end{figure}

For $A_1$ theories, the correspondence between BPS graphs and BPS quivers is rather obvious, given they are both related to ideal triangulations of~$\CC$~\cite{Gaiotto:2009hg, Alim:2011ae}. 
For example, the BPS graph and BPS quiver for the $T_2$ theory are shown in Figure~\ref{T2quiver}. 
In this case, the three edges of the BPS graph lift to one-cycles $\{\gamma_1, \gamma_2,\gamma_3\}$ on $\Sigma$ that intersect twice with each other, once positively and once negatively. This produces a quiver with three nodes connected by pairs of oppositely-oriented arrows. 
We show the BPS graph and BPS quiver for $SU(2)$ gauge theory with $N_f=4$ in Figure~\ref{4pquiver}.
In Appendix~\ref{sec.pantsDec}, we comment on how the method for gluing BPS quivers proposed in~\cite{Alim:2011kw} can be interpreted in terms of BPS graphs.

\begin{figure}
\centering
\begin{minipage}{.25\textwidth}
\centering
\includegraphics[width=\linewidth]{images/4pN2}
\end{minipage}\quad
\begin{minipage}{.34\textwidth}
\centering
\includegraphics[width=\linewidth]{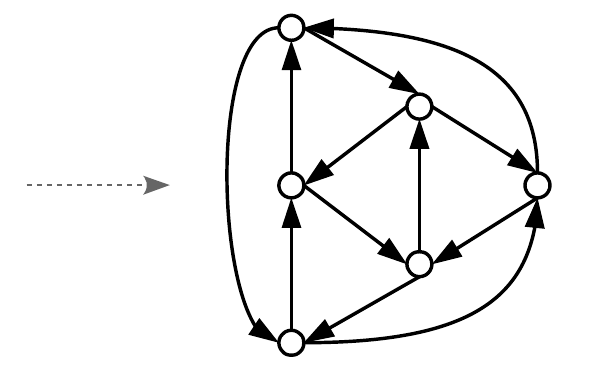}
\end{minipage}
    \caption{From BPS graph to BPS quiver for the 4-punctured sphere. }
    \label{4pquiver}
\end{figure}

\begin{figure}
\centering
\begin{minipage}{.3\textwidth}
\centering
\includegraphics[width=\linewidth]{images/T3graphY}
\end{minipage}\hfill
\begin{minipage}{.33\textwidth}
\centering
\includegraphics[width=\linewidth]{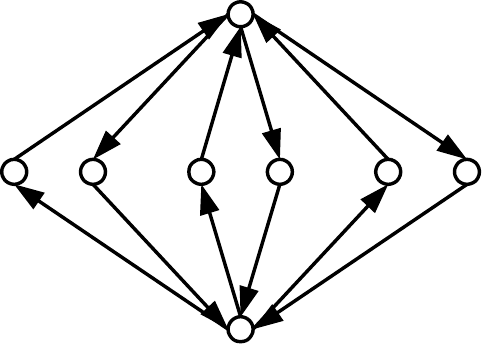}
\end{minipage}\hfill
\begin{minipage}{.2\textwidth}
\centering
\includegraphics[width=\linewidth]{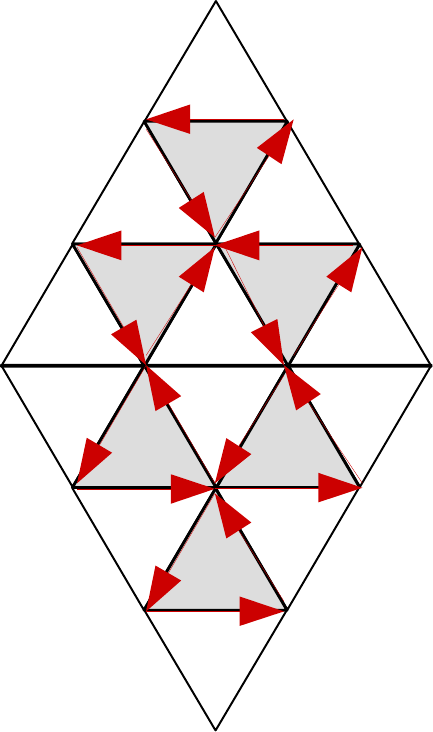}
\end{minipage}

    \caption{BPS graph, BPS quiver, and Fock-Goncharov Poisson tensor for the $T_3$ theory. }
    \label{T3quiver}
\end{figure}

The correspondence is much more interesting for higher-rank $A_{N-1}$ theories, for which not many BPS quivers are known. We show the BPS quiver constructed from the BPS graph for the $T_3$ theory in Figure~\ref{T3quiver} (we omit pairs of cancelling arrows for clarity).
It is mutation-equivalent%
\footnote{To match them exactly, we must perform a mutation at one of the two nodes corresponding to the Y-webs.} 
to the BPS quiver obtained in~\cite{Alim:2011kw} based on the engineering of the $T_N$ theories from M-theory on the Calabi-Yau three-fold $\bC^3/\bZ_N\times \bZ_N$~\cite{Benini:2009gi} (see also Section~7.3.1 in~\cite{Cecotti:2013lda}).
More generally, for $A_{N-1}$ BPS graphs obtained as $N$-lifts of $A_1$ BPS graphs, the relation to $N$-triangulations described in Section~\ref{subsec:N-triangulations} implies that the corresponding BPS quiver agrees with the Poisson structure on the moduli space of flat connections described by Fock and Goncharov~\cite{FockGoncharovHigher}. 
This Poisson structure can indeed be neatly represented by a collection of arrows around the small black triangles of the $N$-triangulation. 
We show the Poisson structure on the 3-triangulation of the 3-punctured sphere on the right of Figure~\ref{T3quiver}, which can be seen to agree with the BPS quiver.
The combinatorial relation between BPS quivers and $N$-triangulations was already conjectured in~\cite{Xie:2012dw, Xie:2012gd}. 
The geometric bridge provided by BPS graphs confirms this conjecture.

\begin{figure} 
    \centering
    \includegraphics[width=0.85\textwidth]{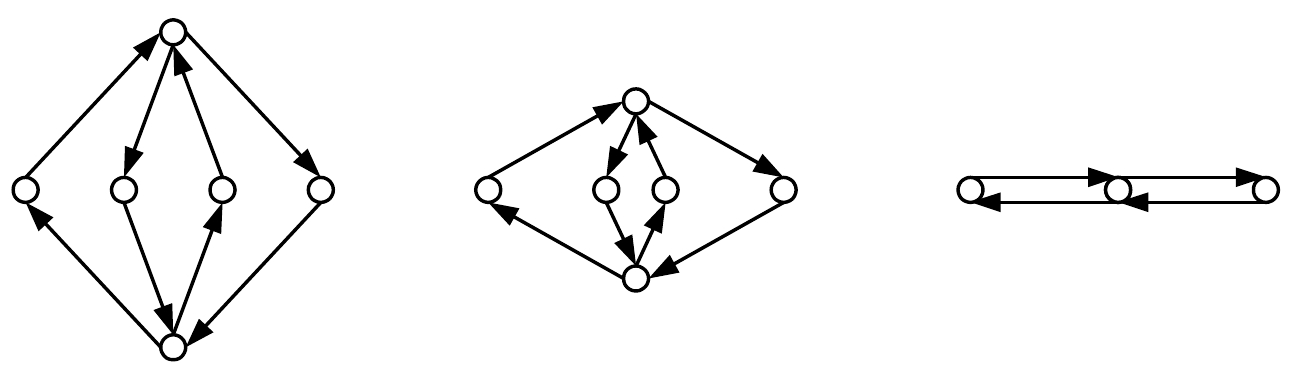}
    \caption{Local transformation of the BPS quiver as a full puncture $[1, 1, 1]$ becomes a simple puncture $[2, 1]$. Compare with Figure~\ref{21puncture}. }
    \label{21quiver}
\end{figure}

We can go further and obtain BPS quivers from BPS graphs for theories of class~$\CS$ with partial punctures. 
In Figure~\ref{21quiver}, we show the \emph{local} transformation of a BPS quiver corresponding to the degeneration of a full puncture $[1,1,1]$ to a simple puncture $[2,1]$ as in Figure~\ref{21puncture}. 
The central 4-cycle in the initial quiver collapses, and the four nodes connected to it merge into a single node. The final quiver has three nodes connected by pairs of oppositely-oriented arrows.

\begin{figure} 
    \centering
    \includegraphics[width=0.99\textwidth]{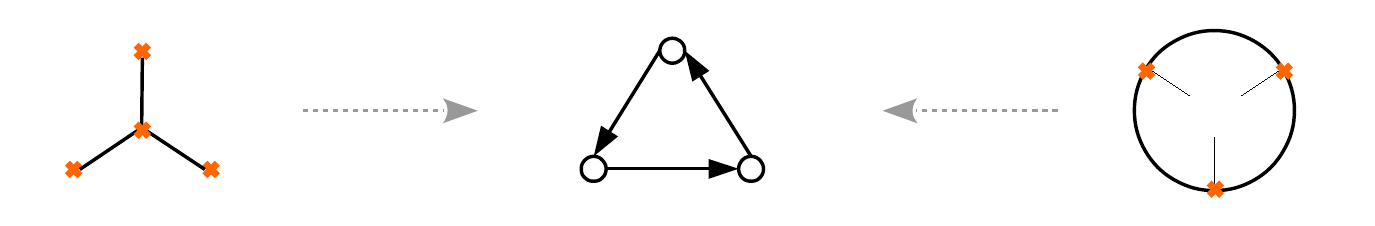}
    \caption{Two different BPS graphs can give the same BPS quiver (thin lines in the BPS graph on the right indicate empty edges).}
    \label{Graphs2quiver}
\end{figure}

In general the map between BPS graphs and BPS quivers cannot be inverted, because different BPS graphs can lead the same BPS quiver (see for example the situation shown in Figure~\ref{Graphs2quiver}). 
This is to be expected given that the topology of a BPS graph contains more information than the corresponding BPS quiver, such as which elementary webs are adjacent.
However, it may be that once the BPS quiver is equipped with a suitable superpotential (see Section~\ref{subsec.superpotential}), all ambiguities can be resolved. There would then be a one-to-one correspondence between topological BPS graphs and BPS quivers with superpotentials
(this is analogous to the relation between brane tilings and quivers with potentials~\cite{Franco:2005rj}).

\subsection{Elementary moves and mutations}\label{subsecMutations}

We now comment on the quiver interpretation of the elementary local transformations of BPS graphs discussed in Section~\ref{subsecMoves}. 

\begin{figure}
\centering
\begin{minipage}{.4\textwidth}
\centering
\includegraphics[width=\linewidth]{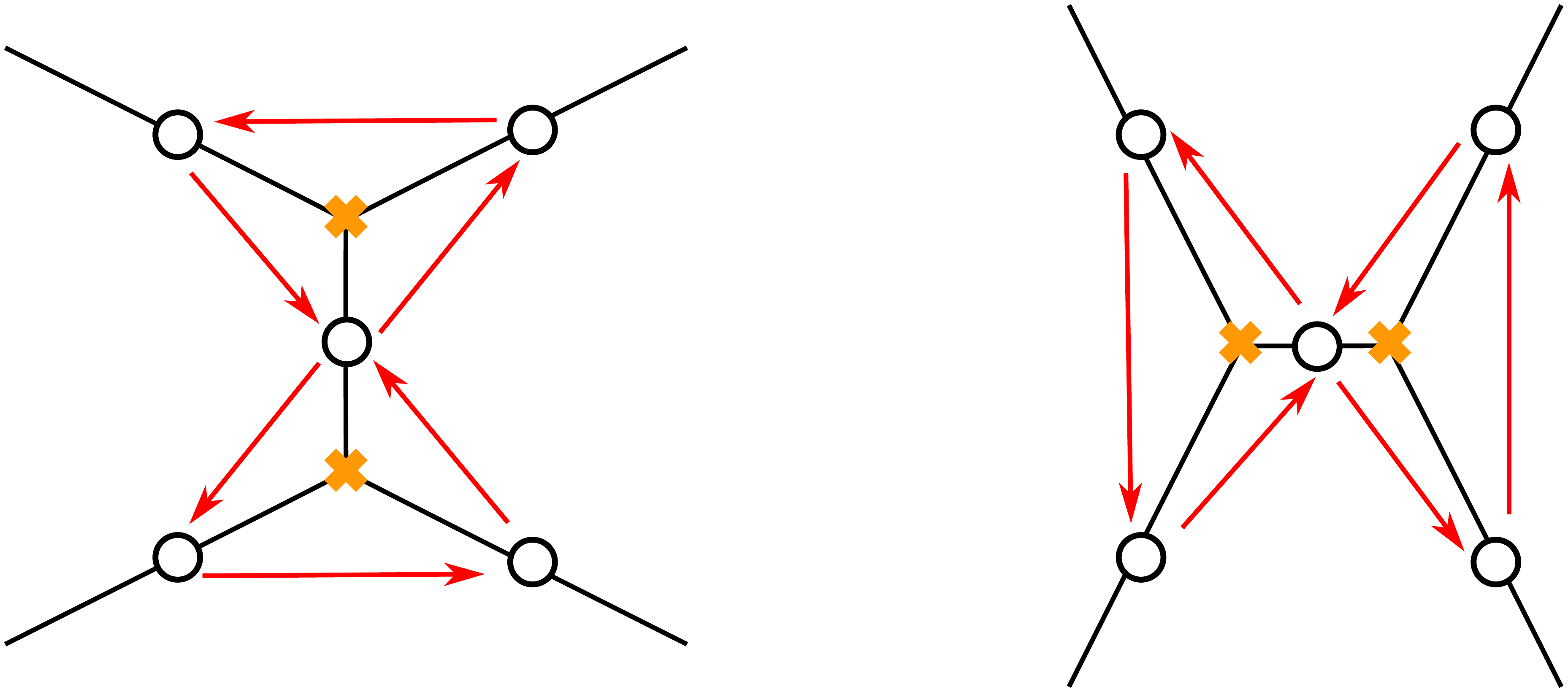}
\end{minipage}\hfill\vline\hfill
\begin{minipage}{.5\textwidth}
\centering
\includegraphics[width=\linewidth]{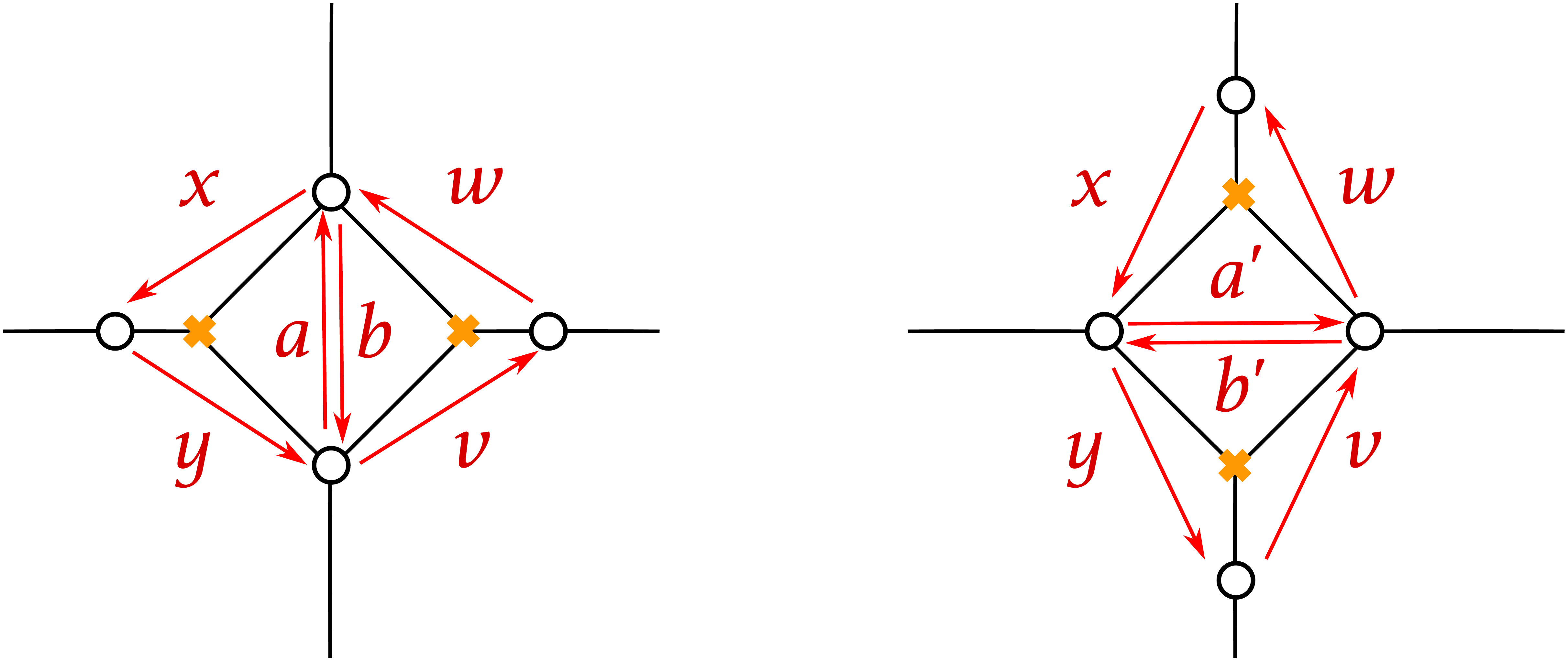}
\end{minipage}
    \caption{\emph{Left}: The flip move corresponds to a mutation of the BPS quiver. \emph{Right}: The cootie move rotates a pair of oppositely-oriented arrows.}
    \label{fig:quiver-mutation}
\end{figure}

As reviewed above, a quiver mutation is induced by crossing a {wall of the second kind}, when some central charge exits the positive half-plane. 
In the context of BPS graphs, since the phases of all central charges are fixed, the only way this can happen is that the central charge of a BPS state goes to zero. This is precisely what happens when two branch points collide! Indeed, at a topological level, the flip move (Figure~\ref{fig:flip}) corresponds to a mutation of the quiver, as shown on the left of Figure \ref{fig:quiver-mutation}.
The change~\eqref{eq:mutation} in the charges of elementary webs agrees perfectly with the change~\eqref{mutationg0} in the charges associated with the nodes.

The second type of elementary transformations of BPS graphs is the cootie move, which corresponds to the operation on quivers shown on the right of Figure~\ref{fig:quiver-mutation}. 
As we will see in Section~\ref{subsec.superpotential}, the quiver superpotential encoded by a BPS graph is always strong enough to cancel the two opposite arrows in the center, so that the cootie move is effectively trivial at the level of the BPS quiver.

\subsection{Superpotentials from BPS graphs}\label{subsec.superpotential}

Given a BPS graph with a set of branch points $\{b\}$ and a set of faces $\{F\}$, we propose that the quiver superpotential is given by
\be\label{eq:superpotential}
	W = \sum_{b} C_b - \sum_{F} C_F ,
\ee
where $C_b$ and $C_F$ are cycles in the quiver path-algebra obtained by composing the arrows which circle around the branch point $b$ and around the face $F$.

For theories with full punctures, $W$ is equivalent to the canonical quiver superpotential proposed by Goncharov in equation~(43) of~\cite{2016arXiv160705228G} (recall the duality between BPS graphs and ideal bipartite graphs discussed in Section~\ref{subsec:ideal-webs}).
It is also equivalent to previous definitions of quiver superpotentials~\cite{2007arXiv0704.0649D, 2009arXiv0904.0676D, 2008arXiv0803.1328L, Alim:2011ae, Alim:2011kw}.
However, we conjecture that our proposal for $W$ also extends to the theories with partial punctures discussed in Section~\ref{PartialP}. 

Formula (\ref{eq:superpotential}) for the superpotential passes some nontrivial consistency checks.
First of all, the superpotential behaves correctly under flips of the BPS graph. 
This follows from the fact that a flip is dual to a mutation of the quiver, and that Goncharov's formula is consistent with mutations~\cite{2016arXiv160705228G}.
Moreover, the superpotential derived from BPS graphs for theories with partial punctures behaves correctly under a flip. 
The reason is that the change of the superpotential does not depend on the global topology of the BPS graph, nor on the topology of the elementary webs which are attached to the edge. More details are provided in Appendix \ref{app:superpotential-flip}.

The behavior of the superpotential under a cootie move can be studied by considering the contribution from the branch points and faces depicted on the right of Figure \ref{fig:quiver-mutation}. Before the cootie move, formula~\eqref{eq:superpotential} gives the following superpotential:
\be\label{eq:W-before-cootie}
	W \supset a y x + b w v - a b \sim  wvyx  \;,
\ee
where $\sim$ denotes equivalence upon integrating out $b$, which does not appear in any other terms of $W$. 
After the cootie move, we have
\be
	W' \supset a' x w + b' v y - a' b' \sim  xwvy \;,
\ee
which is equivalent to (\ref{eq:W-before-cootie}) by cyclicity.

\begin{figure}
    \centering
        \includegraphics[width=.56\textwidth]{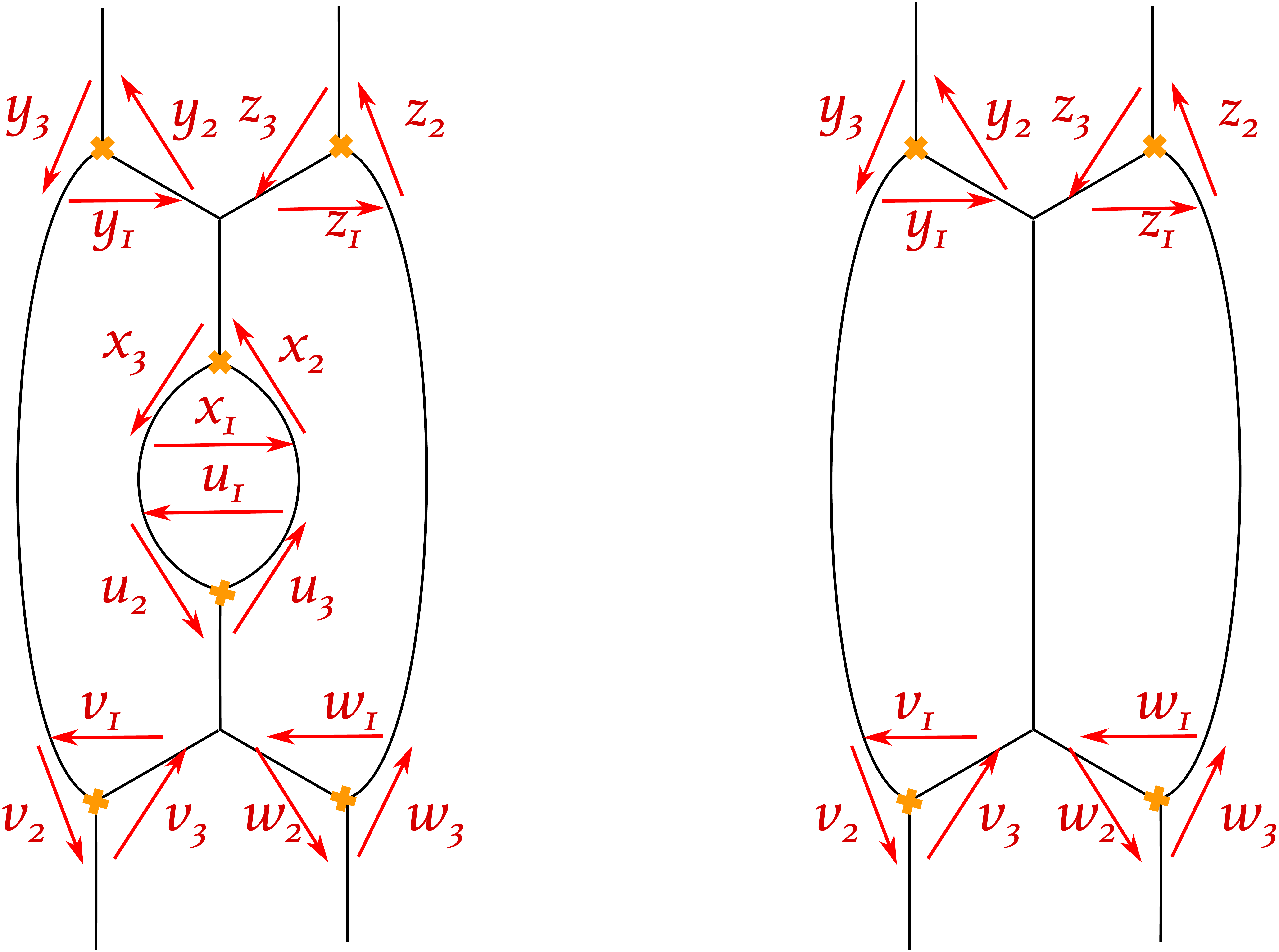}
       \caption{Transformation of the BPS quiver of the $T_3$ theory as a full puncture $[1,1,1]$ degenerates to a simple puncture $[2,1]$.}
    \label{fig:T3-21-quiver}
\end{figure}

Another nontrivial check of (\ref{eq:superpotential}) is to study its behavior under partial degeneration of punctures.
Let us consider the BPS graph of the $T_3$ theory as one of its punctures goes through a degeneration from $[1,1,1]$ to $[2,1]$, as shown in Figure \ref{21puncture}.
The change of the dual BPS quiver is clear from what happens to the elementary webs: the two edges circling the puncture disappear, hence we throw away their quiver nodes; on the other hand, the two Y-webs merge together into an H-web, and so we collapse their dual quiver nodes into a single node.
This results in the BPS quiver dual to the graph shown in Figure \ref{fig:T3-21-quiver}. The final quiver is composed of five nodes (the two upper edges are identified with the two lower ones), and all have vanishing intersection form (all arrows cancel each other). This is consistent with the fact that the sphere with two full punctures and a simple puncture should describe the theory of $9$ free hypermultiplets. The rank of the charge lattice is equal to $f=5$, which is the rank of the flavor symmetry $SU(3)\times SU(3)\times U(1)$.

In the quiver quantum mechanics, the identification of two nodes has a natural interpretation. 
Each node corresponds to a gauge group, and each arrow to a bi-fundamental field. 
Then the arrows connecting those nodes have been \emph{Higgsed}, and only a diagonal subgroup (the node remaining after mutual identification) is preserved.
Let us check whether this intuition is compatible with the construction of the superpotential.
Before Higgsing, the full superpotential of the $T_3$ quiver is
\be
\begin{split}
	W  = &\sum_{\alpha\in\{x,y,z,u,v,w\}}\alpha_1 \alpha_3 \alpha_2 \\
	&  - (x_1 u_1 + y_3 v_2 + z_2 w_3 + y_1 x_3 u_2 v_1 + z_1 w_1 u_3 x_2  + v_3 w_2 z_3 y_2) \;.
\end{split}
\ee
Then to identify the two nodes dual to the Y-webs, we must Higgs the arrows $x_2, x_3, u_2, u_3$, which amounts so setting the vacuum expectation values of these fields equal to 1. This gives
\be
	W'  = \sum_{\alpha\in\{y,z,v,w\}}\alpha_1 \alpha_3 \alpha_2 + x_1 + u_1%
	 - (x_1 u_1 + y_3 v_2 + z_2 w_3 + y_1  v_1 + z_1 w_1  + v_3 w_2 z_3 y_2)\,.
\ee
Noting that the fields $x_1, u_1$ are now decoupled from the rest (in fact, their arrows disappeared from the final quiver), we ignore the terms involving them.\footnote{
One subtlety is that although keeping the terms with $x_1$, $u_1$ would not affect the cohomology of moduli spaces of quiver quantum mechanics, integrating them out would appear to induce a shift of the superpotential, possibly breaking supersymmetry.
} 
What remains is exactly the superpotential obtained from (\ref{eq:superpotential}) for the BPS quiver dual to the BPS graph after the degeneration of the puncture.

From a physical perspective, our proposal for the superpotential should admit an interpretation in terms of brane instantons. 
Nodes of the quiver are identified with elementary webs, which correspond to M2-branes ending on M5-branes~\cite{Klemm:1996bj, Gaiotto:2009hg}.
Likewise, arrows of the quiver arise from bi-fundamental strings connecting nodes at their intersections, in particular at branch points of the BPS graph.
The quiver quantum mechanics describes the worldvolume dynamics of this system of branes. 
Each term $a_1\dots a_k \subset W$ can be thought as being generated by an instanton in the worldvolume theory, arising from a brane bounded by the elementary webs touched by the arrows $a_1,\dots, a_k$.
A more rigorous check of our conjectural formula (\ref{eq:superpotential}) would be to use it to compute BPS spectra, and match with the results of other methods.

\section{Future directions} \label{secApplications}

We hope that this paper has conveyed an initial sense of the usefulness of BPS graphs in the study of theories of class~$\CS$. 
Here we list some open questions and interesting further applications.

\paragraph{Roman locus:} 
Our definition of BPS graphs relies on the existence of the Roman locus~$\cR$, a maximal intersection of walls of marginal stability. Combining the proof of the existence of $\cR$ for (complete) $A_1$ theories with the $N$-lifting procedure of Section~\ref{Lifts} leads us to expect that $\cR$ indeed exists for all theories of class~$\CS$, at least for full punctures. However, the $N$-lifting construction of Roman loci itself is conjectural for high value of $N$. It would be important to find a general proof of the existence of $\cR$, perhaps based on some physical arguments. We have also seen that $\cR$ has an interesting topology, with several components separated by singular domain walls. One open question is whether there can be fully disconnected components of $\cR$.

\paragraph{BPS spectrum:} 
Spectral networks and BPS quivers both provide methods of computing the BPS spectrum in a given chamber on the Coulomb branch~$\cB$. In contrast, BPS graphs seem to only know about elementary BPS states, but not about BPS 
bound states.\footnote{Note that in defining BPS graphs from maximally degenerate spectral networks, we discarded a number of non-elementary webs. Based on the example of the $T_3$ theory, we expect them to correspond to flavor bound states.} In principle, it should be possible to move away from the Roman locus~$\cR$ in a controlled way (maybe based on symmetries of BPS graphs) that could be matched, on one side, to the dismantlement of the maximally degenerate spectral network into double walls appearing at separated critical phases; and on the other side, to the analysis of stable quiver representations. 
An interpretation of double walls corresponding to BPS bound states in terms of quiver representations was recently proposed in~\cite{Eager:2016yxd}.

Another approach to the derivation of the BPS spectrum from BPS graph could be a version of the ``mutation method''~\cite{Alim:2011kw}. The geometric realization of higher mutations in terms of flip and cootie moves could provide a better understanding of the phenomena at play, for examples those discussed in~\cite{Xie:2012gd}. 

Moreover, a description of BPS states in $N$-lifted theories was given in~\cite{Gaiotto:2012db}. They fell into three categories: ``triangle states'' (Y-webs), ``lifted dyons'' (edges, probably in some combination with Y-webs), and ``lifted flavor states'' (edges starting and ending on the same triangular array of branch points).
For the $T_3$ theory, a chamber was found with 24 BPS states, in apparent agreement with the result of the quiver mutation method in~\cite{Alim:2011kw}.
We believe that BPS graphs will be helpful in clarifying this approach.

\paragraph{Tinkertoys:} 
An approach to constructing an $A_1$ BPS graph on a general Riemann surface~$\CC$ is to glue together a collection of BPS graphs for $T_2$ according to a pants decomposition of~$\CC$. This could be achieved 
from the contraction of Fenchel-Nielsen networks~\cite{Hollands:2013qza} mentioned in Section~\ref{StrebelFN} (or alternatively as in Appendix~\ref{sec.pantsDec}). The generalization to higher rank would open up the possibility of constructing general $A_{N-1}$ BPS graphs from building-block BPS graphs on the ``tinkertoys'' of~\cite{Chacaltana:2010ks}. The classification of allowed partial punctures on tinkertoys should find an interpretation in terms of absorption patterns of pairs of branch points described in Section~\ref{PartialP}. This perspective can also be useful in understanding BPS graphs in the presence of more general partial punctures than those of type $[k,1,\ldots, 1]$. 

\paragraph{BPS monodromies:} 
BPS graphs capture wall-crossing invariant information about theories of class~$\CS$. 
The prototypical example of a wall-crossing invariant is the BPS monodromy of Kontsevich and Soibelman~\cite{Kontsevich:2008fj}, which is usually constructed from information about the BPS spectrum in a given chamber on the Coulomb branch. In contrast, as already shown in~\cite{Longhi:2016wtv}, BPS graphs can be used to obtain BPS monodromies in a manifestly wall-crossing invariant way. Finding a factorization into quantum dilogarithms then provides yet another approach to deriving the BPS spectrum. 
This construction of BPS monodromies should prove instrumental in exploring their conjectural relation to the Schur limit of superconformal indices~\cite{Iqbal:2012xm,Cecotti:2010fi,Buican:2015ina,Cordova:2015nma}. The generalization of this relation in the presence of various defects~\cite{Cordova:2016uwk, Cordova:2017ohl} should also find a nice interpretation in terms of BPS graphs, given the relevance of spectral networks for the study of framed and 2d-4d BPS states~\cite{Gaiotto:2010be,Gaiotto:2011tf}.
Closely related questions have been explored recently in~\cite{Cirafici:2017iju, Cirafici:2017wlw, Watanabe:2017bmi}.

\paragraph{Cluster atlas:} 
$N$-lifted BPS graphs provide cluster coordinate systems on the moduli space of rank-$N$ flat connections on~$\CC$. More precisely, the elementary webs lift to closed one-cycles on the $N$-fold cover~$\Sigma$, whose Abelian holonomies agree with Fock-Goncharov coordinates~\cite{FockGoncharovHigher}, as explained in detail in~\cite{Gaiotto:2012rg}. Applying flip and cootie moves on the BPS graph will then generate additional cluster coordinate systems.
This should be compared with the conjecture in~\cite{Gaiotto:2012rg} that spectral networks provide coordinate systems in the ``cluster atlas.''
While such cluster coordinates have been extensively studied for the case of surfaces decorated with complete flags (full punctures),
very little is known for the case with partial flags (partial punctures). This leaves a lot of room for exploration. It is for example interesting to ask whether the H-webs appearing around partial punctures can be related to invariant ratios constructed from partial flags.

\paragraph{Knot invariants:} 
Our results has potential applications to knot invariants.
Consider for example a sphere with $n$ regular punctures, 
and an action of the braid group element permuting the punctures.
This defines a knot inside the mapping torus for the $n$-punctured sphere.
We can then associate the cluster partition function of~\cite{Terashima:2013fg,Gang:2015wya}, which gives partition function for the complex Chern-Simons theory on the knot complement. 
When some of the punctures are partial, it should be possible to understand the braid group action in terms of a sequence of the $N$-flips described in Section~\ref{subsec:generalized-N-flips}.
The resulting partition function would likely be a new invariant of the knot (see~\cite{Gang:2015wya} for a more complete discussion). Thanks to the 3d/3d correspondence~\cite{Terashima:2011qi,Dimofte:2011ju}, this Chern-Simons partition function can also be regarded as the partition function of the associated 3d $\mathcal{N}=2$ theory.

\paragraph{Irregular Punctures:} 
It would be interesting to push the study of $N$-lifted BPS graphs on surfaces with irregular punctures and to consider their degenerations, as we did in Section~\ref{PartialP} for regular punctures. 
The resulting BPS quivers could then be compared to those of generalized Argyres-Douglas theories studied in~\cite{Xie:2012hs}.

\paragraph{Exact WKB analysis:} 
A spectral network essentially coincides with the concept of a Stokes graph
in the exact WKB analysis~\cite{Voros,AKT,DDP}.
Since the connection between WKB analysis and cluster algebra is known for the $A_1$ case~\cite{IwakiNakanishi1}, it is natural to expect that for $A_{N-1}$ with $N>2$, our BPS quivers and the associated cluster algebra structure should have direct counterparts in the WKB analysis, for a differential equation of order $N$ (see~\cite{BerkNevinsRoberts,HondaKawaiTakei}).

\paragraph{Dessins d'enfants:}
In~\cite{1998math.ph..11024M}, Strebel differentials were put in relation with Grothendieck's \emph{dessins d'enfants}. As explained
in \cite{Longhi:2016wtv} and in Section~\ref{StrebelFN}, $A_1$ BPS graphs are special cases of critical graphs for Strebel differentials. It would thus be promising to explore the relation between BPS graphs and \emph{dessins d'enfants}, and to make contact with their recent appearance in the study of theories of class~$\CS$~\cite{He:2015vua,Cecotti:2015qha} (see also the earlier works~\cite{Ashok:2006br,Ashok:2006du}). 
This would be particularly interesting for higher-rank $A_{N-1}$ BPS graphs, which can be viewed as generalizations of Strebel graphs, involving $k$-differentials $\phi_k$ with $k=2,\ldots ,N$.

\paragraph{Legendrian surfaces:}
Cubic planar graphs closely related to $A_1$ BPS graphs were recently shown to determine Legendrian surfaces~\cite{2016arXiv160904892T} (see also~\cite{2015arXiv151208942S}). A better understanding of this relation could lead to a physical interpretation in terms of calibrated M2-branes, and perhaps shed some light on the interpretation of the quiver superpotential~\eqref{eq:superpotential} in terms of instantons.

\acknowledgments
It is a great pleasure to thank Roger Casals, Clay C\'ordova, Tudor Dimofte, Michele del Zotto, Lotte Hollands, Greg Moore, Andy Neitzke, Daniel S.~Park, Boris Pioline, Mauricio Romo, and Kazuya Yonekura for enlightening discussions.
MG's work is supported by the Swiss National Science Foundation (project P300P2-158440).
The research of PL is supported by the grants ``Geometry and Physics'' and ``Exact Results in Gauge and String Theories'' from the Knut and Alice Wallenberg Foundation. 
PL would like to thank CERN and Kavli IPMU for hospitality during completion of this work.
The work of CYP is supported in part by the National Science Foundation under Grant No.~NSF PHY-140405.
CYP would like to thank Kavli IPMU, where part of this work was done, for hospitality and support.
MY is supported in part by the WPI Research Center Initiative (MEXT, Japan), by JSPS Program for Advancing Strategic International Networks to Accelerate the Circulation of Talented Researchers, by JSPS KAKENHI (15K17634), and by JSPS-NRF Joint Research Project.
MY would like to thank the Center for the Fundamental Laws of Nature, Harvard University, where
part of this work was done, for hospitality.

\appendix
 
\section{Identifying elementary webs}\label{AppWebs}

In this appendix, we address in some detail the distinction between the maximally degenerate spectral network ${\cal W}_c$ that appears on the Roman locus (at the critical phase $\vartheta_c$), and the corresponding BPS graph ${\cal G}$.

It may happen that ${\cal W}_c$ includes extra double walls, which should not be retained in~${\cal G}$. 
Let $\{p_i\}$  be the set of double walls appearing in ${\cal W}_c$.
There is a canonical construction of certain formal sums of double walls $L = \sum_i \alpha_i p_i$ with integer coefficients $\alpha_i$, which lift to closed loops on $\Sigma$.
Elementary webs are among these: for example a Y-web involves three edges $p_1+p_2+p_3$.
The construction of closed loops $\{L_k\}$ associated with ${\cal W}_c$ arises naturally from combinatorial properties of spectral networks, and uses the data of 2d-4d BPS states which is encoded by the walls.
However, the set $\{L_k\}$ constructed in this way is typically overcomplete, in the sense that some loops can be expressed as sums of other loops, with positive integer coefficients.
Therefore, only some of the $L_k$ need to be retained to form a positive integral basis for $\Gamma_+$ in (\ref{eq:half-lattice}). The BPS graph ${\cal G}$ is made of the edges $p_i$ which appear in this minimal subset of $L_k$.

The only remaining point to clarify is how the $L_k$ are constructed (details can be found in~\cite[Sec.~3]{Longhi:2016wtv}).
If we slightly resolve the critical phase to $ \vartheta_c\pm\epsilon$, the topology of the spectral network uniquely determines certain combinatorial soliton data  attached to each edge~\cite{Gaiotto:2012rg}. 
Resolving the double walls by going to $\vartheta_c+\epsilon$ determines a generating function $Q(p)$, encoding combinatorial 2d-4d BPS soliton data.
An important fact is that $Q(p)$ is entirely determined by the topology of the spectral network, as first defined in~\cite{Gaiotto:2012rg}.
Moreover, the functions $Q(p)$ factorize into certain universal factors~\cite{Longhi:2016wtv}:
\be
	Q(p) = \prod_i Q_i^{\alpha_i(p)}\,.
\ee
The $L_k$ are then constructed by lifting to $\Sigma$ specific formal linear combinations of the $p_i$, defined as follows:
\be
	L_k = \pi^{-1} \left(\sum_{i} \alpha_k(p_i)\, p_i\right)\,.
\ee

\begin{figure}
    \centering
        \includegraphics[width=0.5\textwidth]{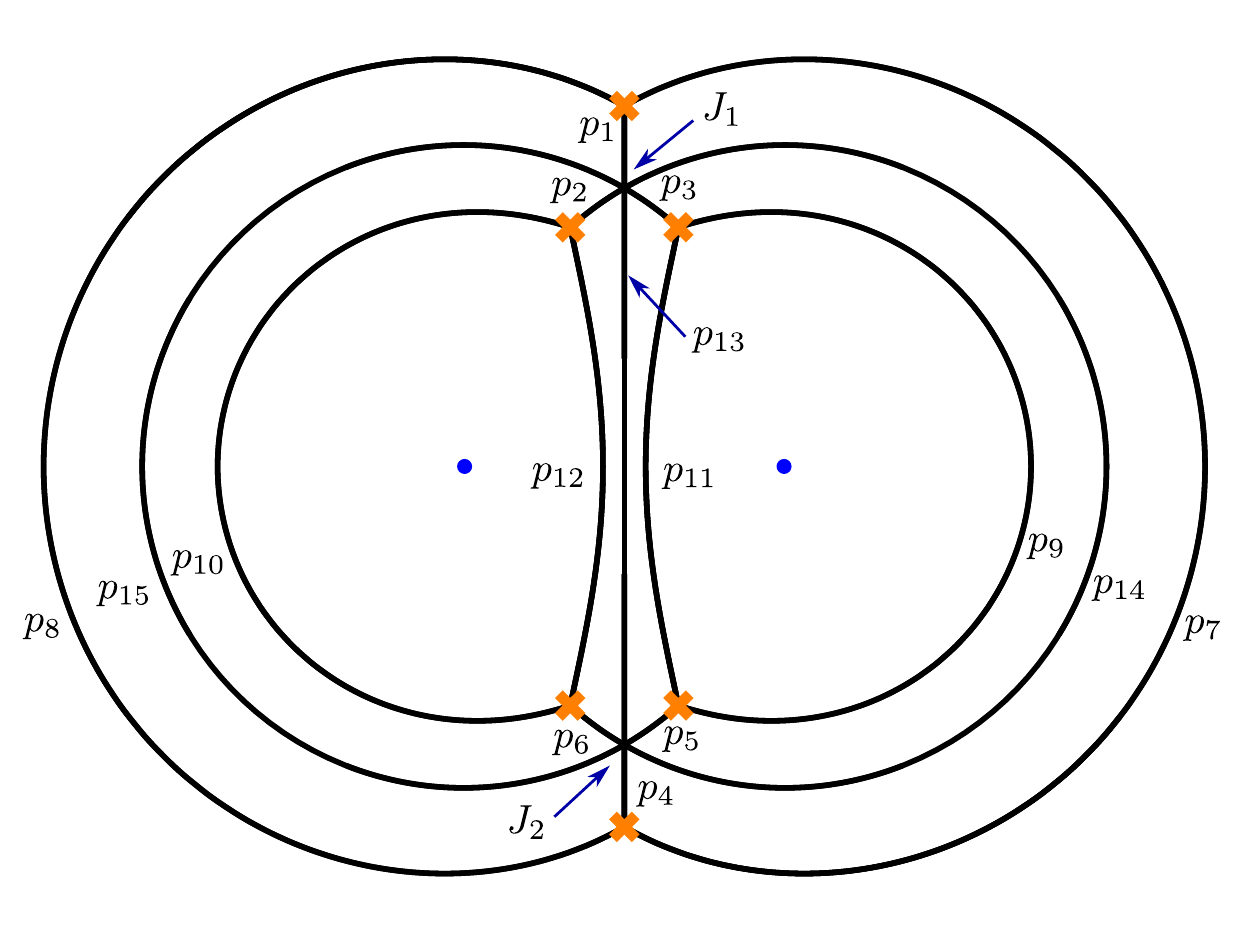}
    \caption{BPS graphs for the $T_3$ theory, with punctures at $z=\pm 1,\infty$.}
    \label{T3graphlabels}
\end{figure}

For example, in the $T_3$ theory, whose maximally degenerate spectral network is shown in Figure~\ref{T3graphlabels}, the combinatorics works as follows. 
The double walls $p_7,\dots, p_{12}$ correspond to a single factor each, respectively $Q_7,\dots, Q_{12}$.
Other double walls, which terminate on either joint of the BPS graph have more structure:  there are six factors naturally associated with each joint $Q_{{\rm odd}}^{J}, Q_{{\rm even}}^{J}, Q_{{\rm all}}^{J} $ and $ Q_{i}^{J}\,,\ i=1,2,3$, for each joint $J=J_1,J_2$. More precisely, these six factors appear naturally on the six walls ending at a joint.
At joint $J_1$ we have:
\be
\begin{split}
	& Q(p_1) = \frac{Q_{1}^{J_1}Q_{{\rm even}}^{J_1}}{Q_{{\rm all}}^{J_1}} \;,  \qquad %
	 Q(p_2) = \frac{Q_{3}^{J_1}Q_{{\rm even}}^{J_1}}{Q_{{\rm all}}^{J_1}} \;, \qquad %
	 Q(p_3) = \frac{Q_{2}^{J_1}Q_{{\rm even}}^{J_1}}{Q_{{\rm all}}^{J_1}} \;, \\
	& Q(p_{13}) = \frac{Q_{1}^{J_1}Q_{{\rm odd}}^{J_1}}{Q_{{\rm all}}^{J_1}} \;, \qquad %
	 Q(p_{14}) = \frac{Q_{3}^{J_1}Q_{{\rm odd}}^{J_1}}{Q_{{\rm all}}^{J_1}} \;, \qquad %
	 Q(p_{15}) = \frac{Q_{2}^{J_1}Q_{{\rm odd}}^{J_1}}{Q_{{\rm all}}^{J_1}} \;.
\end{split}
\ee
At joint $J_2$ instead we have the following decomposition:
\be
\begin{split}
	& Q(p_4) = \frac{Q_{1}^{J_2}Q_{{\rm even}}^{J_2}}{Q_{{\rm all}}^{J_2}}\;,  \qquad %
	 Q(p_5) = \frac{Q_{3}^{J_2}Q_{{\rm even}}^{J_2}}{Q_{{\rm all}}^{J_2}} \;, \qquad %
	 Q(p_6) = \frac{Q_{2}^{J_2}Q_{{\rm even}}^{J_2}}{Q_{{\rm all}}^{J_2}} \;, \\
	& Q(p_{13}) = \frac{Q_{1}^{J_2}Q_{{\rm odd}}^{J_2}}{Q_{{\rm all}}^{J_2}} \;, \qquad %
	 Q(p_{14}) = \frac{Q_{2}^{J_2}Q_{{\rm odd}}^{J_2}}{Q_{{\rm all}}^{J_2}} \;, \qquad %
	 Q(p_{15}) = \frac{Q_{3}^{J_2}Q_{{\rm odd}}^{J_2}}{Q_{{\rm all}}^{J_2}}\;.
\end{split}
\ee
After a somewhat lengthy computation, we find that $Q_1^{J_1} = Q_1^{J_2}$, $Q_2^{J_1} = Q_3^{J_2}$, and $Q_3^{J_1} = Q_2^{J_2}$. The factors $Q^J_{{\rm all}}$ can be dropped, since they correspond to 2d states. 

The $L_k$ are then given by
\begin{align}
	 L_i &= \pi^{-1}(p_i)\;, \qquad i=7,\dots, 12 \;,  &\\
	 L_{1} &= \pi^{-1}(p_1\cup p_2\cup p_3) \;, \qquad  & L_{2} &= \pi^{-1}(p_4\cup p_5\cup p_6) \;,  \\
	 L_{13} &= \pi^{-1}(p_1\cup p_{13}\cup p_4)\;,  \qquad & L_{14} &= \pi^{-1}(p_2\cup p_{14}\cup p_6)\;, \\
	  L_{15} &= \pi^{-1}(p_3\cup p_{15}\cup p_5)\;,   \qquad  & L_{{\rm odd}} &= \pi^{-1}( p_{13} \cup p_{14} \cup p_{15})\;.
\end{align}
Direct inspection shows that $L_{13,14,15}$ and $L_{{\rm odd}}$ are positive integral combinations of the remaining eight cycles. For instance, $[L_{13}] = [L_1] + [L_2] + [L_{11}] + [L_{12}]$, as homology cycles.
Therefore, the natural basis of elementary webs consists of the two Y-webs~$L_{1,2}$ together with the six edges~$L_{7,\ldots, 12}$ that connect pairs of branch points without running through joints, as shown on the right of Figure~\ref{T3graph}.

\section{Gluing and pants decomposition}
\label{sec.pantsDec}

A method for gluing BPS quivers together was given in Section~4.9 (and~6.2) of~\cite{Alim:2011kw}. A global $SU(2)$ symmetry is manifested in the BPS quiver as an $S_2$ symmetry acting on a pair of identical nodes, and gauging it can be described in two steps. The new gauge degrees of freedom are first added in the form of an $SU(2)$ quiver with two nodes. These two nodes are then coupled to one of the two identical nodes in the $S_2$ pair, while the other one is delete since it can now be generated as a bound state. Gluing two BPS quivers together is achieved by gauging the diagonal subgroup of the product of two global $SU(2)$ groups. 

We now describe an analogue procedure of ``gluing'' BPS graphs.
The idea is to construct a BPS graph from a pants decomposition of~$\CC$. We first find a collection of \emph{pants graphs} just like the one we obtained for the $T_2$ theory in Figure~\ref{T2graph3} or~\ref{T2graphFlipped}, which according to Strebel's theorem can be arranged to appear at the same critical phase. This clashes with our definition of BPS graphs because the pants graphs are mutually disconnected. We are indeed not on the Roman locus $\cR$ but rather on a ``Strebel locus,'' where the maximally degenerate spectral network is the disconnected critical graph for a Strebel differential. 
However, we expect that it should be possible to find a distinguished complementary collection of BPS states that precisely connect all the pants graphs.

\begin{figure}
    \centering
    \begin{subfigure}[c]{0.35\textwidth}
        \includegraphics[width=\textwidth]{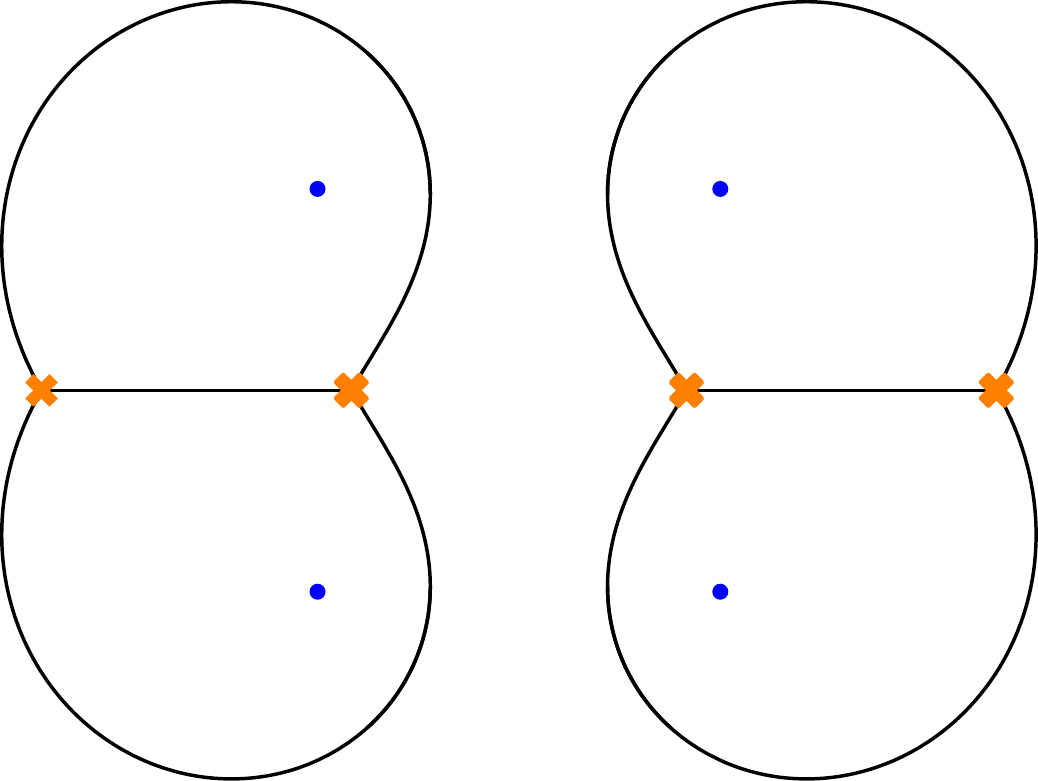}
        \caption{$\vartheta=0$}
    \end{subfigure}
    \hfill
    \begin{subfigure}[c]{0.35\textwidth}
        \includegraphics[width=\textwidth, trim=0 45 0 45]{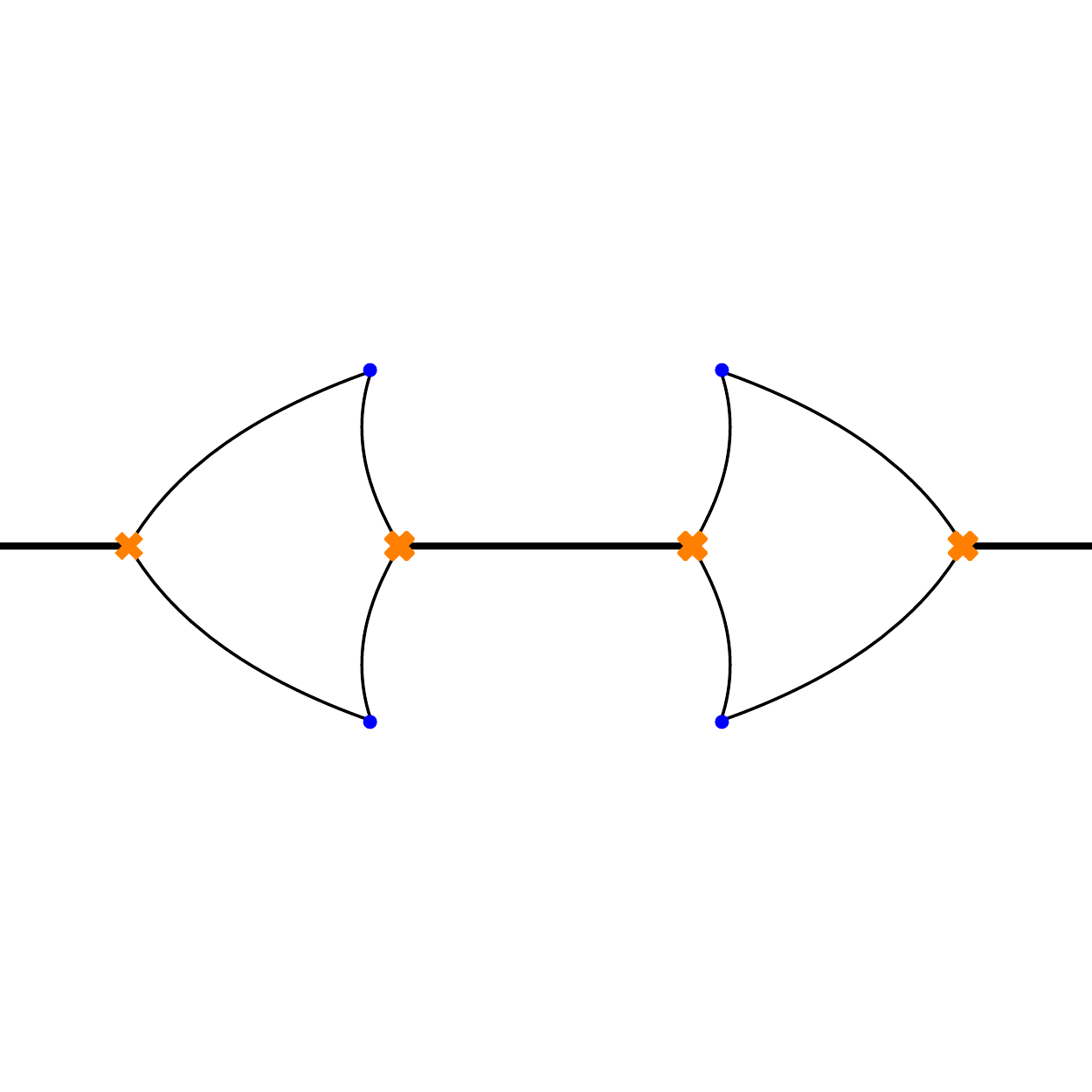}
        \caption{$\vartheta=\pi/2$}
    \end{subfigure}
    \caption{Pants graphs on a pants decomposition of the 4-punctured sphere (a). Additional BPS states used for the gluing (b).}
    \label{4Ppants}
\end{figure}

\begin{figure}
    \centering
        \includegraphics[width=0.4\textwidth]{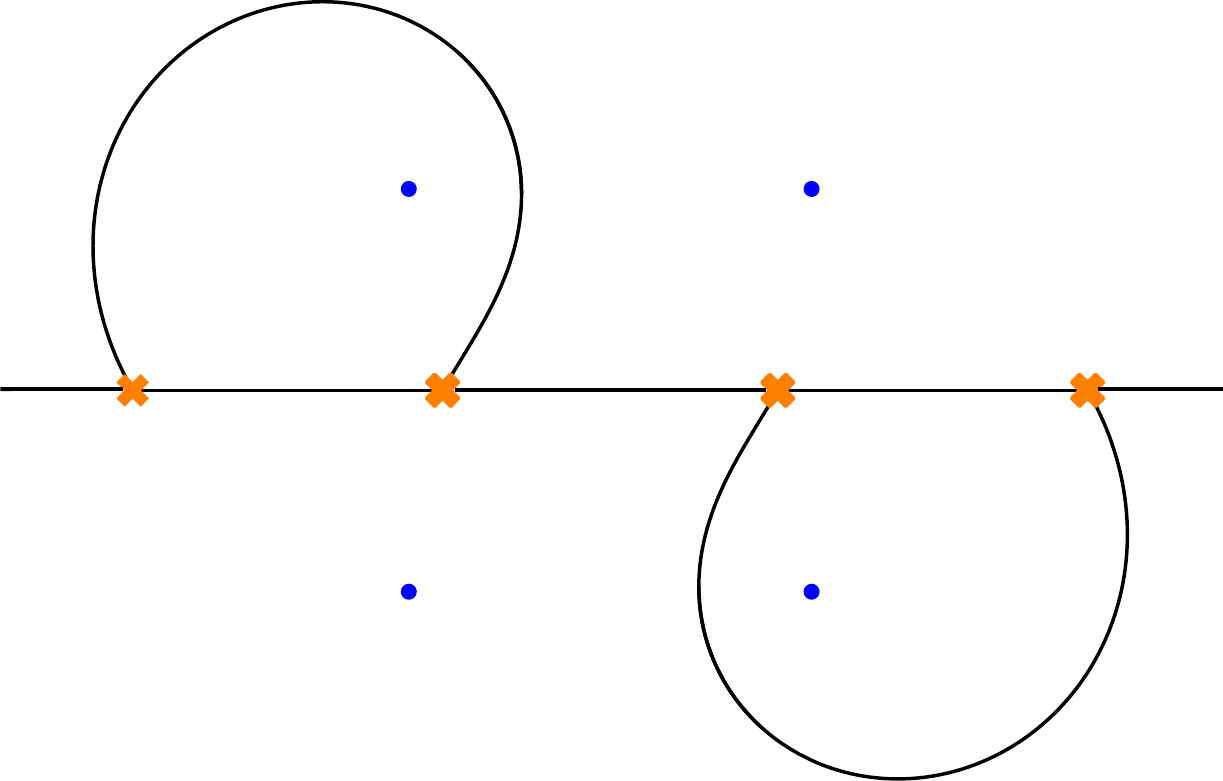}
    \caption{Glued BPS graph obtained by combining the BPS edges in Figure~\ref{4Ppants} and deleting two of them that are combinations of the others. Flipping the central edge produces the BPS graph in Figure~\ref{4pquiver}.}
    \label{4Ppantsquiver}
\end{figure}

As an illustration, let us study the example for the 4-punctured sphere. In Figure~\ref{4Ppants}, we show the two pants graphs that appear at $\vartheta =0$, and the additional two BPS states that appear at $\vartheta= \pi/2$ (we took $M_2^{a,b,c,d}=2$ and $u=20$). 
Together they provide 8 edges, but they are not all independent. Indeed, we can for example express the lower edge in the left pair of pants  and the upper edge in the right pair of pants as combinations of other edges. We are then left with a combined BPS graph that looks like in Figure~\ref{4Ppantsquiver}. Flipping the central edge produces the BPS graph shown in Figure~\ref{4pquiver}.

In conclusion, we see that the addition of two edges on the cylinder between pairs of pants corresponds to the addition of an $SU(2)$ quiver, and the removal of non-elementary edges on the pants graphs to the removal of bound states. 

\section{Superpotential after a flip}\label{app:superpotential-flip}

We examine how the superpotential (\ref{eq:superpotential}) changes under an elementary flip move of the BPS graph.
For $N$-lifted theories, the answer is expected to be consistent with usual Seiberg duality (equivalently, the prescription of Derksen-Weyman-Zelevinsky~\cite{2007arXiv0704.0649D, 2008arXiv0803.1328L}), by manifest duality of BPS graphs and Goncharov's ideal bipartite graphs.
One purpose of this appendix is to show that this process is \emph{local} on the BPS graph, and therefore the superpotential behaves correctly also in theories with partial punctures.

\begin{figure}[h!]
\centering
\includegraphics[width=.45\linewidth]{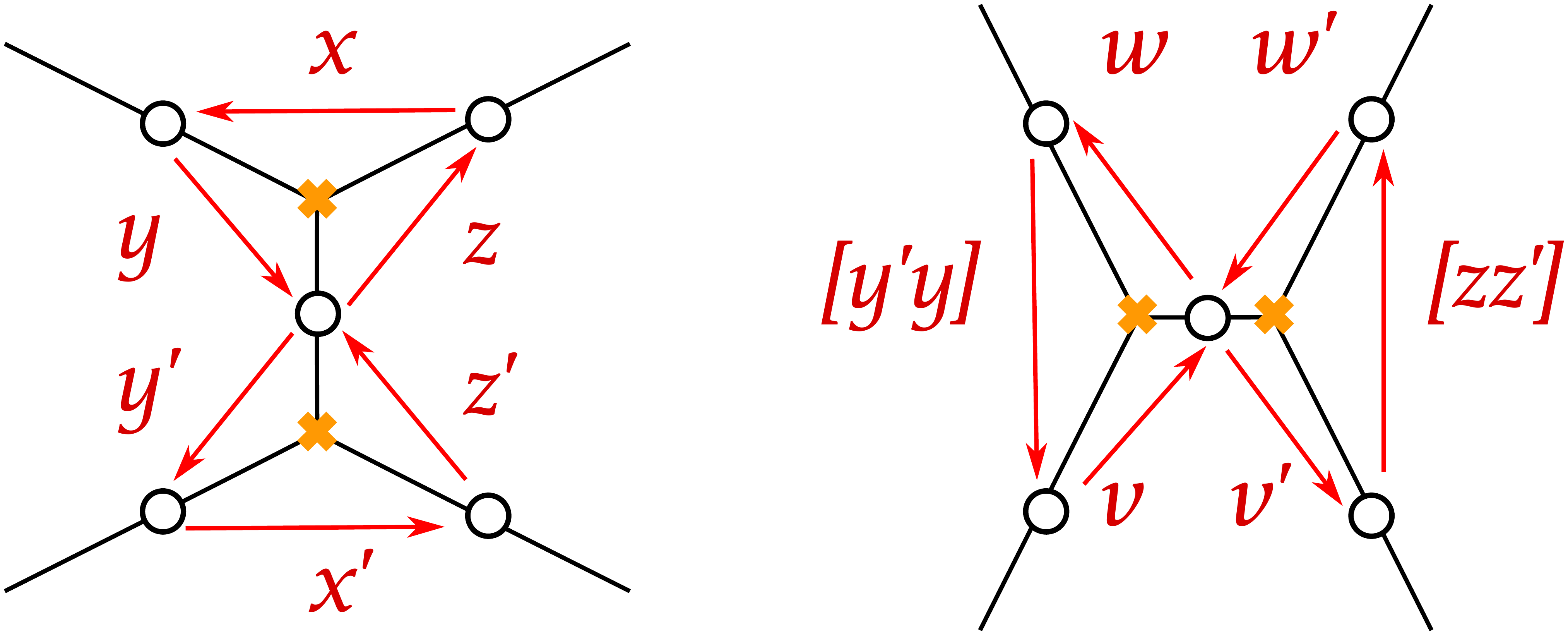}
    \caption{Flip of the BPS graph and mutation of the BPS quiver. }
    \label{fig:W-flip}
\end{figure}

Consider the flip shown in Figure \ref{fig:W-flip}, understood as a local region of a larger BPS graph.
The superpotential before the flip includes the terms
\be
	W \supset x z y + x' y' z' - (A  y' y + B z z' + C x + D x') \;, 
\ee
where $A,B,C,D$ stand for paths from the four faces bounded by this region of the BPS graph.
After the flip, the superpotential includes instead
\be\label{eq:W-after-flip}
	W' \supset [y'y]wv  + [zz'] v'w' - (A  [y' y] + B [z z'] + C ww' + D v'v)\;.
\ee
Note that the change in the superpotential doesn't depend in any way on the global topology of the BPS graph, nor on the topology of the elementary webs ending on the middle edge.

We shall now show that this is compatible with the mutation of a quiver with potential.
We follow the conventions of~\cite{2008arXiv0803.1328L} and identify
\be
	w=y^*\;, \quad%
	v={y'}{}^*\;, \quad%
	w'={z}{}^*\;, \quad%
	v'={z'}{}^*\;, \quad%
	x={[ww']}\;, \quad%
	x'=[v'v]\,.
\ee
The mutation for a quiver with potential involves a ``pre-mutation,'' which produces a superpotential $[W]+\Delta$ with
\be
\begin{split}
	[W] & = x [z y] + x' [y' z'] - (A  y' y + B z z' + C x + D x') \\
	& = x [z y] + x' [y' z'] - (A  y' y + B z z' + C [ww'] + D [v'v]) \\
	\Delta & =  y^* y'{}^* [y'y] + z'{}^* z^* [z z'] \\
	& = w v [y'y] + v' w' [z z'] .
\end{split}
\ee 
There is then a step to produce a reduced superpotential from this, which involves eliminating 2-cycles, in this case the two terms $x [z y] + x' [y' z']$. After reduction we see that $([W]+\Delta)_{\text{red}}$ coincides with (\ref{eq:W-after-flip}), as desired.

\bibliography{mybiblio}{}
\bibliographystyle{JHEP}

\end{document}